\documentclass[fleqn,usenatbib]{mnras}

\usepackage[T1]{fontenc}
\usepackage{newtxtext,newtxmath} 
\usepackage{subcaption} 
\DeclareRobustCommand{\VAN}[3]{#2}
\let\VANthebibliography\thebibliography
\def\thebibliography{\DeclareRobustCommand{\VAN}[3]{##3}\VANthebibliography}

\usepackage{graphicx}      
\usepackage{caption}       
\usepackage{subcaption}    
\usepackage[export]{adjustbox}
\usepackage{threeparttable}
\usepackage{booktabs}      
\usepackage{multirow}
\usepackage{array}
\usepackage{afterpage}
\usepackage{enumitem}  
\usepackage{soul}      
\usepackage{float}         

\title[Biases in stellar masses of high-z quasar hosts]{Quantifying biases in stellar masses of JWST high-$z$ quasar host galaxies caused by quasar subtraction}

\author[S. Berger et al.]{Sabrina Berger,$^{1,2,3}$\thanks{E-mail: sabrina.berger@student.unimelb.edu.au}
Madeline A. Marshall,$^{4}$
J. Stuart B. Wyithe,$^{2,3}$ 
Tiziana di Matteo$^{5}$, 
Yueying Ni$^{6}$, \newauthor Stephen M. Wilkins$^{7}$, and Minghao Yue$^{8}$
\\
$^{1}$School of Physics, University of Melbourne, Parkville, VIC 3010, Australia\\
$^{2}$Research School of Astronomy and Astrophysics, Australian National University, Canberra, ACT 2611, Australia\\
$^{3}$ARC Centre of Excellence for All Sky Astrophysics in 3 Dimensions (ASTRO 3D), Australia\\
$^{4}$Los Alamos National Laboratory, Los Alamos, NM 87545, USA\\
$^{5}$McWilliams Center for Cosmology, Department of Physics, Carnegie Mellon University, Pittsburgh, PA 15213, USA\\
$^{6}$Center for Astrophysics–Harvard $\&$ Smithsonian, 60 Garden Street, Cambridge, MA 02138, USA\\
$^{7}$Astronomy Centre, University of Sussex, Falmer, Brighton BN19QH, UK\\
$^{8}$ MIT Kavli Institute for Astrophysics and Space Research, 77 Massachusetts Ave., Cambridge, MA 02139, USA\\
}

\date{Accepted XXX. Received YYY; in original form ZZZ}

\pubyear{2025}

\begin{document}
\label{firstpage}
\pagerange{\pageref{firstpage}--\pageref{lastpage}}
\maketitle

\begin{abstract} 
JWST has enabled dozens of high-$z$ quasar host galaxy detections. Many of these observations imply galaxies with black holes that are overmassive compared to their low-$z$ counterparts. However, the bright quasar point source removal can cause significant biases in recovered host magnitudes and stellar mass measurements due to the degeneracy in host galaxy and quasar light. We develop a statistical method to disentangle the quasar host galaxy stellar mass measurements from observational biases during the point source removal assuming the PSF is modelled perfectly. We use the BlueTides simulation to generate mock images and perform point source removal on thousands of simulated high-$z$ quasar host galaxies, constructing corrected host magnitude posteriors. We find that removing a bright quasar in JWST photometry tends to either correctly recover or modestly misestimate host magnitudes, with a maximum magnitude underestimate of 0.2 mag. With our corrected magnitude posteriors, we perform SED fitting on each quasar host galaxy and compare the stellar mass measurement before and after the correction. We find that stellar mass estimates are generally robust, or misestimated by <0.3 dex. We also find that the stellar masses of a subset of hosts (J0844-0132, J0911+0152, and J1146-0005) remain unconstrained, as key photometric bands provide only flux upper limits. Accounting for observational biases does not resolve the apparent mismatch between black hole and host galaxy growth at high-$z$, where some quasars appear to host overmassive black holes while others reside in relatively massive galaxies.
\end{abstract}




\section{Introduction}
In the age of JWST, the high-$z$ Universe has presented many challenges to our understanding of galaxy evolution built from observations of the low-z Universe. One of the many issues that has emerged is the relationship between black holes with masses greater than $10^5 M_{\odot}$ and the galaxies in which they reside. These are supermassive black holes (SMBHs) that are found in the center of galaxies. Those that are actively accreting are known as active galactic nuclei (AGN), and the brightest SMBHs in the AGN population are quasars. Quasars are the subset of AGN with bolometric luminosities greater than approximately $10^{10} L_{\odot}$.

One of the key difficulties in detecting the host galaxies of high-$z$ quasars in the rest frame UV-optical is that the quasar can far outshine the host galaxy by approximately 100 times or more \citep{Mechtley_2012}. The ideal way to measure stellar mass of the host galaxy is by direct detection of the host's stellar emission. Before JWST was launched, attempts were made to detect the hosts of high-$z$ quasars with the Hubble Space Telescope (HST). However, only upper limits on flux were set for quasars at $z \gtrsim 4$ \citep{Mechtley_2012, McGreer_2014, 10.3847/1538-4357/abaa4c}. Even with careful point source subtraction of high-$z$ quasars, HST lacked the necessary near-infrared wavelength coverage, resolution, and sensitivity to make these detections. At $z \gtrsim
6$, the galaxy stellar emission in the rest-frame UV/optical is redshifted into the near infrared (NIR), which makes this detection a prime use case for JWST's NIRCam instrument. 

The launch of JWST has recently enabled the first photometric detections of high-$z$ quasar hosts, thanks to its significantly higher resolution compared to HST. The first host galaxy detections were made in \citet{10.48550/arxiv.2211.14329}, followed by a handful more detections in \citet{stone2023detection}, \citet{yue2023eiger}, and \citet{marshall2024ganifseigermerging}, and 9 more new detections (with 1 upper limit) in \citet{ding2025shellqsjwstunveilshostgalaxies}. In \citet{li2025dichotomynuclearhostgalaxy}, 30 new quasar host galaxy detections are presented, which we explore in future work.

Despite the quasar far outshining its host, high-$z$ host detections can be made with proper modeling and subtraction of the point spread function (PSF). The PSF varies with frequency and is extremely sensitive to spatial variation. Each high-$z$ quasar host detection has proved to need slightly different PSF modeling. The PSF is typically formed by combining multiple point sources (stars) within the same NIRCam image used in the host galaxy detection, ensuring the same exposure time for all sources. The PSF used is then either averaged from all observed PSF sources (as in \citealt{10.48550/arxiv.2211.14329}) or first optimized, subsampled, and then averaged (as in \citealt{10.48550/arxiv.2211.14329}). In some cases, the PSF star and quasar are positioned at the same position on the detector, to minimize the effects of spatial variation \citep{stone2023detection}. Even with very little PSF mismatch, part of the central galaxy light is inevitably lost due to the intrinsic degeneracy between the AGN and host components in photometric images. We summarize the current list of high-$z$ quasars with JWST detected hosts and black hole measurements from \citet{10.48550/arxiv.2211.14329}, \citet{yue2023eiger}, \citet{marshall2024ganifseigermerging}, and \citet{ding2025shellqsjwstunveilshostgalaxies} in Table \ref{tab:quasar_host_comprehensive} in chronological order. We only include hosts that were detected through the Bayesian detection method we use in the analysis for this work. Such detections have proved very difficult, and many host galaxy magnitudes are still lower limits \citep{stone2023undermassive, ding2025shellqsjwstunveilshostgalaxies, yue2023eiger} or are likely detections of quasar outflows \citep{liu2024fastoutflowhostgalaxy}. 

We can also estimate galaxy mass using dynamical measurements with $\rm [CI],~[CII]$ or $\rm CO$ emission from high-$z$ quasar host galaxies. This emission at high-$z$ has been detected in the sub-mm with instruments such as the Plateau de Bure Interferometer (PdBI)/Northern Extended Millimeter Array (NOEMA) from the Institut de Radioastronomie Millimetrique (IRAM) \citep{Venemans_2012, Wang_2013, Decarli_2022_a} or the Atacama Large Millimeter/sub-millimeter Array (ALMA) \citep{2016ApJ...816...37V, Decarli_2018, Yue_2021}. However, this only traces the dusty interstellar medium and gives measurements of the dynamical mass of the system, not the stellar mass. For example, \citet{marasco2025photometricvsdynamicalstellar} found a systematic offset of 0.22 dex in dynamical versus photometric stellar mass estimates in the nearby Universe. Complementary studies of dynamical mass and stellar mass measurements could be useful in constraining high-$z$ host stellar mass estimates, but they will almost certainly deviate more significantly from photometric measurements.  

Alongside a stellar mass measurement, measuring the galaxy's central SMBH is essential to place constraints on galaxy and black hole co-evolution. At high-$z$, black hole mass measurements are made by combining single epoch emission lines such as H$\beta$, $H\alpha$, Mg II, and CIV and continuum emission to estimate the velocity and radius of the quasar's broad line region (BLR) \citep{2002MNRAS.337..109M}. Although black hole measurements using the single epoch $H\beta$ line (typically thought to exceed the precision of other emission lines used in black hole mass estimates as seen in \citealt{2006ApJ...641..689V}) have a 0.43 dex systematic error, there tends to be more confidence in the black hole mass measurement compared to the stellar mass measurement. At $z\sim2$, the BLR of quasar SDSS J092034.17+065718.0 was recently resolved with the interferometric precision of GRAVITY+, and the black hole dynamical mass measurement was shown to be within less than half a dex of the empirical relations \citep{nature_bh_dynamical}. This is approximately within the systematic uncertainty of most empirical black hole mass relations \citep{Vestergaard_2006}, supporting the reliability of these mass measurements.

It has been postulated that the SMBH and its host galaxy evolve symbiotically. There is evidence for this in correlations found between galaxy stellar mass, black hole mass, stellar velocity dispersion, radius, and luminosity (e.g., \citealt{kormendy_1977, Magorrian_1998, 2000ApJ...539L...9F, Croton_2006, 2013ARA&A..51..511K, Munari_2013, Reines_2015, Schutte_2019, Danieli_2019, Sharma_2021}) mostly in the local universe. The black hole to stellar mass scaling relation: $M_{\bullet}$--$M_{*}$ or the black hole to galactic bulge mass: $M_{\bullet}$--$M_{\rm bulge}$ were once thought to be tightly correlated even up to high-$z$ accounting for some redshift evolution. For example, \citet{2003ApJ...595..614W} claimed a scaling proportional to $(1+z)^{3/2}$ due to quasar and supernova feedback. 

\begin{figure*}
    \centering
    \includegraphics[width=\linewidth]{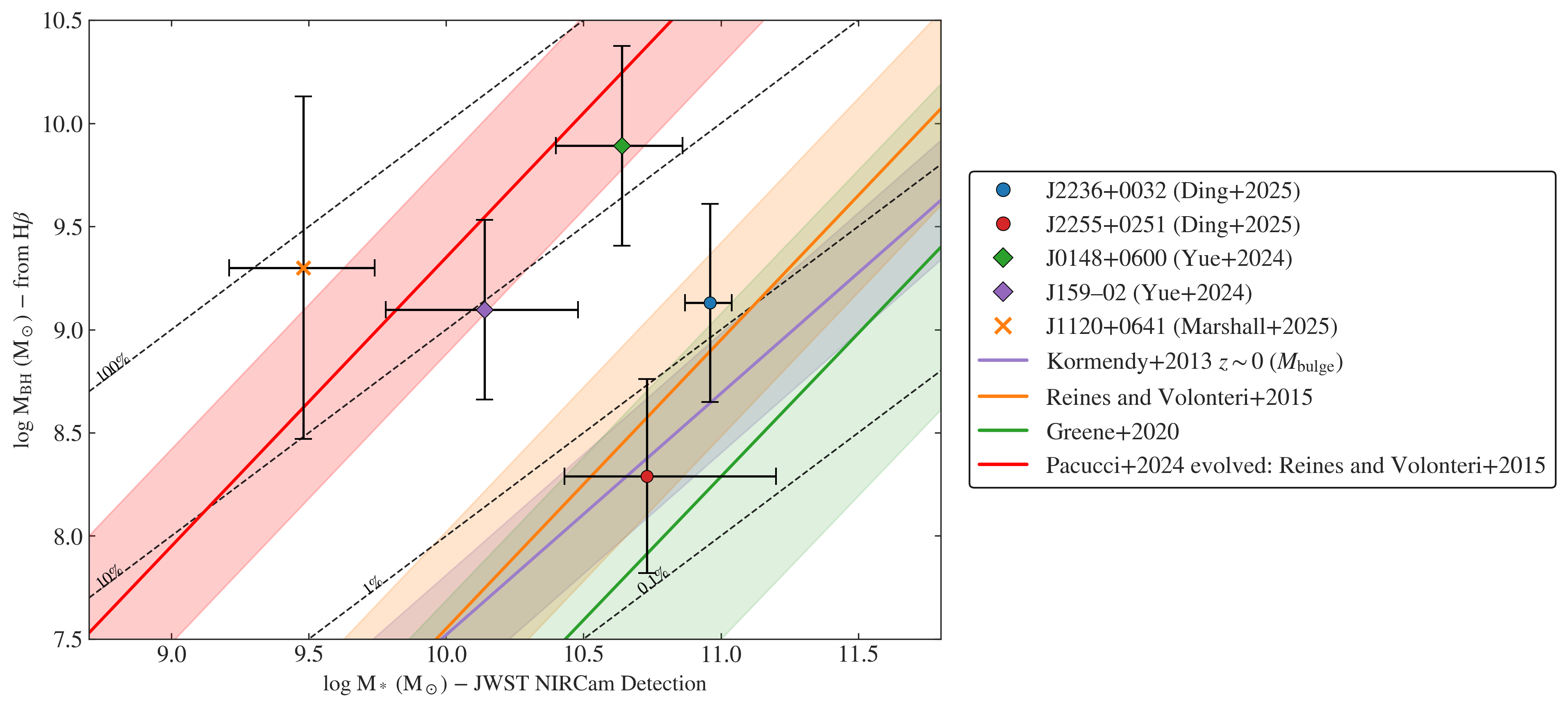}
    \caption{All of the high-$z$ quasars with both quasar host detections from JWST and black hole mass measurements explored in this work, alongside the \citet{2013ARA&A..51..511K}, \citet{Reines_2015}, \citet{2020ARA&A..58..257G}, and evolved \citet{Pacucci_2024} ($z \sim 6$) black hole to stellar mass relation. The \citet{2020ARA&A..58..257G} relation adds additional black holes to the samples used in \citet{2013ARA&A..51..511K} and \citet{Reines_2015}. The shading indicates $\sim$ 0.5 dex error in the scaling relation. The error bars also include the black hole empirical scatter of 0.43 dex. The tabulated version of the parameters from this plot can be seen in Table \ref{tab:quasar_host_comprehensive}.}
    \label{fig:bh_sm}
\end{figure*}

As more detections are made at higher-$z$, a discrepancy has emerged in the masses of the host galaxies and their central SMBHs expected from low-$z$ galactic scaling relations.  We plot the high-$z$ quasar host galaxy stellar mass measurements from \citet{10.48550/arxiv.2211.14329, yue2023eiger, marshall2024ganifseigermerging} over the low-z \citep{2013ARA&A..51..511K, Reines_2015, 2020ARA&A..58..257G} black hole to stellar mass relation in Figure \ref{fig:bh_sm}. We also include the high-$z$ evolution of \citet{Reines_2015} from \citet{Pacucci_2024}. The offset can be seen most significantly in the three quasars with detected hosts and measured black hole masses from the Emission-line galaxies and Intergalactic Gas in the Epoch of Reionization (EIGER) quasar sample (J0148+0600, J159-02, and J1120+0641, \citet{yue2023eiger, marshall2024ganifseigermerging}). The hosts of the low luminosity Subaru High-$z$ Exploration of Low-Luminosity Quasars \citep[SHELLQs;][]{Matsuoka_2016, Matsuoka_2018, Matsuoka_2019, Matsuoka_2019b, Matsuoka_2022} quasars detected with JWST in \citet{10.48550/arxiv.2211.14329} and \citet{ding2025shellqsjwstunveilshostgalaxies} are considered faint quasars with luminosities less than $10^{47}~\rm \rm ergs/s$ (J2236+0032 and
J2255+0251). As seen in Figure \ref{fig:bh_sm}, these two quasar host lie closer to the low-z $M_{\bullet}$--$M_{*}$ relation.

This discrepancy was seen even before the photometric detection of high-$z$ quasar hosts. \citet{Habouzit_2022} claimed that a discrepancy with the low-z relation would not be seen with lower luminosity quasars (faint quasars with $L_{\rm bol} = 10^{45}$--$10^{46} \rm ergs/s$), and that the problem was a selection bias.  However, the quasar host detection in \citet{stone2023detection} is an even dimmer SHELLQs quasar than those observed in \citet{10.48550/arxiv.2211.14329}. This quasar mass and its host mass are still far above the low-$z$ relation.  In addition to this, an increasingly large sample of mysterious highly obscured AGN at high-$z$, i.e., "little red dots (LRDs)" ($M_{\bullet} \sim 10^{6-8} M_{\odot}$) \citep{labbe2023uncovercandidateredactive, matthee2024littlereddotsabundant, 2024ApJ...968...38K, akins2024cosmosweboverabundancephysicalnature} have emerged. These also seem to lie significantly above the low-$z$ black hole to stellar mass relation \citep{Pacucci_2023}. 

The question of whether the black hole to stellar mass scaling relation evolves over cosmic time remains highly debated (e.g., see \citealt{Pacucci_2023, 2025ApJ...981...19L, 2025ApJ...984..122D, silverman2025shellqsjwstperspectiveintrinsicmass}). Adding further complexity, \citet{li2025dichotomynuclearhostgalaxy} recently reported a ``dichotomy'' among JWST quasar hosts, finding that the typical $M_{\bullet}/M_{*}$ ratio decreases from $\sim4.7\%$ in lower-luminosity quasars ($L_{5100} \lesssim 10^{45}\,\mathrm{erg\,s^{-1}}$) to $\sim2.3\%$ in higher-luminosity systems. Determining whether this diversity reflects genuine physical differences in early black hole and galaxy growth or instead arises from observational biases remains a key challenge. In this work, we test whether biases in quasar host galaxy recovery could drive these results by quantifying how the quasar removal process affects the inferred host galaxy properties.

Considering the effects of PSF mismatch and observational uncertainty due to point source removal are essential to ensure that the measured undermassive hosts are purely physical and not biased by observation. The PSF effects have already caused issues around the detection of high-$z$ quasar hosts. For example, when further PSF calibration and removal methods were performed for J1120+0641, \citet{marshall2024ganifseigermerging} detected a companion galaxy to a $z = 7.08$ quasar host detected initially in \citet{yue2023eiger}. The F115W host magnitude reported by \citet{marshall2024ganifseigermerging} is $25.94 \pm 0.20$ mag—more than two magnitudes fainter than the value obtained when the companion galaxy is excluded ($23.48 \pm 0.24$ mag; \citealt{yue2023eiger}). This resulted in a decrease in the original \citet{yue2023eiger} mass of $10^{9.81} M_{\odot}$ down to $10^{9.47} M_{\odot}$. Ensuring actual host detection and disentangling the observational bias of the point source removal from high-$z$ quasar hosts is the main goal of this work. 

Hydrodynamical simulations with box sizes large enough to allow for quasar formation from black hole seeds are essential to study the ongoing output of JWST. They provide the stellar and gas particles necessary for us to make both mock images and spectra that we can compare to JWST observations. Hydrodynamical simulations, including BlueTides \citep{Feng_2015, feng_2016}, Magneticum \citep{Steinborn_2015}, Illustris \citep{Vogelsberger_2014}, IllustrisTNG \citep{Pillepich_2017}, ASTRID \citep{Bird_2022}, and FLARES \citep{FLARES-I, FLARES-II, FLARES-XV, FLARES-XVII, FLARES-XVIII} have been used to study the high-$z$ black hole host galaxy co-evolution. Due to its large volume with a box size of $400h^{-1}~\rm cMpc^{3}$, BlueTides is one of the best large scale hydrodynamic simulations for studying rare high-$z$ quasars. As shown in \citet{2024MNRAS.530.4765B}, there are an array of readily available simulated analogs of the high-$z$ quasar hosts detected with JWST within BlueTides. BlueTides has also provided insight into the evolutionary history of quasars as in \citet{Tenneti_Matteo_Croft_Garcia_Feng_2017}, which described the first halos hosting high-$z$ quasars. In the BlueTides sister simulation, ASTRID, with a slightly smaller box size of $250 h^{-1} \rm Mpc$, \citet{dattathri2024redshiftevolutionmrmbhm} studied the redshift evolution of the $M_{\bullet}$--$M_{*}$ and compared it to a semi-analytic model. Populations such as LRDs could be more suitably explored in smaller scale simulations such as ASTRID. LRD analogs within ASTRID were found and studied in \citet{lachance2025propertieslittlereddot}. Hydrodynamical simulations remain key in helping decipher the formation of the full mass range of the first SMBHs. 

In this paper, we statistically quantify the observational bias due to PSF removal on high-$z$ quasar host stellar masses through Bayesian statistics. We divide our methods into observational (preparatory imaging and point source removal) and statistical methods (simulated posteriors). In Section \ref{sec:mock_methods}, we describe the BlueTides simulation and the mock imaging and point source removal pipeline. In Section \ref{sec:stat_methods}, we describe the construction of updated posteriors and stellar mass measurements for measured host galaxy magnitudes from an observed host galaxy magnitude measurement. We perform spectral energy distribution (SED) fitting on the median values from our updated photometric magnitude posteriors to compute corrected stellar masses. In Section \ref{sec:results}, we describe the effects on magnitude and stellar mass as interpreted from the updated posteriors for the magnitudes of detected high-$z$ quasar hosts. We also describe magnitude requirements for PSFs used in future quasar point source removal. In Section \ref{sec:discussion}, we discuss a comparison to similar works at varying redshifts, describe further potential stellar mass bias, and future extensions. In Section \ref{sec:conclusion}, we conclude with a summary of our results. We assume a $\Lambda \rm CDM$ cosmology with parameters from WMAP9 (\citealt{Hinshaw_2013}) as follows: $H_0= 100~\rm (km/s)/Mpc \times \textit{h} = 69.7~(km/s)/Mpc$, $\Omega_m=0.2814$, $\Omega_b=0.0464$, $\Omega_{\Lambda}=0.7186$, $\sigma_8=0.820$, and $n_s=0.9710$. We express all magnitudes in AB magnitude.

\section{Mock Observing Methods}
\label{sec:mock_methods}
\subsection{The BlueTides simulation}
The BlueTides simulation (\url{http://bluetides-project.org/}) is one of the largest hydrodynamic simulations to date with a volume of $(400/h \approx 577)^3 \rm cMpc^3$ \citep{Feng_2015, feng_2016} and has been run from $z = 99$ through $z=6.5$. The volume of BlueTides allows the simulation of the largest structures in the early universe. This enables the creation of analogs similar to the largest black holes and their host galaxies as seen in Figure \ref{fig:bh_sm}. BlueTides used the Smoothed Particle Hydrodynamics code MP-GADGET \citep{2018zndo...1451799F} and the AGN formation and feedback model from the MassiveBlack I and II simulations (growth by either gas accretion or mergers) \citep{Di_Matteo_2012, 10.1093/mnras/stv627} developed in \citet{2005Natur.433..604D}. The additional subgrid physics includes a modified multiphase star formation model \citep{2003MNRAS.339..289S, 2013MNRAS.436.3031V}, patchy reionization \citep{2013ApJ...776...81B}, and the type II supernovae wind feedback model used in Illustris \citep{Nelson_2015}.

BlueTides uses a black hole seed mass of $10^{5.8} M_{\odot}$ (also the mass resolution of the simulation), which is a rough prediction for the seed mass from the direct collapse black hole formation scenario \citep{Yue_2014}. Black holes are only seeded in halos with masses greater than $5 \times 10^{10}h^{-1}M_{\odot}$. There are $2 \times 7040^3$ particles in the BlueTides cube with a gravitational smoothing length of $\epsilon=1.5\rm h^{-1}kpc$. The BlueTides dark matter particle has a mass of $1.2 \times 10^{7}h^{-1}M_{\odot}$, and the mass of a gas particle is $2.36 \times 10^6 h^{-1} M_{\odot}$. BlueTides contains approximately 160,000 galaxies having a stellar mass greater than $10^{8}~M_{\odot}$ \citep{Huang_Matteo_Bhowmick_Feng_Ma_2018}. At z = 6.5, the largest black hole is $1.4 \times 10^{9}~M_{\odot}$, which is slightly more than an order of magnitude below the brightest quasar ever found ($\rm log (M_{BH} / M_{\odot}) = 10.24\pm0.02$ in \citealt{wolf2024}). We limit our BlueTides sample to 108,000 galaxies containing the highest black hole accretion rate. The galaxies in this sample have grown at least half a dex above the black hole seed mass, or have masses $>10^{6.3} M_{\odot}$.
\begin{figure*}
    \centering
    \includegraphics[width=\linewidth]{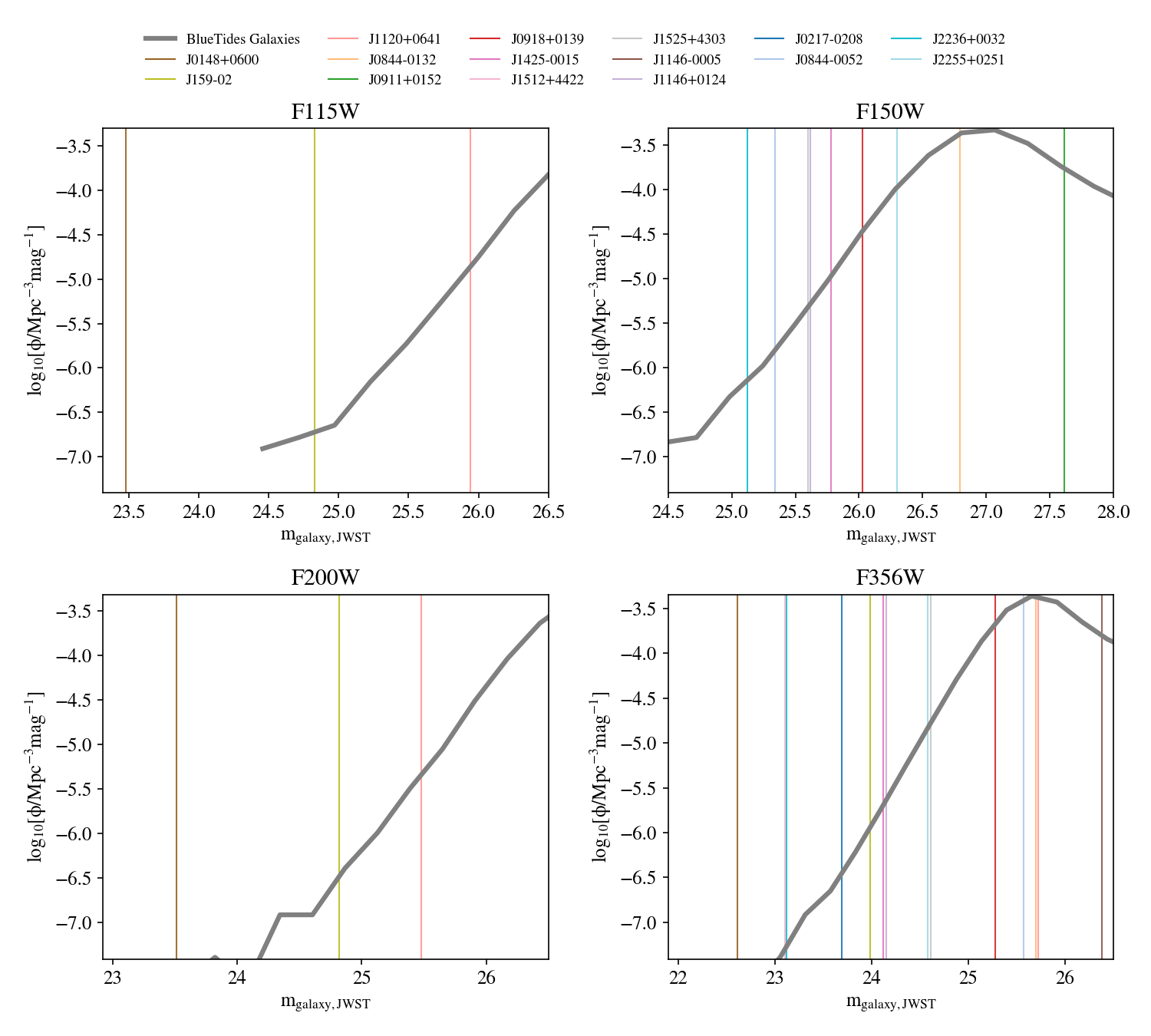}
    \caption{The photometric filter galaxy functions for BlueTides galaxies in the same filters as those for the high-$z$ quasars detected. The colored vertical lines correspond to the galaxy magnitude detected in those filters. See Table \ref{tab:quasar_host_comprehensive} for an overview of the observed quasar and host magnitudes for each filter. We are unable to generate a bright enough sample to include J0148+0600 in this work. \label{fig:photo_mags_with_quasars}}
\end{figure*}
We use the BlueTides z = 6.5 snapshot for the second time (first used and described in \citealt{2024MNRAS.530.4765B}). 

\begin{table*}
\centering
\caption{All observed and inferred parameters for high-$z$ quasars ($z \gtrsim 6$) with JWST host detections \textit{and} black hole mass measurements as of June 2025, using a Bayesian point source removal technique. Each reference describes the detections (or reanalysis, as in \citealt{marshall2024ganifseigermerging}) of the high-$z$ host associated with the quasar. Black hole masses do not include the 0.43 dex uncertainty from the scaling relations \citep{2006ApJ...641..689V}.}
\label{tab:quasar_host_comprehensive}
\begin{adjustbox}{max width=\textwidth}
\begin{threeparttable}
\renewcommand{\arraystretch}{1.3}
\small
\begin{tabular}{@{}lcccccccc@{}}
\toprule
\bsp
\textbf{References} & \textbf{Quasar} & \textbf{z} & \textbf{Filter /} \newline \textbf{Exposure Time} & \textbf{$\rm m_{QSO}$} & \textbf{$\rm m_{gal}$} & \textbf{$r_{1/2}$ [kpc]} & \textbf{$\log_{10}M_{*}$} \newline \textbf{[$M_{\odot}$]} & \textbf{$\log_{10}M_{\rm BH}$} \newline \textbf{[$M_{\odot}$]} \\
\midrule
\multirow{2}{*}{\citet{ding2025shellqsjwstunveilshostgalaxies}} & \multirow{2}{*}{J2236+0032} & \multirow{2}{*}{6.40} & F150W / 3100s & $22.74 \pm 0.01$ & $25.00 \pm 0.15$ & $0.5 \pm 0.2$ & \multirow{2}{*}{$10.96^{+0.08}_{-0.09}$} & \multirow{2}{*}{$9.13^{+0.05}_{-0.05}$} \\
 &  &  & F356W / 3100s & $22.00 \pm 0.13$ & $23.01 \pm 0.28$ & $0.7 \pm 0.1$ & & \\
\hline
\multirow{2}{*}{\citet{ding2025shellqsjwstunveilshostgalaxies}} & \multirow{2}{*}{J2255+0251} & \multirow{2}{*}{6.34} & F150W / 3100s & $22.89 \pm 0.02$ & $>26.3$\textsuperscript{*} & -- & \multirow{2}{*}{$10.73^{+0.47}_{-0.30}$} & \multirow{2}{*}{$8.29^{+0.04}_{-0.04}$} \\
 &  &  & F356W / 3100s & $22.15 \pm 0.03$ & $24.58 \pm 0.30$ & $1.5 \pm 1.1$ & & \\
\hline
\multirow{3}{*}{\citet{yue2023eiger}} & \multirow{3}{*}{J0148+0600} & \multirow{3}{*}{5.977} & F115W / 4381s & $19.522 \pm 0.003$ & $23.48 \pm 0.24$ & -- & \multirow{3}{*}{$10.64^{+0.22}_{-0.24}$} & \multirow{3}{*}{$9.892^{+0.053}_{-0.055}$} \\
 &  &  & F200W / 5959s & $18.912 \pm 0.001$ & $23.51 \pm 0.15$ & -- & & \\
 &  &  & F356W / 1578s & $19.109 \pm 0.003$ & $22.61 \pm 0.07$ & $2.23 \pm 0.11$ & & \\
\hline
\multirow{3}{*}{\citet{yue2023eiger}} & \multirow{3}{*}{J159-02} & \multirow{3}{*}{6.381} & F115W / 4381s & $20.146 \pm 0.003$ & $24.83 \pm 0.06$\textsuperscript{*} & -- & \multirow{3}{*}{$10.14^{+0.34}_{-0.36}$} & \multirow{3}{*}{$9.096^{+0.007}_{-0.005}$} \\
 &  &  & F200W / 5959s & $19.680 \pm 0.002$ & $24.82 \pm 0.23$ & -- & & \\
 &  &  & F356W / 1578s & $19.543 \pm 0.003$ & $23.98 \pm 0.16$ & $2.64 \pm 0.16$ & & \\
\hline
\multirow{3}{*}{\citet{yue2023eiger}; \citet{marshall2024ganifseigermerging}} & \multirow{3}{*}{J1120+0641} & \multirow{3}{*}{7.085} & F115W / 4381s & $20.366 \pm 0.003$ & $25.94 \pm 0.20$ & $1.93 \pm 0.03$ & \multirow{3}{*}{$9.48^{+0.26}_{-0.27}$} & \multirow{3}{*}{$9.3^{+0.4}_{-0.4}$} \\
 &  &  & F200W / 5959s & $19.886 \pm 0.002$ & $25.48 \pm 0.37$\textsuperscript{*} & -- & & \\
 &  &  & F356W / 1578s & $19.632 \pm 0.003$ & $25.72 \pm 0.47$\textsuperscript{*} & -- & & \\
 \hline
\multirow{2}{*}{\citet{ding2025shellqsjwstunveilshostgalaxies}} & \multirow{2}{*}{J0844-0052} & \multirow{2}{*}{6.01} & F150W / 3100s & $22.64 \pm 0.02$ & $25.34 \pm 0.25$ & $0.39 \pm 0.06$ & & \\
 & & & F356W / 3100s & $22.03 \pm 0.02$ & $25.57 \pm 0.51$ & $1.27 \pm 0.79$ & $9.49^{+0.52}_{-0.42}$ & \\
\hline
\multirow{2}{*}{\citet{ding2025shellqsjwstunveilshostgalaxies}} & \multirow{2}{*}{J0844-0132} & \multirow{2}{*}{6.07} & F150W / 3100s & $24.06 \pm 0.02$ & $>26.79$\textsuperscript{*} & & & \\
 & & & F356W / 3100s & $22.13 \pm 0.01$ & $25.70 \pm 0.34$ & $3.15 \pm 1.49$ & $10.03^{+0.53}_{-0.45}$ & \\
\hline
\multirow{2}{*}{\citet{ding2025shellqsjwstunveilshostgalaxies}} & \multirow{2}{*}{J0911+0152} & \multirow{2}{*}{6.64} & F150W / 3100s & $24.00 \pm 0.01$ & $>27.61$\textsuperscript{*} & & & \\
 & & & F356W / 3100s & $24.10 \pm 0.02$ & $26.56 \pm 0.30$ & $0.89 \pm 0.56$ & $9.72^{+0.49}_{-0.38}$ & \\
\hline
\multirow{2}{*}{\citet{ding2025shellqsjwstunveilshostgalaxies}} & \multirow{2}{*}{J0918+0139} & \multirow{2}{*}{6.70} & F150W / 3100s & $23.05 \pm 0.01$ & $>26.03$\textsuperscript{*} & & & \\
 & & & F356W / 3100s & $22.44 \pm 0.02$ & $25.28 \pm 0.25$ & $1.30 \pm 0.69$ & $10.21^{+0.52}_{-0.37}$ & \\
\hline
\multirow{2}{*}{\citet{ding2025shellqsjwstunveilshostgalaxies}} & \multirow{2}{*}{J1425-0015} & \multirow{2}{*}{6.13} & F150W / 3100s & $23.17 \pm 0.02$ & $25.78 \pm 0.29$ & $0.73 \pm 0.27$ & & \\
 & & & F356W / 3100s & $22.44 \pm 0.07$ & $24.12 \pm 0.29$ & $0.61 \pm 0.16$ & $10.54^{+0.36}_{-0.37}$ & \\
\hline
\multirow{2}{*}{\citet{ding2025shellqsjwstunveilshostgalaxies}} & \multirow{2}{*}{J1512+4422} & \multirow{2}{*}{6.62} & F150W / 3100s & $23.63 \pm 0.09$ & $24.07 \pm 0.17$ & $0.44 \pm 0.04$ & & \\
 & & & F356W / 3100s & $22.46 \pm 0.17$ & $23.10 \pm 0.28$ & $1.05 \pm 0.23$ & $10.57^{+0.29}_{-0.41}$ & \\
\hline
\multirow{2}{*}{\citet{ding2025shellqsjwstunveilshostgalaxies}} & \multirow{2}{*}{J1525+4303} & \multirow{2}{*}{6.30} & F150W / 3100s & $23.69 \pm 0.03$ & $25.60 \pm 0.21$ & $0.38 \pm 0.07$ & & \\
 & & & F356W / 3100s & $23.02 \pm 0.05$ & $24.61 \pm 0.24$ & $0.89 \pm 0.20$ & $10.00^{+0.30}_{-0.38}$ & \\
\hline
\multirow{2}{*}{\citet{ding2025shellqsjwstunveilshostgalaxies}} & \multirow{2}{*}{J1146-0005} & \multirow{2}{*}{6.37} & F150W / 3100s & $24.83 \pm 0.02$ & $>28.04$\textsuperscript{*} & & & \\
 & & & F356W / 3100s & $22.90 \pm 0.02$ & $>26.38$\textsuperscript{*} & $0.70 \pm 0.30$ & < 9.92 & \\
\hline
\multirow{2}{*}{\citet{ding2025shellqsjwstunveilshostgalaxies}} & \multirow{2}{*}{J1146+0124} & \multirow{2}{*}{6.43} & F150W / 3100s & $22.96 \pm 0.01$ & $25.62 \pm 0.20$ & $1.03 \pm 0.35$ & & \\
 & & & F356W / 3100s & $22.43 \pm 0.02$ & $24.15 \pm 0.16$ & $2.04 \pm 0.65$ & $10.38^{+0.22}_{-0.30}$ & \\
\hline
\multirow{2}{*}{\citet{ding2025shellqsjwstunveilshostgalaxies}} & \multirow{2}{*}{J0217-0208} & \multirow{2}{*}{6.49} & F150W / 3100s & $24.78 \pm 0.10$ & $24.29 \pm 0.13$ & $0.34 \pm 0.00$ & & \\
 & & & F356W / 3100s & $24.62 \pm 0.35$ & $23.69 \pm 0.18$ & $0.68 \pm 0.11$ & $10.06^{+0.28}_{-0.32}$ & \\

\bottomrule
\end{tabular}
\begin{tablenotes}\footnotesize
\item[*] Non-detections with lower magnitude limits.
\end{tablenotes}
\end{threeparttable}
\end{adjustbox}
\end{table*}

\subsection{Mock imaging}
\label{sec:methods_mock_imaging_specific}
Determining the effects of the point source removal on high-$z$ quasars requires building a large statistical sample of the process with simulated galaxies. We make mock images of all BlueTides galaxies in the JWST NIRCam filters used in \citet{10.48550/arxiv.2211.14329, ding2025shellqsjwstunveilshostgalaxies, yue2023eiger, marshall2024ganifseigermerging} (F115W, F150W, F200W, and F356W). Our sample is not limited to quasar hosts, given that the central black hole mass is not directly utilized in this work. We then identify samples of galaxies with similar apparent magnitudes to those measured with JWST. This allows us to form photometric filter specific galaxy luminosity functions, which we plot in these four filters in Figure \ref{fig:photo_mags_with_quasars} and overlay the detected quasar host magnitudes in that filter. As described in \citet{marshall_2022_dust} and \citet{2024MNRAS.530.4765B}, we use 6 $\times$ the half light radius of the galaxy as the aperture radius in the photometric mock image. This ensures accurate photometric magnitude calculations covering the entire galaxy. However, we discuss the difference and implications for differences in photometric and Sérsic magnitudes in Appendix \ref{appendix:sersic}.

With a hydrodynamical simulation such as BlueTides, we can make a mock image of any galaxy within the simulation. The mock imaging is a multistep process which involves both modeling the physical structure and instrumental effects such as exposure time, PSF, and background noise. We outline the full pipeline for adding physical and instrumental effects below.

\begin{enumerate}[labelwidth=*, leftmargin=4mm]
\item \textit{Stellar spectra} --- Using the \texttt{SynthObs} Python package \citep{synthobs}, we assign an SED to each star particle based on its age, mass, and metallicity using the Binary Population and Spectral Synthesis (BPASS) models \citep{Eldridge_2017, 2018MNRAS.479...75S} processed through \texttt{Cloudy} \citep{2017RMxAA..53..385F}. These models adopt a Salpeter initial mass function (IMF; \citealt{1955ApJ...121..161S}), extending up to initial stellar masses of $300~M_{\odot}$. We assume 90\% of Lyman-continuum photons escape. Dust attenuation is then applied to each star particle's spectra.
   \item \textit{Dust attenuation} -- Each star particle in BlueTides is assigned a dust model based on its age and surrounding environment. To simulate dust in the interstellar medium (ISM), we apply a dust attenuation to all star particles and calculate its optical depth ($\tau$) as follows:  

\begin{equation}  
    \tau_{\rm ISM} = \kappa \left(\frac{\lambda}{5500 \text{\AA}} \right)^{\gamma} \int_{d=0}^{d} \rho_{\rm metal}(x,y,z)~dz,  
\end{equation}  

where $\kappa = 10^{4.6}$, $\gamma = -1$, $\lambda$ is the wavelength of light passing through, and $\rho_{\rm metal}$ is the metallicity density along the line of sight, integrated up to the distance to the star, $d$ \citep{Wilkins_2017}. The parameters  $\gamma$ and $\kappa$ are calibrated to match the observed UV luminosity function at $z=7$ using data from \citet{Bouwens_2015} as seen in \citet{marshall_2022_dust}. Additionally for stars younger than 10 Myr, an exponential dust model accounts for the stellar natal cloud, with the optical depth computed as:  

\begin{equation}  
    \tau_{\rm birth~cloud} = 2 \left( \frac{Z}{Z_{\odot}} \right) \left( \frac{\lambda}{5500 \text{\AA}} \right)^{\gamma},  
\end{equation}  

where $Z$ is the metallicity of the birth cloud and $\gamma = -1$. 
\item \textit{Image generation} --- 
As shown in \citet{2024MNRAS.530.4765B}, adding both the physical and instrumental effects changes the raw input galaxy magnitude by approximately 0.4 mag. For all filters, we simulate a 25x25 kpc field around each galaxy and assign each stellar particle to a pixel in this grid. The pixel scale is 0.031 arcsec/pixel for the shorter wavelength filters: F115W, F150W, and F200W. For the longer wavelength filter, F356W, the native pixel scale is 0.063 arcsec/pixel, but we super sample to increase the effective resolution of our image to match the shorter wavelength filter resolution of 0.031 arcsec/pixel. At redshift z = 6.5, 1 pkpc corresponds to 0.18 arcsec or a few pixels. 

\item \textit{Adding a JWST-like quasar} --- When imaging galaxies with the quasar, we add a quasar with an apparent magnitude from Table \ref{tab:quasar_host_comprehensive}. In this way, we do not assume anything about the quasar spectra, and the quasar is a mock point source. This part of the pipeline is not self-consistent with the rest of the simulation as we do not use the black hole that exists in the specific BlueTides galaxy. However, we are more accurately able to constrain the exact effects of the PSF removal by pasting on a mock quasar with the exact magnitude of those observed with JWST. This also provides a much larger sample that allows for this statistical study.

\item \textit{Mock PSF convolution of the galaxy and quasar} ---
Filter-specific fluxes for each star particle are derived by convolving each SED with the corresponding JWST filter transmission curves. Finally, these stellar spectra maps and the quasar point source (if included) are convolved with mock NIRCam PSFs generated using \texttt{WebbPSF} \citep{2015ApJ...798...68G} to generate the final mock image. 

\item \textit{Exposure time and background noise} --- To simulate observational noise, we add a background derived from the 10\(\sigma\) flux limit of each JWST filter for the corresponding exposure time. The pixel specific background flux is calculated as:  

\begin{equation}  
\label{eq:background}  
    \rm background (\rm pixel) =  N(0,  (F_{\rm limit}/10)),  
\end{equation}  

where \( F_{\rm limit} \) is the 10\(\sigma\) aperture flux limit (from \citealt{2017JATIS...3c5001G} using the 2024 ETC v4 version), \( A_{\rm aperture} \) is the JWST aperture area ($2.5''$), and \( N \) is a Gaussian distribution. We simulate exposure time by adding shot noise to each image (number of counts = $\rm t_{int} \times flux(\rm pixel)$) as follows:
\begin{equation}
    \label{eq:shot}  
    \rm flux(\rm pixel,~t_{int}) =   Pois(t_{int} \times flux(\rm pixel)) / t_{int},
\end{equation}
where $\rm t_{int}$ is the exposure time. This ensures a realistic noise profile for testing point source removal and host galaxy recovery. 
\end{enumerate}
\subsection{Mock PSF and IVM Used in PSF Removal}
\label{appendix:psf}

The recovered JWST host-galaxy magnitude is highly sensitive to the PSF adopted in the PSF subtraction procedure (we use \textit{psfmc} as seen in Section~\ref{sec:psf_removal}). In addition to the PSF image itself, \textit{psfmc} requires an inverse variance map (IVM) associated with the PSF in order to account for uncertainties in the PSF model during the quasar removal.

In this work, we adopt an idealized mock PSF in order to isolate biases arising specifically from the quasar removal process. We therefore do not attempt to model mismatches between the true and assumed PSF. To avoid introducing additional systematics, the PSF used to generate the mock images is identical to the PSF used in the removal step, ensuring that any measured bias arises from the quasar subtraction itself. The mock PSFs are generated with the same magnitude (17\,mag) and matched integration time as the quasar host under consideration. Each mock PSF is constructed following the procedure described in Section~\ref{sec:methods_mock_imaging_specific}.

The IVM is a function of the background variance term described in Section~\ref{sec:methods_mock_imaging_specific} and an additional variance term near the PSF center. The background variance is modeled as $(\sigma_{\mathrm{bkg}} \times ss)^2$, where $\sigma_{\mathrm{bkg}}$ is the standard deviation of the background we add ($F_{\rm limit}/10$) and $ss$ is the supersampling factor used in the modeling. We model the additional central variance of the PSF as a two-dimensional Gaussian of the form

\begin{equation}
G(x,y) = A \exp\left[
-\left(
\frac{(x-x_0)^2}{2\sigma_x^2} +
\frac{(y-y_0)^2}{2\sigma_y^2}
\right)
\right],
\end{equation}
where $A = F_{\mathrm{PSF,max}}/5000$ is the amplitude, $(x_0, y_0)$ is the PSF center, and $\sigma_x = \sigma_y = 1.5$ pixels. This Gaussian term increases the modeled variance near the quasar position, downweighting central pixels where PSF subtraction residuals are expected to be largest. The inverse variance map (IVM) is therefore defined as

\begin{equation}
\label{eq:IVM}
\mathrm{IVM}(x,y) =
\left[(\sigma_{\mathrm{bkg}}\times  ss)^2 + G(x,y)\right]^{-1}.
\end{equation}

\begin{figure*}
    \centering
    \includegraphics[width=0.8\linewidth]{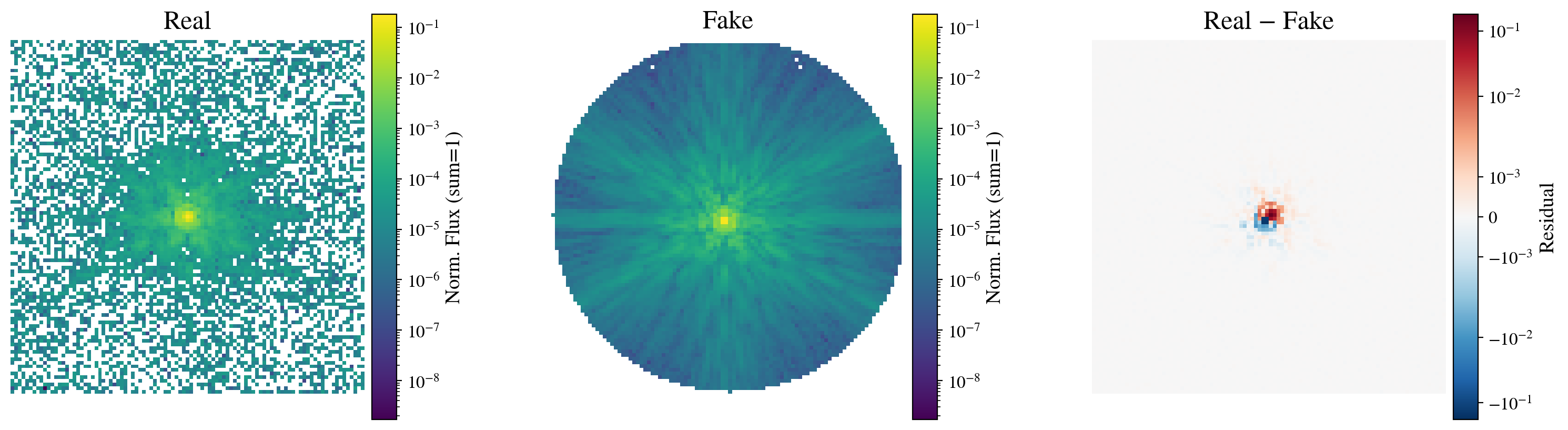}
    \caption{Comparison between a \textbf{real} JWST F150W PSF (left) and the \textbf{fake/mock} JWST F150W PSF used in this work (middle) in the F150W filter, with residuals shown in the right panel. All PSFs are normalized to sum to unit total flux.}
    \label{fig:psf_ivm_comparison_f150w}
\end{figure*}

\begin{figure*}
    \centering
    \includegraphics[width=0.8\linewidth]{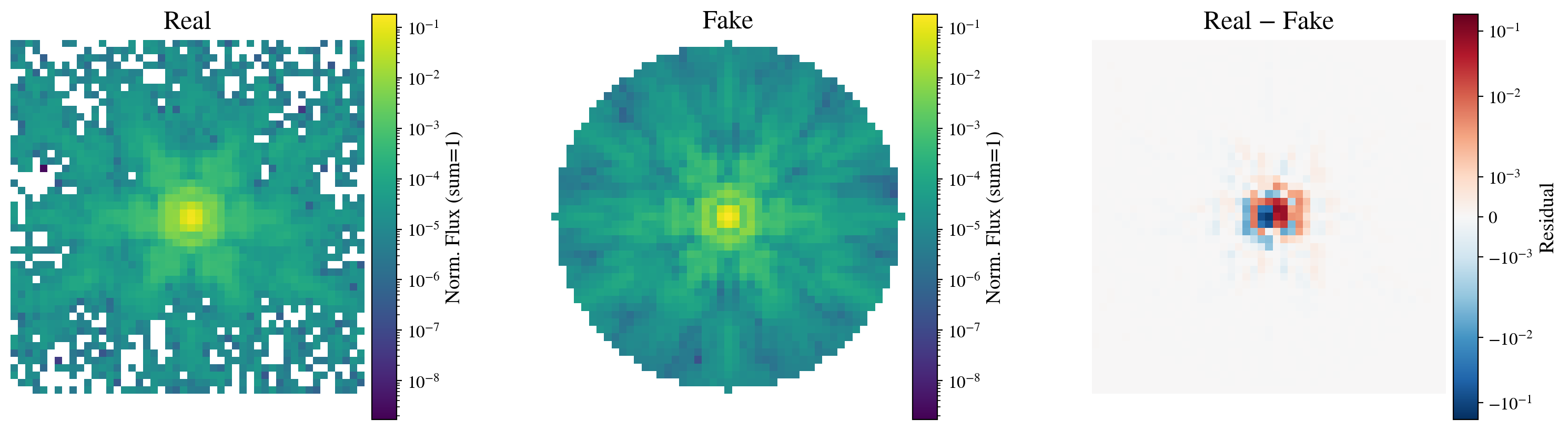}
    \caption{Same as Figure \ref{fig:psf_ivm_comparison_f150w} except in F356W.}
    \label{fig:psf_ivm_comparison_f356w}
\end{figure*}

Figures~\ref{fig:psf_ivm_comparison_f150w} and \ref{fig:psf_ivm_comparison_f356w} show a comparison between a real JWST PSF and our mock PSF. The observed PSFs are generated using the ePSF algorithm implemented within Photutils, which models PSFs from stars in an input image with exposure times of 625s and 940s for F150W and F356W, respectively. The input images are NIRCam individual exposures of the quasar SDSS J2054-0005 from the program GO \#1813 (PI Marshall). Full details on how these ePSFs were created to best match the quasar PSF will be given in Marshall et al. (in prep). 

All PSFs are normalised such that the total flux sums to unity. To quantify the difference between the mock and observed PSFs, we compute the mean absolute residual of the real $-$ mock image within the PSF core, or less than approximately $0.2''$ from the center of the image in both filters. In the core for both filters, we find mean absolute residuals of $\sim1\%$ and a residual in a surrounding ring falling below $\sim0.005\%$ in both filters. Lastly, the adopted mock IVMs calculated in Equation \ref{eq:IVM} are larger than those typically achieved observationally. This deliberately increases the assumed PSF uncertainty relative to real observations and makes convergence more challenging for the sampler. In our final recovered posteriors shown in Figures \ref{fig:weighted_images_1} through \ref{fig:weighted_images_5}, we show that this does not have a large effect by showing results for IVMs at 10\% our nominal value. We also explored IVMs 10x as large and found that convergence very rarely happened. This indicates the ideal range of IVM maps cannot be much larger than those explored in this work.

\subsection{Point source removal}
\label{sec:psf_removal}
\noindent\hspace*{1em}One of the key potential biases in understanding high-$z$ quasar hosts arises from inaccurate point source removal. Once we have created mock images, we can perform a mock point source removal such as the one used in the JWST high-$z$ quasar host detections. There are many algorithms to perform a point source removal, including \texttt{galight} software \citep{2020ApJ...888...37D} used by \citet{10.48550/arxiv.2211.14329} and \citet{ding2025shellqsjwstunveilshostgalaxies} and \textit{psfmc} \citep{2014PhDT.........1M} used by \citet{yue2023eiger} and \citet{marshall2024ganifseigermerging}. Both use a Bayesian approach to decompose a point source and underlying host galaxy Sérsic profile from an input quasar image. We use \textit{psfmc} since it was designed specifically for quasar and host decomposition to remain consistent with \citet{2024MNRAS.530.4765B}. \citep{2014PhDT.........1M} \textit{psfmc} uses the Affine Invariant Markov chain Monte Carlo (MCMC) ensemble sampler from \texttt{emcee} \citep{Foreman_Mackey_2013}. From this, we can extract galaxy Sérsic parameters for both the host galaxy with (observed) and without the point source (updated or truth). 

\noindent\hspace*{1em}The inputs to \textit{psfmc} include the mock JWST PSF, mock quasar image described in Section \ref{sec:methods_mock_imaging_specific}, and an IVM for both the PSF and quasar image.  We use a mock PSF star with a magnitude of approximately $\rm m_{AB}  = 17~mag$ ($\rm 6\times 10^5~nJy$) simulated following the same process as the galaxy image, with the same exposure time (See Section \ref{appendix:psf}). This ensures that the mock PSF star is always much brighter than the quasar with high signal to noise, which is essential for accurate PSF removal (See Section \ref{subsec:results_psf}).
In \citet{10.48550/arxiv.2211.14329}, \citet{yue2023eiger}, and \citet{marshall2024ganifseigermerging}, the PSF used in the point source removal for each quasar image is varied to get the best detected quasar host fit. \citet{10.48550/arxiv.2211.14329} tries PSFs using individual stars and combined averaged stars. Similarly, \citet{yue2023eiger} used averaged PSFs of stars in the quasar fields with about 10-20 stars available per JWST NIRCam module\footnote{The JWST NIRCam modules have nearly identical optics and detectors and are used together to make up the full FOV.} and filter. The PSF varies with detector location and so using multiple measured PSFs helps to find the best fit. We do not consider this spatial variation, using the same WebbPSF model to create both the quasar and PSF image as described in Section \ref{appendix:psf}. Emulating this more complex, realistic PSF generation process is necessary and will likely increase the errors of the PSF removal process. More realistic PSFs were explored in \citet{Zhuang_2024}, which is discussed in Section \ref{sec:discussion}.

\noindent\hspace*{1em}As described in \citet{2014PhDT.........1M}, inverse variance, or $1/\sigma^2$ must be provided to indicate the weight of each pixel used in the sampling. We show a mock JWST PSF and its corresponding mock inverse variance map in Figure \ref{fig:psf_ivm_comparison_f150w} and \ref{fig:psf_ivm_comparison_f356w}. The mock inverse variance maps for both the PSF and quasar image are formed with the background described in Equation \ref{eq:background} and a 2D Gaussian centered at the center of each image (See \citealt{10.1093/mnras/staa2982}) as described more fully in Section \ref{appendix:psf}.

\noindent\hspace*{1em}We maintain consistency with the JWST observations we consider by fitting a Sérsic profile to the host galaxies. Although this may not describe each galaxy perfectly, it allows for a more similar comparison to JWST high-$z$ quasar host detections. The Sérsic profile is defined as follows
\begin{equation}
\label{eq:sersic}
    I(R) = I_e \exp \left \{-b_n \left[\left(\frac{R}{R_{1/2}}\right)^{1/n} - 1\right]\right\},
\end{equation}
where $R_{1/2}$ is the radius containing half the light in the galaxy, which we define as the Sérsic radius, n is the Sérsic index defining the relation in log brightness and log radius space, $b_n \sim 2n - \frac{1}{3}$, and $I_e$ is the intensity at the half light radius. We discuss the difference between the Sérsic magnitude and photometric magnitude for various filters in Appendix \ref{appendix:sersic}. To maintain consistency with images with an added quasar, we also use \textit{psfmc} for our truth galaxy Sérsic profile modeling. This allows us to maintain the Bayesian approach to obtain the Sérsic magnitude posterior for the galaxy without an added quasar. In \textit{psfmc}, we sample from the following joint Gaussian likelihood where we do not assume correlations between pixels:

\begin{multline}
P(\text{image(pixel)} \mid \theta) = 
\frac{1}{\sqrt{2 \pi \sigma(\text{pixel})^2}} \\
\times \exp\left(-
\frac{(\text{image(pixel)} - \text{image}_{\text{model}}(\text{pixel}))^2}{2 \sigma(\text{pixel})^2}
\right)
\label{eq:likelihood}
\end{multline}

where $\theta$ are the Sérsic parameters listed below and $\sigma$ is a function of the pixel noise derived from our mock background. We ensure consistent sampling hyperparameters across all quasars by using 128 walkers with 2000 samples each such that each posterior is sampled 256k times.
We inform our posterior with the following priors similar to the JWST observations we compare to: 

\begin{enumerate}[labelwidth=*, leftmargin=2em]
    \item $m_{\rm filter, quasar}$ --- [$m_{\rm quasar, filter}-1$, $m_{\rm quasar, filter}+1$]
    \item $m_{\rm filter, S\acute{e}rsic}$ --- [20, 30]
    \item $x_{\rm quasar}, y_{\rm quasar}$ [pixels] --- [$x_{\rm center}-6$, $x_{\rm center}+6$], [$y_{\rm center}-6$, $y_{\rm center}+6$]
    \item $x_{\rm S\acute{e}rsic}, y_{\rm S\acute{e}rsic}$ [pixels] --- [$x_{\rm center}-6$, $x_{\rm center}+6$], [$y_{\rm center}-6$, $y_{\rm center}+6$]
    \item $r_{\rm 1/2}$ [pixels] --- [1, 21]
    \item $r_{\rm 1/2, b}$ [pixels] --- [1, 21]
    \item Sérsic index --- fixed at 1, or testing in Section \ref{app:free_sersic} and Appendix \ref{app:more_sersic} with [0.5, 5.5]
    \item Sérsic angle [deg] --- [0, 180]
\end{enumerate}

We note that there can be Sérsic index variability that we are not considering in our modeling. We choose to set the Sérsic index at 1 to remove any bias in the fitting from using different Sérsic indices for comparison to observation, as \citet{10.48550/arxiv.2211.14329}, \citet{yue2023eiger}, and \citet{marshall2024ganifseigermerging} set their Sérsic indices to 1. However, \citet{ding2025shellqsjwstunveilshostgalaxies} explored the effects of the Sérsic index values of 1, 2, 3, and 4 on radius and magnitude. They chose to explore Sérsic index values less than 4 since this is where most of the change in light profile occurs. The light profile shape change is negligible at index values greater than 4 \citep{Graham2013}. In \citet{ding2025shellqsjwstunveilshostgalaxies}, only J2236+0032 is shown to have a better fit with n = 4 whereas J2255+0251 still finds a best fit at n = 1. They find that increasing n may lead to brighter inferred hosts, by up to $\sim$0.5 mag. We do not consider other higher Sérsic indices for J2236+0032 to maintain consistency with the other quasar host detections. 

\subsection{SED fitting}
\label{subsec:sed}
We expect a clear correlation between quasar host magnitude and stellar mass, as the stellar mass largely determines the normalization of the SED \citep{yue2023eiger}. However, SED fitting is needed to infer information about stellar populations and an accurate updated stellar mass of galaxies. We redo the SED fitting for all corrected and measured host magnitudes to maintain consistency in any systematic biases introduced by the fitting procedure, allowing us to better isolate biases arising \textit{specifically} from PSF removal. We refit the masses from the \citet{10.48550/arxiv.2211.14329}, \citet{yue2023eiger}, \citet{marshall2024ganifseigermerging}, and \citet{ding2025shellqsjwstunveilshostgalaxies} from their original magnitudes. We follow an identical SED fitting for the \textit{corrected} magnitudes from our analysis. We fit an observed spectra to infer stellar mass, dust attenuation, star formation history, and other parameters. However, stellar mass is typically the only parameter that can be reliably constrained, as seen in most high-$z$ quasar host stellar mass estimates. By performing an SED fit, we make a proper comparison of the offset in magnitudes we infer with the offset in stellar mass. 

We perform the SED fitting with the \texttt{PROSPECTOR} package \citep{2021ApJS..254...22J}. This is used by \citet{yue2023eiger} and \citet{marshall2024ganifseigermerging} to infer stellar masses. We use the same \textit{emcee} setup including the priors and initialization as described in \citet{yue2023eiger}. We adopt the same delayed-$\tau$ model from \citet{marshall2024ganifseigermerging} for simplicity, as the primary goal of the SED fitting is to assess potential offsets in stellar mass. However, \citet{marshall2024ganifseigermerging} found that a delayed-$\tau$ plus a starburst star formation history (SFH) best matches the fluxes they found. The redshift value is fixed at the observed value of the quasar. We briefly summarize the SED fitting setup below :
\begin{enumerate}[labelwidth=*, leftmargin=2em]
\item \textbf{Stellar mass} — Modeled with a log-uniform prior between $10^{8},M_{\odot}$ and $10^{12},M_{\odot}$, with the sampler initialized at $5 \times 10^{9},M_{\odot}$.

\item \textbf{Stellar metallicity} --- A uniform prior is adopted for $\log(Z/Z_{\odot})$ in the range $[-2, 0.2]$.

\item \textbf{Star formation history (SFH)} --- We assume a delayed-$\tau$ SFH of the form $\text{SFR} \propto t\,e^{-t/\tau}$, along with a Chabrier initial mass function \citep{2003PASP..115..763C}.
\begin{itemize}[leftmargin=2em]
    \item \textit{Onset time of star formation} ($t_0$) --- Uniform prior between $t = 0$ and the age of the universe at the quasar's redshift.
    \item \textit{Exponential timescale} ($\tau$) — Uniform prior between $0.01$ Myr and $20$ Myr.
\end{itemize}

\item \textbf{Dust attenuation} ($\tau_{5500}$)--- Optical depth at $5500\,\text{\AA}$ is modeled with a uniform prior between 0 and 2. We adopt the attenuation law from \citet{2000ApJ...533..682C}.

\item \textbf{Gas-phase metallicity} --- Modeled using a uniform prior on $\log(Z_g/Z_{\odot})$ in the range $[-2, 0.5]$.

\item \textbf{Ionization parameter} ($\log U$) --- Modeled with a uniform prior between $-3$ and $1$.
\end{enumerate}

We also recalculate the stellar masses for all \citet{10.48550/arxiv.2211.14329} quasar hosts from the measured magnitudes to ensure consistency in SED fitting software, so that all stellar masses compared are derived using the same method. For this reason, we do not directly compare our stellar mass estimates to the intrinsic stellar masses of the simulated galaxies in BlueTides. The intrinsic mass of all stellar particles in a simulated galaxy is not necessarily the same as the observed mass measured from the stellar luminosity \citep{narayanan2023outshiningrecentstarformation, cochrane2024highzstellarmassesrecovered}, and our goal is to measure the bias in the luminosity-based mass estimate due to the quasar subtraction process. Instead, we compare recovered host luminosities to the true host luminosities in the simulations. This allows us to isolate the observational bias introduced by quasar subtraction prior to any SED modeling assumptions.

There is ongoing debate about the recovery of accurate stellar mass using SED fitting. \citet{cochrane2024highzstellarmassesrecovered} claimed that most high-$z$ stellar masses are accurately recovered with SED fitting models up to within $\sim$0.5 dex, with slightly overestimated masses for lower mass galaxies ($<10^8 M_{\odot}$) and underestimated masses for higher mass galaxies ($>10^{9} M_{\odot}$). However, \citet{narayanan2023outshiningrecentstarformation} found that stellar masses of high-$z$ galaxies can only be estimated to an order of magnitude. The effects of both the PSF removal and the intricacies of the SED fitting must be accounted for in high-$z$ quasar host stellar mass recovery. While the cumulative effects of all modeling choices must ultimately be considered, here we focus solely on the bias introduced by quasar subtraction.

We note that our overall SED fitting setup is very similar to that used for the SHELLQs sample \citep{ding2025shellqsjwstunveilshostgalaxies}. However, the updated stellar mass estimates in \citet{li2025dichotomynuclearhostgalaxy} adopt a \citet{2001MNRAS.322..231K} initial mass function and include an AGN SED component. A comparison of the resulting stellar mass estimates is presented in Section \ref{sec:results} and seen in Figure~\ref{fig:stellar_mass_resid}.

\section{Statistical Methods}
\label{sec:stat_methods}
\begin{figure*}
    \centering
\includegraphics[width=0.68\textwidth]{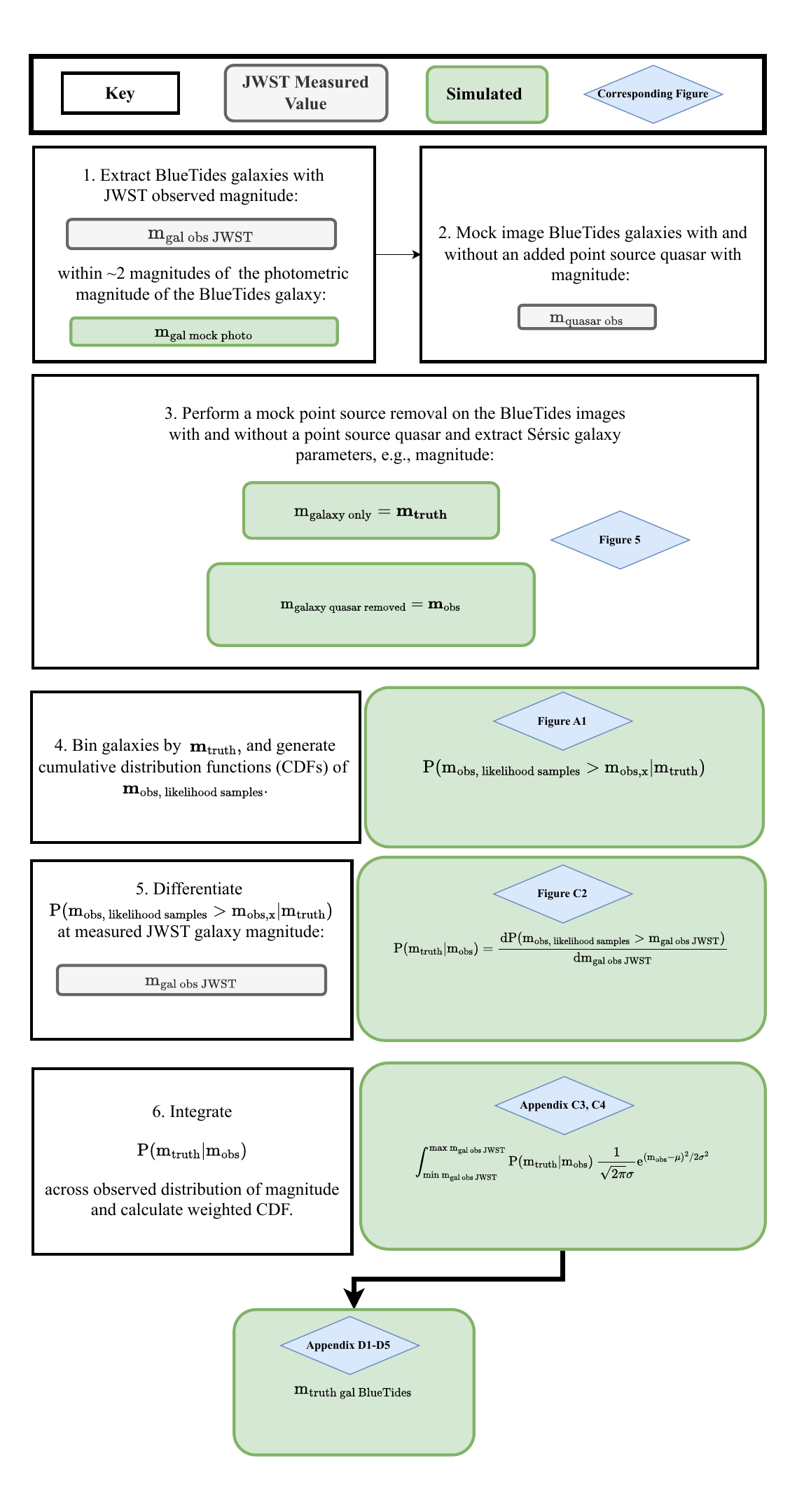}
    \caption{\label{fig:flowchart} An overview of all mock imaging, point source removal, and statistical steps to generate the an updated posterior: $\rm P(m_{truth}|m_{obs})$ from an observed JWST galaxy magnitude. If we only want an updated posterior for one observed magnitude, we stop at Step 5. Otherwise, we can integrate across all potential observed values (weighted by a Gaussian distribution around the measured value) through to Step 6.}    
\end{figure*}
\subsection{\texorpdfstring{Forming the posterior: $\rm P(m_{truth}|m_{obs})$}{Deriving the posterior: P(m|m)}}

With observed magnitudes from newly detected high-$z$ quasar host galaxies, we apply a statistical correction to infer their intrinsic magnitudes. To achieve this, we develop a simulation-based inference (or SBI, see \citealt{Cranmer_2020} for a review) pipeline that incorporates intermediate Bayesian sampling, as detailed in Sections \ref{sec:psf_removal} and \ref{subsec:sed}. This framework allows us to recover the corrected distribution of host galaxy properties by leveraging mock imaging of BlueTides galaxies and is the first application of SBI to understanding high-$z$ host galaxy biases. We do not use any machine learning to perform the SBI, but this remains an interpretable SBI application since we do not define an explicit likelihood to go from a truth magnitude to an observed magnitude. Note that there are explicit likelihoods internally defined in \textit{psfmc} (See Equation \ref{eq:likelihood}) and in the SED fitting. We forward model BlueTides galaxies through a mock observation (described in Section \ref{sec:methods_mock_imaging_specific}) and point source removal (described in Section \ref{sec:psf_removal}). This allows us to take a JWST NIRCam detection of a quasar host galaxy and determine the corrected posterior of the host magnitude. Using our desired parameters, Bayes theorem is defined as:
\begin{equation}
    P(\rm m_{truth}|\rm m_{\rm gal~obs~JWST}) = \frac{P(\rm m_{obs}|\rm m_{truth}) P(\rm m_{truth})}{P(\rm m_{obs})}, 
 \end{equation}
where $P(\rm m_{truth}|\rm m_{\rm gal~obs~JWST})$ is the final posterior we seek, $P(\rm m_{obs}|\rm m_{truth})$ is the likelihood (from our forward model through BlueTides to a mock observation), $P(\rm m_{truth})$ is the prior (we set this to 1 for the remainder of the paper), and $P(\rm m_{obs})$ is the evidence or a normalization factor to ensure the posterior sums to 1. We do not have an analytic expression for the likelihood but determine this posterior through the SBI pipeline. A flowchart summary of the full pipeline is shown in Figure \ref{fig:flowchart}, but we describe each step in depth below. Each step corresponds to the same number as in Figure \ref{fig:flowchart}.

\begin{enumerate}[labelwidth=*, label=\arabic*., leftmargin=4mm] 
    \item We find galaxies in BlueTides with magnitudes within approximately 2 magnitudes to the observed quasar hosts in the detected JWST filters. We select galaxies from BlueTides randomly in bins spanning this range on either side of the observed magnitude. We ensure that each final corrected magnitude posterior uses at least 50 galaxies in its estimate. We are limited by the magnitude lower limits of the BlueTides simulation as seen in Figure \ref{fig:photo_mags_with_quasars}. We select only quasars with a sufficient number of galaxies in BlueTides to use in our posterior derivations---we are unable to include J0148+0600 in our analysis as there are not enough bright galaxies in the BlueTides sample. We limit the BlueTides galaxies that we include to those with magnitude residuals (true host magnitude $-$ host magnitude with PSF removal) less than 3 magnitudes. However, this does not limit the range of higher magnitudes that we are able to consider. As seen in Figure \ref{fig:residuals}, there are galaxies with residuals smaller than this at all truth magnitudes. 
    \item Next, we perform mock imaging of these analogs as described in Section \ref{sec:methods_mock_imaging_specific}. We produce the leftmost two images from the panel in Figure \ref{fig:panel_sample} that include the true galaxy by itself and the galaxy with a convolved point source (i.e., mock quasar) with magnitude equal to the observed quasar we are simulating. Despite the minimal PSF mismatch, there remains an unavoidable ``missing core" effect arising from the intrinsic degeneracy between the AGN point source and the central regions of the host galaxy in photometric images. Even with an accurate PSF, the quasar dominates the inner pixels, making it impossible to perfectly disentangle the true central galaxy light from the AGN contribution. 

    \begin{figure}
    \centering
    \includegraphics[width=1.1\linewidth]{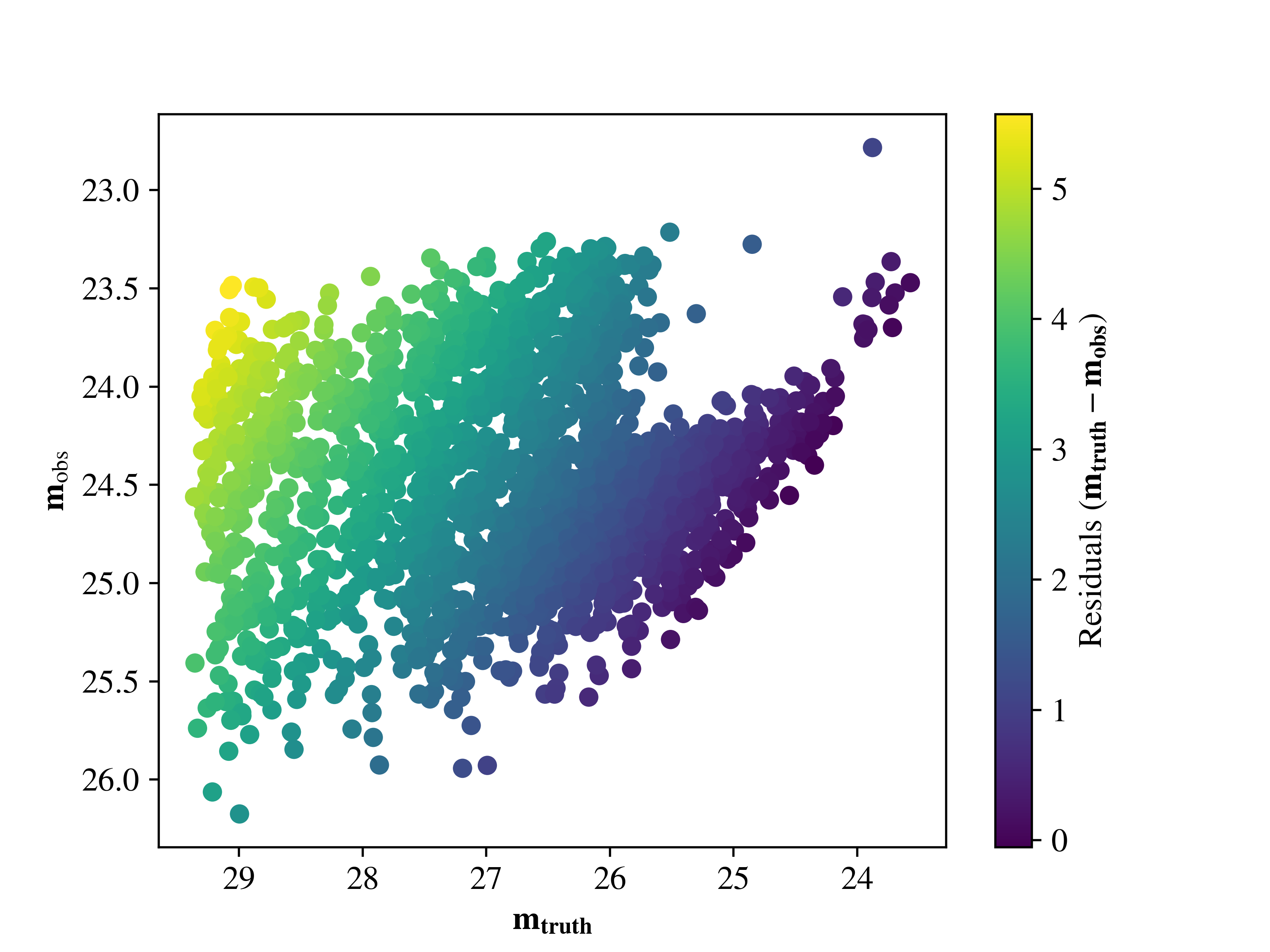}
    
    \caption{Example residuals plot for BlueTides galaxies going into the BlueTides truth galaxy magnitude posterior generation. The error between the observed (Sérsic magnitude with quasar added) and the truth (Sérsic magnitude without quasar added) magnitude increases for dimmer galaxies as expected. More than 2,000 BlueTides galaxies are sampled with a point source with $\rm m_{F356W}$ = 19.5 mag (approximately the F356W magnitude of the J159-02 quasar) added. Axes are inverted so that brighter magnitudes appear toward the right and top of the plot, opposite to the convention used elsewhere in this paper.\label{fig:residuals}}
    \end{figure}
    
    \begin{figure*}
    \centering
    \includegraphics[width=\linewidth]{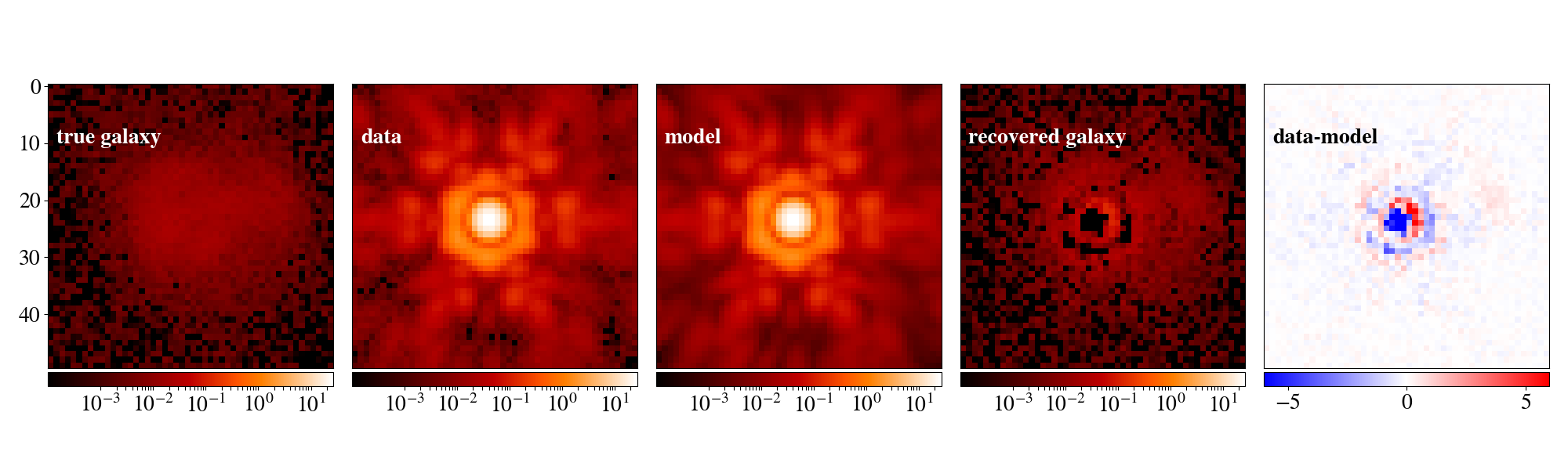}
    \caption{An example panel of the full array of mock imaged data products in our pipeline. We conduct the same point source removal and magnitude estimation described in Section \ref{sec:mock_methods} for each galaxy  in BlueTides within the magnitude sample range. This panel highlights the central challenge of this work. Even with nearly perfect PSF matching, the innermost host galaxy light cannot be fully recovered beneath the quasar, due to the intrinsic AGN and host degeneracy in the core and imperfections in the Sérsic profile modeling.\label{fig:panel_sample}}
    \end{figure*}

    \item We perform point source removal on the added quasar images as described in Section \ref{sec:methods_mock_imaging_specific} and extract Sérsic parameters from the posteriors. We also perform the same source modeling on the galaxy-only images to determine the \textit{true} Sérsic properties. We use the median sampled value within the posterior as the nominal value. A sample corner plot with all galaxy parameters sampled in \textit{psfmc} is shown in Figure \ref{fig:corner_sample}.
    \item We bin galaxies by their Sérsic truth magnitude in that filter and generate an empirical cumulative distribution function (CDF) for each bin. We use 20 bins of truth magnitude over the entirety of the truth magnitude range, which means we are limited in sensitivity in our inferred magnitudes to approximately 0.1--0.3 mag. We tabulate the bin sizes of each posterior in Table \ref{tab:bin_sizes} as the bin size of the posterior determines the precisions of our updated magnitude estimates. Because of the uneven distribution of truth magnitude bins, we have variability in counts of galaxies in each bin. In Figure \ref{fig:hist_mags_example}, we show this binning using the host galaxy of J1120+0641 as an example. We have an especially small sample for brighter galaxies (lower magnitudes). To be able to include brighter galaxies in our sample, we still include bins that have just one galaxy. We use an empirical CDF generated with \citet{seabold2010statsmodels} to avoid binning issues in generating a probability density function (PDF). This CDF is described by \begin{equation}
        \rm P(m_{obs,~likelihood~samples} > m_{obs, x} | m_{truth}).
    \end{equation}

    \item $\rm P(m_{truth}|m_{obs})$: We differentiate each CDF at $\rm m_{obs}$ for each truth bin to produce the PDF of the observed magnitude. This allows us to compare to the posterior of the measurement found with JWST.
     \begin{equation}
     \label{eq:m_t_m_}
         \rm P(m_{truth}|m_{\rm gal~obs~JWST}) = \frac{dP(m_{obs,~likelihood~samples} >  m_{\rm gal~obs~JWST})} {dm_{\rm gal~obs~JWST}},
     \end{equation}
where this derivative is performed along multiple CDFs as shown in Figure \ref{fig:cdfs_one}.

An example of this PDF for J1120+0641 with the JWST observed $m_{\rm gal}$ is shown in Figure \ref{fig:pdf_one}. If we want the posterior of one observed value with JWST, we can stop here and use the non-weighted posterior. However, this does not consider the entire observed distribution. To do this, we derive a weighted posterior for a sample of values from within the uncertainty range of the observed magnitudes. If we integrate across the range of observed values, we continue through to Step 6: Marginalizing the posterior.
\end{enumerate}

\subsection{Marginalizing the posterior:}
\label{subsec:convolve}
6. $\rm \Sigma_i P(m_{truth}|m_{obs,i})~N(m_{obs, JWST}, \sigma)$: To determine the most likely truth value within $1\sigma$ of the nominal observed value, we can construct an integral of Equation \ref{eq:m_t_m_}, $\rm P(m_{truth}|m_{obs})$, weighted on the distribution of the observed magnitude values:
 \begin{equation}
 \label{eq:conv_post}
\int_{\rm min~m_{gal~obs~JWST}}^{\rm max~m_{gal~obs~JWST}} \rm P(m_{truth}|m_{obs}) ~\frac{1}{\sqrt{2\pi} \sigma}e^{(m_{obs} - \mu)^2/2\sigma^2} dm_{obs},
 \end{equation}
 where $\mu$ is the median of the posterior of the observed magnitude observation and $\sigma$ is the error reported on the observation\footnote{Here we assume a Gaussian distribution of the posterior from the measured values. The actual distribution likely varies from Gaussian, but we expect this likely provides a reasonable approximation.}.
This allows us to better compare our posteriors to those detected by JWST as we must sum our individual posteriors for $\rm m_{obs}$ where each posterior is weighted by $\rm N(m_{obs, JWST}, \sigma)$. 

We take a JWST magnitude observation ($\rm m_{obs} \pm \sigma$) and assuming a Gaussian distribution of measured values, we infer the corrected posterior of the measurement. We use 10 bins from within $\pm 5\sigma$ of the nominal measure host magnitude with JWST. After the marginalization of the posterior, we choose not to assume a distribution of the inferred posteriors when determining our new measured values. We calculate a CDF from Equation \ref{eq:conv_post}. We then take the new inferred nominal value of the host magnitude as the median of the CDF and calculate the uncertainty with the 16th and 84th percentiles, i.e., $\sigma_{\rm top} = \rm  CDF(0.84) - CDF(0.5)$ and $\sigma_{\rm bottom} = \rm  CDF(0.5) - CDF(0.16)$.

The posterior is shown in Figure \ref{fig:weighted_images_4} and \ref{fig:weighted_images_5} for the same quasar and galaxy combination as shown in Figure \ref{fig:cdfs_one} (J1120+0641 in F200W).
We show and describe an example of the individual posteriors that go into the weighted sum in Figure \ref{fig:sum_figure}.

\section{Results}
\label{sec:results}
\subsection{PSF magnitude selection}
\label{subsec:results_psf}

\begin{figure*}
\centering
\includegraphics[width=0.7\linewidth]{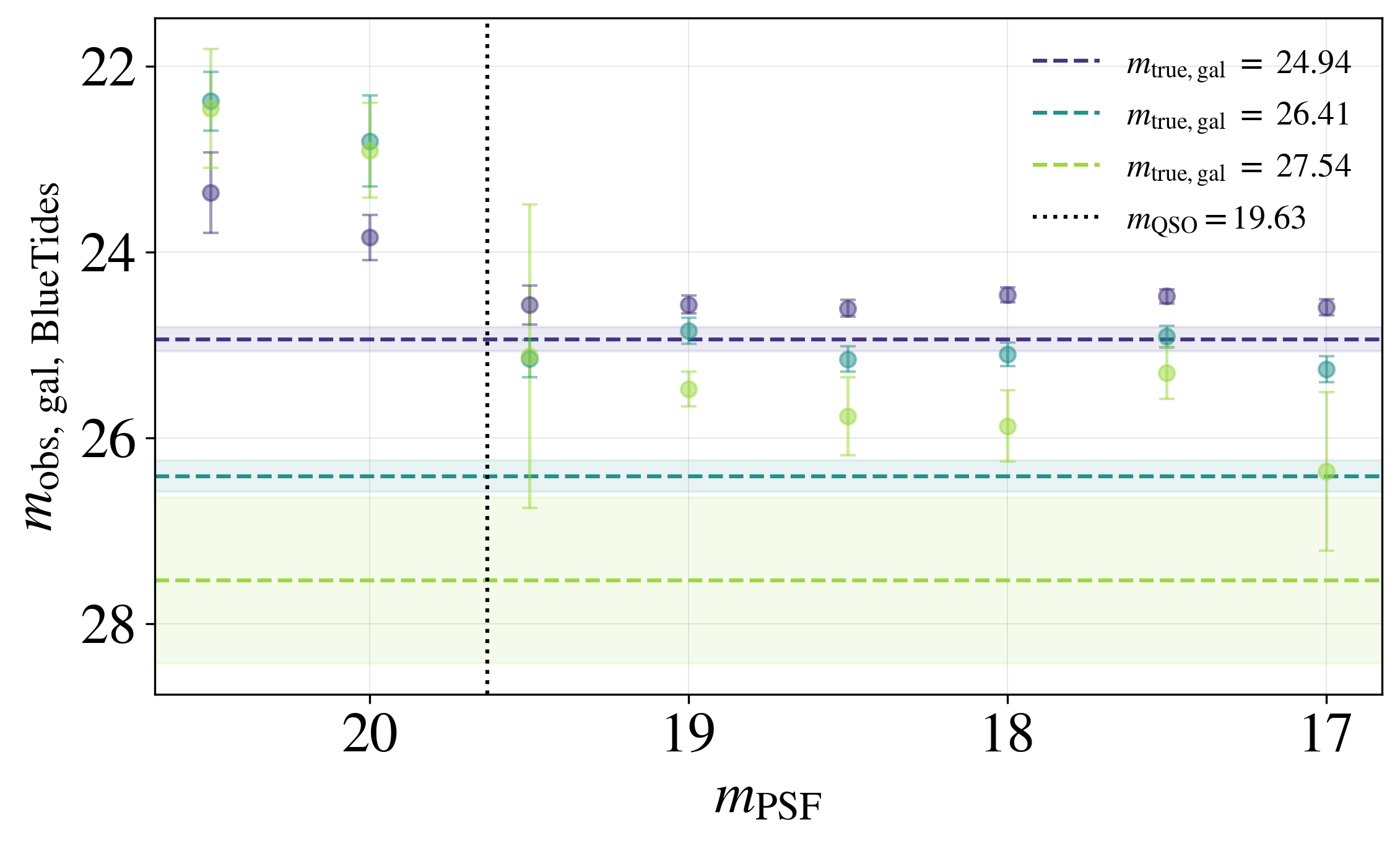}
\caption{We compare a few galaxies recovered host magnitude for several different runs of the quasar removal with mock PSF star magnitudes less than and greater than that of the quasar. The PSF used in the host magnitude recovery must be at least as bright as the quasar to ensure convergence to a consistent magnitude value. For the faintest galaxy (light green), the PSF may need to be significantly brighter than the quasar.}\label{fig:psf_comp}
\end{figure*}

We perform a series of tests to determine how the input model PSF magnitude used in the point source removal affects the quasar and host decomposition. We test convergence to the same magnitude for the removal of a mock quasar for varying mock input PSF magnitudes. This is analogous to varying the SNR of the input PSF model, as we assume the same exposure time for each PSF star. Brighter PSF stars have higher SNR due to reduced fractional Poisson noise. To test this, we use mock input PSFs with apparent magnitudes in the F356W filter between 16 and 22 mag, assuming the same exposure time for the PSF star and quasar. We then compare the recovered host galaxy magnitudes for three representative BlueTides galaxies. In Figure \ref{fig:psf_comp}, we show the recovered F356W galaxy magnitudes (24.94 – 27.54 mag) for systems with a quasar of magnitude 19.63 mag at an exposure time of 1578s.

We find that the PSF used in the quasar removal must be at least brighter than the quasar, assuming the same exposure time (i.e., a signal to noise ratio higher than the quasar). For the brightest galaxy, we detect a host magnitude that converges to a consistent value when the PSF star is at least as bright as the quasar. However, for the dimmer galaxies, the host galaxy magnitudes start to converge when the PSF is 2 mag brighter than the quasar. This is likely because lower SNR PSF stars provide poorer constraints on the PSF wings and produce a sharper PSF model that fails to fully capture the extended quasar light profile. For PSF stars that are dimmer than the quasar, we consistently recover brighter host galaxy magnitudes across all galaxies. This indicates that part of the quasar flux is misattributed to the host during PSF subtraction. 

In \citet{yue2023eiger}, all model PSFs used in quasar host subtraction are at least as bright as the quasar itself. In \citet{10.48550/arxiv.2211.14329} and \citet{ding2025shellqsjwstunveilshostgalaxies}, the final PSF magnitudes used are not reported. As described in Section \ref{sec:mock_methods}, we use a PSF with a magnitude of 17 mag for all quasars when performing a point source removal, which is at least 2 mag larger than all observed quasar magnitudes considered.

\subsection{Galaxy magnitude biases}
\label{subsec:galaxy_mag_biases}
We now investigate how the PSF removal affects the measured host galaxy magnitudes. The weighted sum posterior described in Section \ref{subsec:convolve} is our corrected observed magnitude posterior for each quasar and filter combination with JWST. Using the methodology described in Sections \ref{sec:mock_methods} through \ref{sec:stat_methods}, we generate a corrected posterior for every quasar with sufficient (at least 50) BlueTides galaxies to generate a large enough sample within the range of the observed magnitude and 5 times the reported uncertainty, i.e., $\rm m_{obs} \pm 5\sigma$. We apply a strict radial and quasar error tolerance on the galaxies included. Inferred radial estimates for the galaxies must be within $\pm 1 \sigma$ and quasar magnitudes must be within 0.05 mag as described in Appendix \ref{app:quasar}. All of the observed host magnitudes and updated host magnitudes with BlueTides are tabulated in Table \ref{tab:magnitude_offsets}. We show our simulated posteriors in the subfigures of Figure \ref{fig:weighted_images_1} and \ref{fig:weighted_images_5} and describe the results for specific hosts below. This quasar host magnitude correction allows us to then perform an SED fit to determine updated stellar mass estimates.

We determine four different kinds of updated measurements in our work from the shape of the updated posterior CDF and offset. We classify a magnitude as a lower limit (i.e., a non-detection) if its posterior CDF is approximately linear beyond the median, indicating the absence of a peak and therefore no well-constrained detection. We list the categories below for both measurements and lower limits and refer to them later according to their category in the subsequent sections.  Each category is determined from the quasar host galaxy magnitude posterior seen in Figures \ref{fig:weighted_images_1} through \ref{fig:weighted_images_5}. \begin{enumerate}[labelwidth=*, leftmargin=4mm]
    \item \textbf{Category A} (measurements confirmed) --- In this category, we confirm the detection and magnitude measurement of the high-$z$ quasar host galaxy robustly in this work. The CDF of the quasar host magnitude posterior peaks and flattens out, showing convergence to a Gaussian-like posterior.
    \item \textbf{Category B} (measurements biased) --- We find significant offset (near or exceeding the observational uncertainty) from the original measured host galaxy magnitude. The CDF of the quasar host magnitude posterior peaks at an offset value within observational uncertainties and flattens out, showing convergence to a Gaussian-like posterior.
    \item \textbf{Category C} (lower limits confirmed as measurements) --- We find that the lower limits measured for the high-$z$ quasar host magnitude are actually likely detections. The CDF of the quasar host magnitude posterior is not linear (which indicates lack of convergence) and peaks and flattens out. This indicates that the lower limit may be an actual measurement with slight bias.
    \item \textbf{Category D} (lower limits confirmed) --- We find that the lower limits measured for the high-$z$ quasar host magnitudes are likely lower limits. The CDF of the quasar host magnitude is linear and does not peak. The magnitude measurements or stellar mass measurements should not be used for these quasar hosts. In this case, we quote our lower magnitude limits as the value that 84\% of our posterior probability lies above.
\end{enumerate}

\begin{table*}
    \centering
\begin{tabular}{l l c c c c}
    \hline
    \textbf{Quasar} & \textbf{Filter} & \textbf{$\rm m_{obs, gal, JWST}$} & \textbf{$\rm m_{corrected, gal, BlueTides}$} & $\rm m_{obs, gal, JWST} - m_{corrected, gal, BlueTides}$ & \textbf{Category} \\
    \hline
J0844-0132 & F150W$^{*\dagger}$ & $>26.79$ & -- & -- & D \\
           & F356W & $25.70^{+0.34}_{-0.34}$ & $25.68^{+0.32}_{-0.27}$ & 0.02 & A \\
J0911+0152 & F150W$^{*\dagger}$ & $>27.61$ & -- & -- & D \\
           & F356W & $26.56^{+0.30}_{-0.30}$ & $26.54^{+0.65}_{-0.41}$ & 0.02 & A \\
J0918+0139 & F150W$^{*}$ & $>26.03$ & $25.96^{+0.95}_{-0.63}$ & 0.07 & C \\
           & F356W & $25.28^{+0.25}_{-0.25}$ & $25.19^{+0.25}_{-0.24}$ & 0.09 & A \\
J1425-0015 & F150W & $25.78^{+0.29}_{-0.29}$ & $25.72^{+0.27}_{-0.27}$ & 0.06 & A \\
           & F356W & $24.12^{+0.29}_{-0.29}$ & $23.96^{+0.25}_{-0.22}$ & 0.16 & A \\
J1525+4303 & F150W & $25.60^{+0.21}_{-0.21}$ & $25.56^{+0.22}_{-0.29}$ & 0.04 & A \\
           & F356W & $24.61^{+0.24}_{-0.24}$ & $24.51^{+0.20}_{-0.23}$ & 0.10 & A \\
J1146-0005 & F150W$^{*\dagger}$ & $>28.04$ & -- & -- & D \\
           & F356W$^{*}$ & $>26.38$ & $26.31^{+1.07}_{-0.64}$ & 0.07 & C \\
J1146+0124 & F150W & $25.62^{+0.20}_{-0.20}$ & $25.60^{+0.23}_{-0.32}$ & 0.02 & A \\
           & F356W & $24.15^{+0.16}_{-0.16}$ & $24.01^{+0.16}_{-0.14}$ & 0.14 & A/B \\
J0217-0208 & F150W & $24.29^{+0.13}_{-0.13}$ & $24.21^{+0.16}_{-0.08}$ & 0.08 & A \\
           & F356W & $23.69^{+0.18}_{-0.18}$ & $23.59^{+0.10}_{-0.11}$ & 0.10 & A \\
J0844-0052 & F150W & $25.34^{+0.25}_{-0.25}$ & $25.28^{+0.29}_{-0.25}$ & 0.06 & A \\
           & F356W & $25.57^{+0.51}_{-0.51}$ & $25.51^{+0.43}_{-0.40}$ & 0.06 & A \\
J2236+0032 & F150W & $25.00^{+0.15}_{-0.15}$ & $24.94^{+0.15}_{-0.14}$ & 0.06 & A \\
           & F356W & $23.01^{+0.28}_{-0.28}$ & $22.68^{+0.24}_{-0.07}$ & 0.33 & B \\
J2255+0251 & F150W$^{*}$ & $>26.66$ & $26.59^{+1.28}_{-0.68}$ & 0.07 & C \\
           & F356W & $24.56^{+0.14}_{-0.14}$ & $24.47^{+0.20}_{-0.17}$ & 0.09 & A \\
J159-02    & F115W$^{*}$ & $>24.83$ & $24.86^{+0.12}_{-0.11}$ & -0.03 & C \\
           & F200W & $24.82^{+0.23}_{-0.23}$ & $24.94^{+0.13}_{-0.20}$ & -0.12 & A \\
           & F356W & $23.98^{+0.16}_{-0.16}$ & $24.16^{+0.40}_{-0.20}$ & -0.18 & B \\
J1120+0641 & F115W & $25.94^{+0.20}_{-0.20}$ & $25.88^{+0.55}_{-0.20}$ & 0.06 & A \\
           & F200W$^{*}$ & $25.48^{+0.37}_{-0.37}$ & $25.70^{+0.70}_{-0.37}$ & -0.22 & C \\
           & F356W$^{*\dagger}$ & $>25.72$ & -- & -- & D \\
    \hline
\end{tabular}
    \caption{We show the JWST observed galaxy host magnitudes, our corrected magnitudes from BlueTides, the difference between the two, and the category of the measurement (described thoroughly in Section \ref{sec:results}). Category A and B indicate detections which are robust or exceeding observational uncertainties and the bin width tolerances outlined in Appendix \ref{tab:bin_sizes}, respectively. The observational lower magnitude limits are denoted in the same way as Table \ref{tab:quasar_host_comprehensive} with a $^*$. Quasar host magnitude measurements with a $^*$ and not a $\dagger$ may actually be detections and not lower limits (Category C). We denote our confirmed lower limits with a $\dagger$ (Category D), which correspond to the 16\% confidence interval—indicating that 84\% of the posterior probability lies above this value.}
    \label{tab:magnitude_offsets}
\end{table*}

\begin{figure*}
    \centering
    \includegraphics[width=\linewidth]{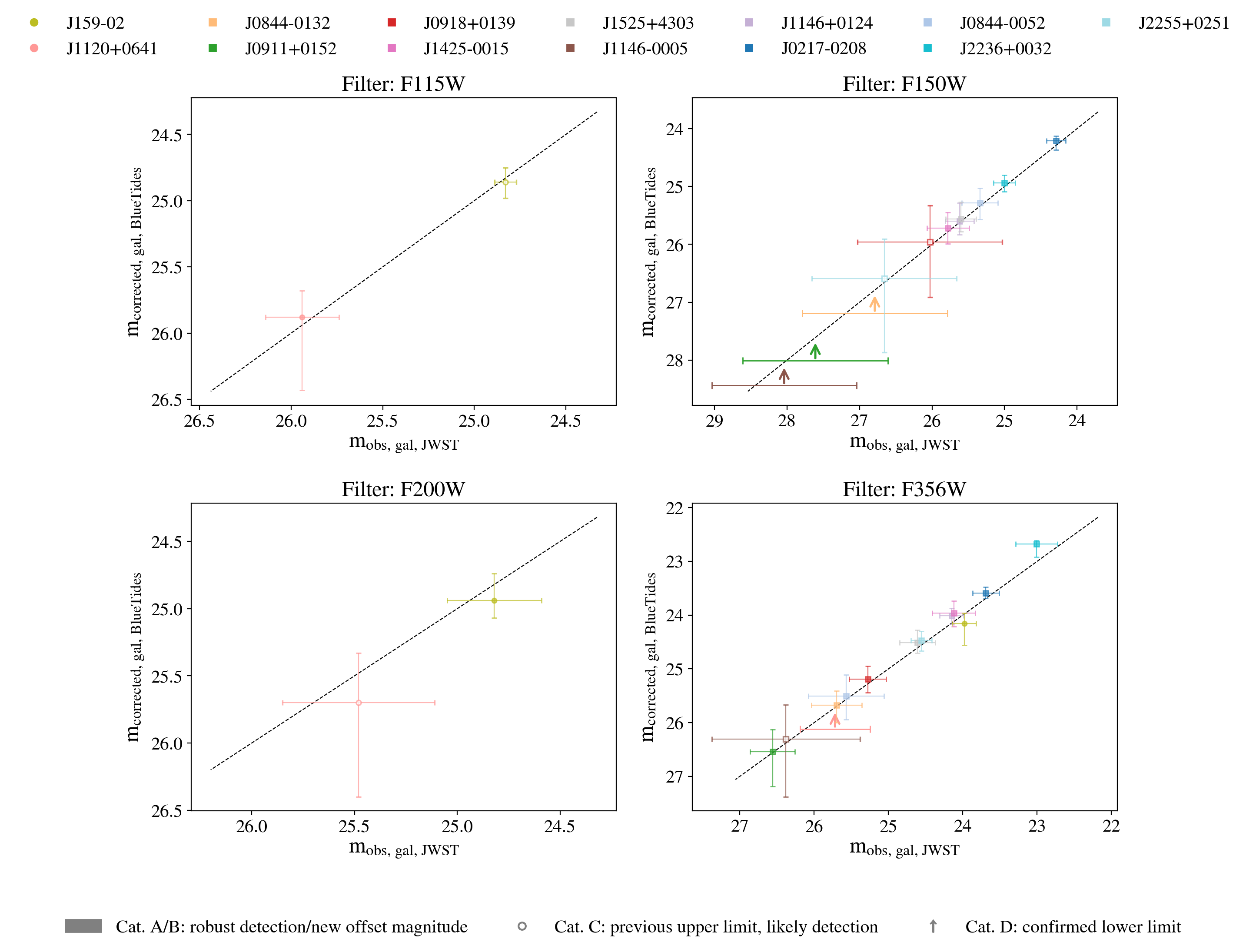}
    \caption{We show the magnitude biases for each quasar host galaxy in their four filters for the host galaxies of the quasars detected in \citet{10.48550/arxiv.2211.14329} (squares), \citet{ding2025shellqsjwstunveilshostgalaxies} (squares), \citet{yue2023eiger} (circles), and \citet{marshall2024ganifseigermerging} (circles). Filled in points with error bars on both axes represent Category A or B measurements—that is, robust detections that are either accurate or only slightly biased. Category C measurements are shown as open circles with error bars, indicating they likely reflect true host magnitudes rather than strict lower limits. Category D measurements correspond to confirmed lower limits in both our analysis and the JWST observations and are indicated with upwards arrows. While most quasar host magnitudes appear to be well recovered, a few show offsets comparable to the size of the observational uncertainties.
}
    \label{fig:mag_biases}
\end{figure*}

In Figure \ref{fig:mag_biases}, we show the magnitude biases for each filter and quasar host. 
The average magnitude offsets (observed minus corrected) vary by filter and quasar group.
 In the F115W filter, only EIGER quasars had data, with a very small average offset of 0.02 mag. 
 For the F150W filter, only SHELLQs quasars contributed, yielding an average offset of 0.06 mag. The F150W filter produces only confirmed upper limits for $\rm m_{F150W} \gtrsim 26.5$ mag. In the F200W filter, again only EIGER quasars were present, showing a larger negative offset of $-$0.2 mag. 
The F200W filter tends to overestimate the truth host flux for both \citet{yue2023eiger} quasar hosts discussed in this work. 
Finally, the F356W filter had data from both groups: the SHELLQs quasars showed a modest average offset of 0.1 mag, while the only EIGER quasar with a measurement exhibited a negative offset of $-$0.2 mag.
For the two observed lower limits in F356W, we confirm one as a measurement (J1146-0005 in Category C) and one as a true lower limit (J1120+0641 in Category D). Other than in the F200W filter, these are within our bin width resolution of $\sim$0.2 mag.

\subsubsection{\citet{10.48550/arxiv.2211.14329} and \citet{ding2025shellqsjwstunveilshostgalaxies} Quasar Hosts}
We infer posteriors from BlueTides for 11 of the 12 positive and tentative host detections in \citet{10.48550/arxiv.2211.14329} and \citet{ding2025shellqsjwstunveilshostgalaxies}. Of these, five quasar hosts—J2255+0251, J0844-0132, J0911+0152, J0918+0139, and J1146-0005—were reported as non-detections in the F150W band and were thus treated as lower magnitude limits. We do not infer quasar host magnitudes for J1512+4422 as there were not enough recovered quasar host magnitudes that meet the radial and quasar magnitude tolerance criteria we apply. We also note that for J2236+0032 we do not enforce the quasar magnitude accuracy cut and allow hosts exceeding the 0.05 mag threshold to ensure we have enough galaxy analogs in the sample. Combined with the limited number of host samples, this likely contributes to the larger uncertainty for this quasar host. 

Across most sources, our inferred posteriors align well with the observed JWST magnitudes: the medians and uncertainties are typically consistent within our bin width resolution of $\sim$0.2 mag. This indicates that the majority of the F150W and F356W detections in \citet{10.48550/arxiv.2211.14329} and \citet{ding2025shellqsjwstunveilshostgalaxies} are robust and not significantly biased (Category A and B).

We confirm that three of the five F150W non-detections—J0844-0132, J0911+0152, and J1146-0005—should indeed be treated as lower limits. The corrected posteriors of these host galaxies are truncated within five times the observational uncertainty and do not reveal a clear posterior peak. These are categorized as Category D in Table \ref{tab:magnitude_offsets}, indicating that both JWST and our analysis support the lower-limit interpretation.

Conversely, for J2255+0251 and J0918+0139 in F150W, the corrected posteriors suggest that these may be true detections. While these were originally flagged as lower limits, the posterior distributions flatten out, indicative of a detection. These are therefore classified as Category C. J1146-0005 is a mixed case: while its F150W detection is confirmed as a lower limit (Category D), its F356W measurement likely reflects a real detection (Category C), with a small magnitude offset of 0.1 mag.

Overall, our corrected posteriors reinforce the credibility of most reported SHELLQs measured host galaxy magnitudes and help clarify which should be treated as limits versus true measurements. The consistency in median magnitude estimates and the small bias values across filters suggest that the host detections in \citet{10.48550/arxiv.2211.14329} and \citet{ding2025shellqsjwstunveilshostgalaxies} are generally reliable, with exceptions appropriately reclassified using our pipeline.

\subsubsection{EIGER Quasar Hosts}
\label{subsec:eiger}
We infer host galaxy magnitude posteriors for two quasars from the EIGER sample \citep{yue2023eiger, marshall2024ganifseigermerging}, shown in Figure \ref{fig:weighted_images_5}\footnote{As shown in Table \ref{tab:quasar_host_comprehensive} and the photometric galaxy luminosity function in Figure~\ref{fig:photo_mags_with_quasars}, we do not have a sufficient number of bright comparison galaxies (i.e., with magnitudes $<$ 23 mag in each filter) to reliably infer posteriors for J0148+0600.}. For J159-02, our residuals suggest that the observed host magnitudes in all three filters—F115W, F200W, and F356W—are likely underestimates. The corrected magnitudes are 0.1 mag and 0.2 mag fainter than the observed values in F200W and F356W. These differences are roughly $1\sigma$ deviations in F200W and F356W, and are about the bin width tolerance reported in Table \ref{tab:bin_sizes}. The F115W measurement was reported as a lower limit in \citet{yue2023eiger}, but we find the updated quasar host magnitude posterior converges and falls into Category C (a lower limit that is better represented as a measurement). 

The F115W host detection for J1120+0641 appears extremely robust, with no offset between the observed and corrected magnitude. In contrast, the F200W and F356W host magnitudes were considered lower limits in \citet{marshall2024ganifseigermerging}, due to fitting uncertainties exceeding 0.3 mag. While the F200W correction suggests a mild dimming and may still correspond to a real detection (Category C), the F356W correction likely reflects a true lower limit (Category D). This is evident in Figure \ref{fig:weighted_images_5}, where the linear shape of the J1120+0641 CDF indicates a lack of convergence in the posterior. Overall, our corrections reinforce the interpretation from \citet{marshall2024ganifseigermerging} that the F356W measurement is a lower limit, while the F200W host may represent a marginal detection. Overall we find that the EIGER quasar hosts have a larger measurement bias than the SHELLQs quasars. This is likely due to the EIGER quasars being more luminous, making the quasar subtraction more challenging.

\begin{table*}
\centering
\begin{tabular}{l c c c c c}
\hline
\textbf{Quasar} &
\textbf{$\log_{10}(M_*/M_\odot)$} &
\textbf{$\log_{10}(M_*/M_\odot)$} &
\textbf{$\log_{10}(M_*/M_\odot)$} &
\textbf{$\Delta M_{*, \text{rerun} - \text{corrected}}$} &
\textbf{$\Delta M_{*, \text{rerun} - \text{lit}}$} \\
&
\textbf{[corrected]} &
\textbf{[JWST, rerun]} &
\textbf{[JWST, literature]} &
[dex] & [dex]\\
\hline
J2236+0032              & $11.37^{+0.23}_{-0.26}$  & $11.15^{+0.24}_{-0.31}$  & $10.96^{+0.08}_{-0.09}$  & $+0.22$  & $+0.19$ \\
J2255+0251              & $10.81^{+0.37}_{-0.37}$  & $10.77^{+0.38}_{-0.32}$  & $10.73^{+0.47}_{-0.30}$  & $+0.04$  & $+0.04$ \\
J159-02                 & $9.93^{+0.43}_{-0.79}$   & $10.15^{+0.29}_{-0.68}$  & $10.14^{+0.34}_{-0.36}$  & $-0.22$  & $+0.01$ \\
J1120+0641$^{*\dagger}$ & $9.13^{+0.79}_{-0.69}$   & $9.39^{+0.49}_{-0.64}$   & $9.48^{+0.26}_{-0.27}$   & $-0.26$  & $-0.09$ \\
J0844-0052              & $9.66^{+0.52}_{-0.76}$   & $9.53^{+0.47}_{-0.70}$   & $9.49^{+0.52}_{-0.42}$   & $+0.13$  & $+0.04$ \\
J0844-0132$^{*\dagger}$ & --                        & --                        & $10.03^{+0.53}_{-0.45}$  & --       & --      \\
J0911+0152$^{*\dagger}$ & --                        & --                        & $9.72^{+0.49}_{-0.38}$   & --       & --      \\
J0918+0139              & $10.44^{+0.46}_{-0.49}$  & $10.41^{+0.45}_{-0.49}$  & $10.21^{+0.52}_{-0.37}$  & $+0.03$  & $+0.20$ \\
J1425-0015              & $10.72^{+0.29}_{-0.32}$  & $10.64^{+0.30}_{-0.36}$  & $10.54^{+0.36}_{-0.37}$  & $+0.08$  & $+0.10$ \\
J1512+4422              & $10.42^{+0.39}_{-0.21}$  & $10.45^{+0.38}_{-0.20}$  & $10.57^{+0.29}_{-0.41}$  & $-0.03$  & $-0.12$ \\
J1525+4303              & $10.35^{+0.34}_{-0.44}$  & $10.20^{+0.30}_{-0.53}$  & $10.00^{+0.30}_{-0.38}$  & $+0.15$  & $+0.20$ \\
J1146-0005$^{*\dagger}$ & --                        & --                        & $<9.92$                   & --       & --      \\
J1146+0124              & $10.73^{+0.30}_{-0.26}$  & $10.56^{+0.23}_{-0.27}$  & $10.38^{+0.22}_{-0.30}$  & $+0.17$  & $+0.18$ \\
J0217-0208              & $10.40^{+0.30}_{-0.68}$  & $10.34^{+0.33}_{-0.70}$  & $10.06^{+0.28}_{-0.32}$  & $+0.06$  & $+0.28$ \\

\hline
\end{tabular}
\caption{Comparison of inferred stellar masses with the corrected magnitudes, the \texttt{PROSPECTOR} SED fit with the original magnitudes, and the original measured values reported in \citet{yue2023eiger}, \citet{marshall2024ganifseigermerging}, and \citet{ding2025shellqsjwstunveilshostgalaxies}. The stellar mass values for quasar hosts remeasured in \citet{li2025dichotomynuclearhostgalaxy} are not included in this Table but are included in Figure \ref{fig:stellar_mass_resid}. Column 2 gives the BlueTides corrected mass, column 3 shows our rerun JWST-based estimates, and column 4 lists values from the literature. Columns 5 and 6 show the differences between the rerun JWST values and the corrected or literature values, respectively. We note that stellar masses could not be determined for certain quasar hosts; these are indicated with a $\dagger$ and $^*$, respectively. The stellar masses of these host galaxies are not yet constrained.}
\label{tab:stellar_mass_offsets}
\end{table*}

\subsection{Corrected stellar masses from quasar host removal}
The true physical interpretation of our updated magnitude posteriors lies in the stellar mass measurement. We re-perform the SED fitting on the original JWST host galaxy magnitudes, to ensure a direct comparison of the masses before and after being corrected by our approach. This makes a significant difference for the stellar masses of the \citet{10.48550/arxiv.2211.14329} and \citet{ding2025shellqsjwstunveilshostgalaxies} quasar host galaxies, which were calculated using the \textit{gsf} fitting software instead of \texttt{PROSPECTOR}. We find that changing from \textit{gsf} to \texttt{PROSPECTOR} can change the stellar mass by up to 0.22 dex, as seen in the rightmost column of Table \ref{tab:stellar_mass_offsets}. 

\begin{figure*}
    \centering
    \includegraphics[width=\linewidth]{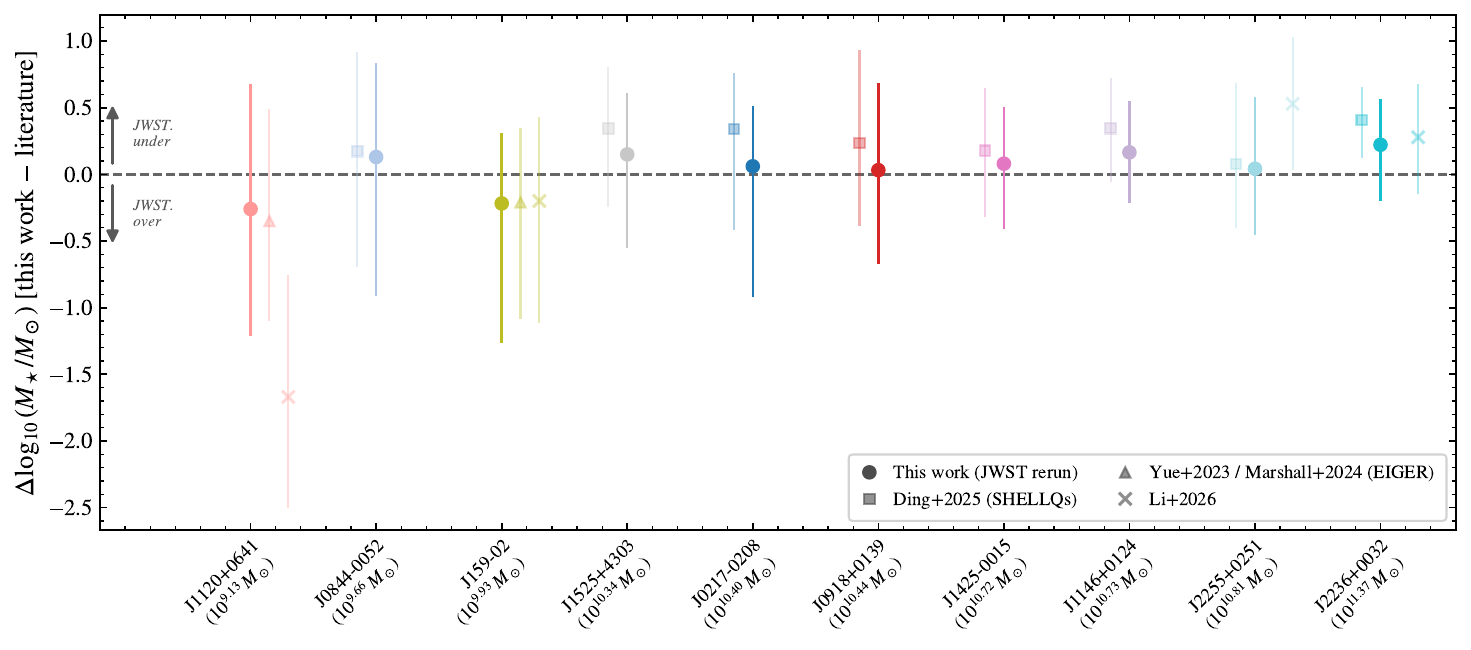}
    \caption{Residuals in stellar mass between this work and literature measurements for our sample of quasar host galaxies, sorted by increasing BlueTides-corrected stellar mass. For each quasar, the BlueTides-corrected mass (this work) is subtracted from each literature estimate: our JWST rerun (circles), \citet{ding2025shellqsjwstunveilshostgalaxies} (SHELLQs; squares), \citet{yue2023eiger} and \citet{marshall2024ganifseigermerging} (EIGER, triangles), and \citet{li2025dichotomynuclearhostgalaxy} (crosses). Literature markers are shown at slightly offset x positions for clarity. Error bars reflect the quadrature sum of uncertainties from both measurements. Positive residuals indicate that that the stellar mass is underestimated relative to our BlueTides-corrected values, and negative residuals indicate an overestimate. The BlueTides-corrected stellar mass for each quasar host is indicated on the x-axis next to the name of the quasar.}
    \label{fig:stellar_mass_resid}
\end{figure*}

In Figure~\ref{fig:stellar_mass_resid} and within the stellar mass range shown in Figure~\ref{fig:corrected}, we also include the stellar mass values reported and remeasured in \citet{li2025dichotomynuclearhostgalaxy} for J1120+0641, J159$-$02, J2255+0251, and J2236+0032. These measurements were obtained using the newly developed code \texttt{GALFITs} (R. Li \& L. C. Ho 2026, in preparation), which performs multiband image decomposition. For J1120+0641, \citet{li2025dichotomynuclearhostgalaxy} report a stellar mass approximately 1 dex higher than both our inferred value and that reported in \citet{marshall2024ganifseigermerging}. This discrepancy is likely driven by the significantly fainter F356W host magnitude found ($23.43 \pm 0.21$), which is $\sim$2 magnitudes lower than the value recovered in this work.

 Different SED fitting codes and assumptions are known to result in different stellar mass estimates (see \citealt{mizener2025cluesiiiuserchoices, narayanan2023outshiningrecentstarformation}), which is likely to be a particular issue for these galaxies which only have photometric measurements in two or three filters. However, the SED fitting bias is a complex topic (e.g., \citealt{cochrane2024highzstellarmassesrecovered} finds high-$z$ host stellar masses are likely robustly recovered using correct assumptions). We leave a detailed investigation of how these effects may impact high-$z$ quasar host mass measurements to future work. Here we focus on the \textit{difference} between stellar masses caused by the magnitude biases as calculated above, produced by the same SED fitting code and assumptions.

These mass estimates are listed in Table \ref{tab:stellar_mass_offsets} and compared in Figure \ref{fig:stellar_mass}.
We find that PSF subtraction leads to an average stellar mass overestimation of 0.051 dex, with individual offsets ranging from –0.079 dex (underestimation) to 0.314 dex (overestimation). J1146+0124 has the largest overestimation, which is likely due to the large error bar on the F356W corrected magnitude. However, it is still approximately within the observational uncertainties. Many of the stellar masses remain robust. While the magnitude estimates themselves have changed little, the updated uncertainty ranges used in the SED fitting likely contribute to these shifts—introducing more or less flexibility in the inferred mass depending on the photometric error. We find some evidence for mild overestimation of stellar masses, although most offsets remain within the observational uncertainties. At this stage, uncertainties arising from the SED fitting appear to have a greater impact than those introduced by PSF removal.

In general, we find that host galaxy magnitudes $\lesssim  26$ mag and stellar masses above $10^{9.5}~M_\odot$ can be reliably recovered. Outside of these regimes, the constraints become significantly less robust. J0911+0152 appears to be a potential exception since this source is one of the faintest known high-$z$ quasars, with $m_{F356W}=24.10\pm0.02$ mag, $\sim$2 mag fainter than the brighter SHELLQs quasars, and $\sim$5 mag fainter than the EIGER quasars. The \citet{ding2025shellqsjwstunveilshostgalaxies} F356W modeling of J0911+0152 required a fixed radius for the host galaxy model to converge, and the Bayesian Information Criterion (BIC) comparison was over an order of magnitude lower than that of other detections in the SHELLQs sample. This very faint $m_{F356W}=26.56\pm0.30$ mag host galaxy appears to push the limits of the quasar subtraction method. Given its non-detection in F150W, its stellar mass could not be reliably constrained. Overall, a magnitude of $\lesssim  26$ mag is generally required for a successful detection of a quasar host following these observing strategies.

\subsubsection{How low could the stellar mass of J1120+0641 be?}
\citet{marshall2024ganifseigermerging} finds that J1120+0641 could host one of
the largest black hole to stellar mass ratios ever reported for a quasar, with
$M_{\bullet}/M_{\star} = 0.63^{+0.54}_{-0.31}$. However, previous studies did
not use upper limits in the SED fitting for the tentative detections in filters, instead treating them like detections in \texttt{PROSPECTOR}. We
address this by setting the flux to zero and the $1\sigma$ errors to the upper
limit value, following Appendix~A of \citet{2012PASP..124.1208S}, to properly account for upper limits.

To ensure convergence of \texttt{PROSPECTOR}, we include additional upper limits
from \citet{stone2023undermassive} in the F210M, F360M, and F480M filters (0.48, 0.80, 0.62 $\mu Jy$, respectively).
While these measurements do not account for the nearby companion galaxy, they
serve as reasonable upper limits. We follow the SED fitting procedure of
Section~\ref{subsec:sed}, excluding a starburst component and emission line detections as in
\citet{marshall2024ganifseigermerging}. The resulting stellar mass is consistent with \citet{marshall2024ganifseigermerging}, confirming our SED fitting procedure before using our the upper limit corrections. We do not recover the stellar mass reported by
\citet{li2025dichotomynuclearhostgalaxy} for J1120+0641
($\log M_{\star}/M_{\odot} = 10.80 \pm 0.45$) likely due to differences in the
measured photometry. Specifically,
\citet{li2025dichotomynuclearhostgalaxy} report a detection in F356W
($m_{\mathrm{F356W}} = 23.43 \pm 0.21$), whereas our analysis yields only an
upper limit nearly one magnitude fainter. 

As for all quasar hosts with limited photometric magnitude measurements, this remains a statistically challenging case. The number of free parameters
exceeds the number of secure detections, making a reduced $\chi^2$ difficult to interpret. We therefore compare absolute $\chi^2$ values, since both models
share the same number of free parameters and magnitudes. The BlueTides updated mass based fit yields a
slightly lower $\chi^2$, though both solutions are consistent within the
uncertainties. 

\begin{table}
\centering
\caption{Possible black hole to stellar mass ratios for J1120+0641. This work's derived stellar masses are 
derived from SED fitting with upper limits treated following 
\citet{2012PASP..124.1208S}. $M_{\bullet} = (1.9^{+2.9}_{-1.1})\times10^9\,M_\odot$ 
from \citet{marshall2024ganifseigermerging}.}
\label{tab:j1120_masses}
\begin{threeparttable}
\begin{tabular}{lccc}
\hline
Method & $\log M_{\star}/M_{\odot}$ & $M_{\bullet}/M_{\star}$ & $\chi^2$ \\
\hline
\citet{marshall2024ganifseigermerging} & $9.48^{+0.26}_{-0.27}$ & $0.63^{+0.54}_{-0.31}$ & -- \\
JWST (this work)      & $9.39^{+0.49}_{-0.64}$ & $0.77^{+2.6}_{-0.52}$ & 1.48 \\
BlueTides (this work) & $9.13^{+0.79}_{-0.69}$ & $1.41^{+5.5}_{-1.19}$ & 0.49 \\
\hline
\end{tabular}
\end{threeparttable}
\end{table}

Adopting
$M_{\bullet} = (1.9^{+2.9}_{-1.1})\times10^9\,M_\odot$ from
\citet{marshall2024ganifseigermerging}, our results span from a median value of $0.63$ to $1.41$. This places inferred black hole to stellar mass ratios for J1120+0641 place this object in an extreme regime, approaching those reported for LRD systems (e.g., \citealt{2025ApJ...983...60C}).  However, we show here that even small magnitude uncertainties and the treatment of photometric upper limits can shift the inferred stellar mass by more than an order of magnitude. Robust constraints on stellar mass therefore require maximizing the number of photometric detections and properly incorporating upper limits across the available photometric filters.

\subsubsection{Lack of stellar mass constraints for J0844-0132, J1146-0005, and J0911+0152}
\label{subsec:J1120}
The F150W host galaxy magnitudes for J0844-0132, J0911+0152, and J1146-0005 are likely lower limits (Category D), as noted in both \citet{10.48550/arxiv.2211.14329} and \citet{ding2025shellqsjwstunveilshostgalaxies}. Because of this, any stellar masses inferred from these magnitudes should be treated with extreme caution and can not yet be reliably constrained. For a reliable stellar mass estimate with meaningful constraints on stellar age, we require detections in two filters at minimum, one on either side of the Balmer break. Although we note that the F356W filter, which probes the rest-frame optical at $z \sim 6$, provides the strongest constraints on the stellar mass since it traces the older stellar populations that dominate the galaxy’s mass budget. \texttt{PROSPECTOR} fails to provide a reliable stellar mass estimate with only one detection and upper limits on either side of the Balmer break. We leave these stellar mass constraints to future observations.

We conclude that we cannot reliably constrain the stellar masses of J0844-0132, J0911+0152, and J1146-0005 due to the presence of lower-limit magnitudes. Any reported stellar masses for these sources should be considered highly uncertain, and future observations with secure detections across the Balmer break are needed to obtain physically meaningful estimates.

\begin{figure}
    \centering
    \includegraphics[width=\linewidth]{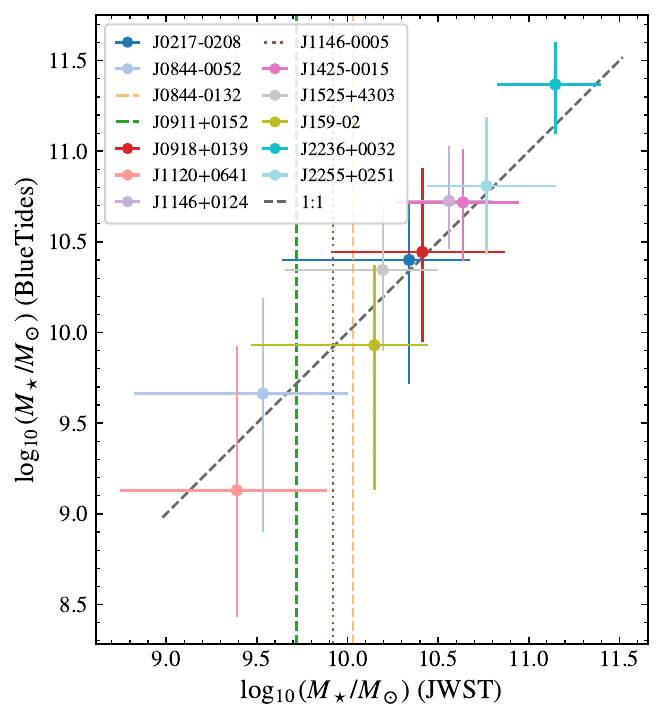}
    \caption{Overview of measured versus updated stellar masses with our pipeline. The stellar masses shown as vertical lines indicate that they are not well constrained with the currently available JWST data. These masses are generated from the SED fitting pipeline described in Section \ref{subsec:sed}. Note that we recalculate the stellar masses to ensure that the stellar mass is calculated in the same way for both the observed and inferred magnitudes. Dashed vertical lines indicate measurements that we find are unconstrained.}
    \label{fig:stellar_mass}
\end{figure}

\begin{figure*}
    \centering
    \includegraphics[width=\linewidth]{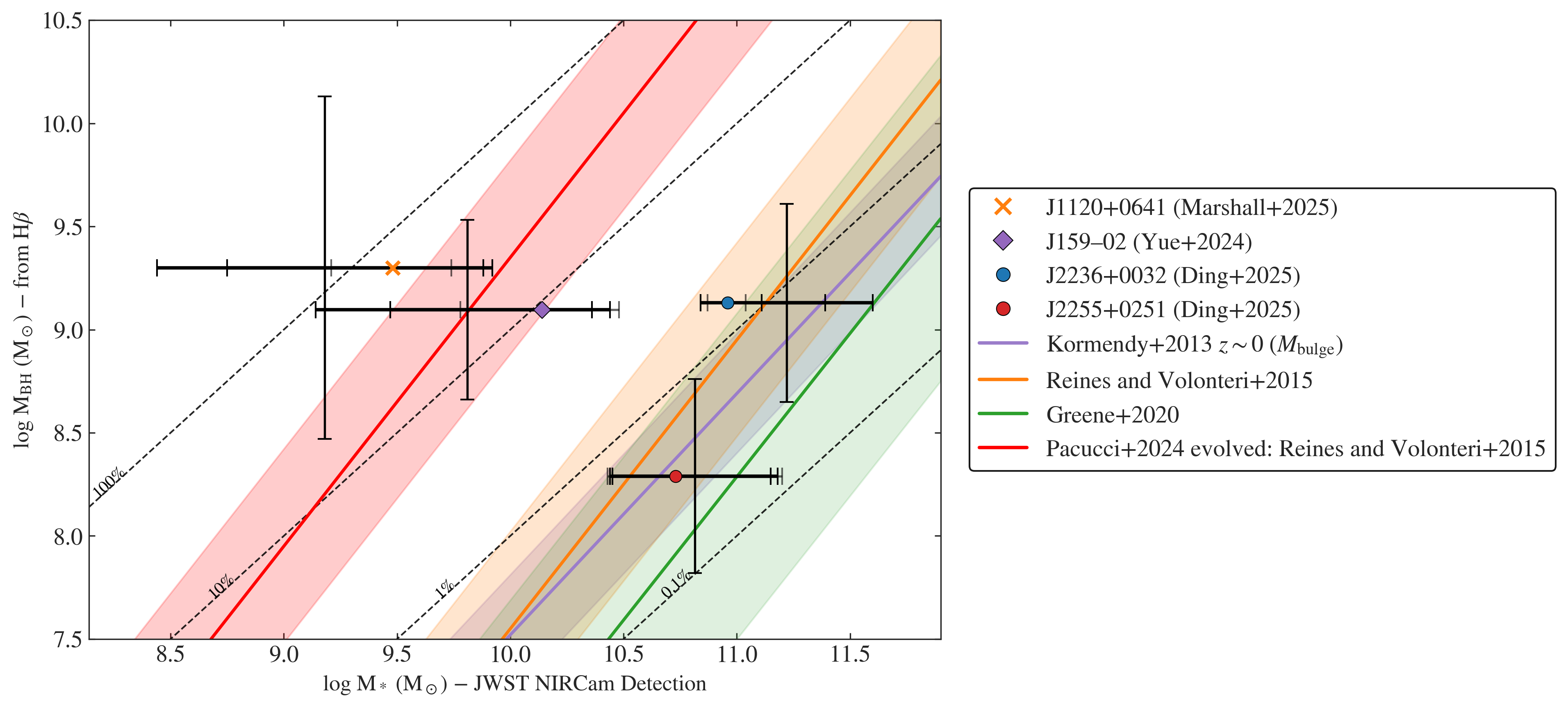}
    \caption{We overlay stellar mass ranges from this work on top of all available observed JWST stellar mass measurements from the literature (\citet{yue2023eiger}, \citet{ding2025shellqsjwstunveilshostgalaxies}, \citet{marshall2024ganifseigermerging}, or \citet{li2025dichotomynuclearhostgalaxy})  as seen in Figure \ref{fig:bh_sm} for all quasar hosts with published black hole masses. Horizontal error bars span the reported $\pm1\sigma$ stellar mass uncertainty from each source with caps marking the endpoints. We also overplot both the low-z relations and evolved \citet{Reines_2015} low-z relation using the prescription in \citet{Pacucci_2024}. Most stellar mass updates lie within the observational uncertainty but show potential for slight misestimation by JWST within 0.3 dex.}
    \label{fig:corrected}
\end{figure*}

\section{Discussion}
\label{sec:discussion}
JWST has opened a new window into the high-$z$ Universe, enabling unprecedented exploration of early galaxy and quasar evolution. However, observational biases remain critical in distinguishing genuine physical trends from spurious results. In this work, we develop a model to correct for the effects of PSF subtraction in high-$z$ quasars observed with JWST. With larger-scale simulations, this framework could be extended to even brighter host galaxies.

Recent studies have reported discrepancies in the high-$z$ $M_{\bullet}$–$M_{*}$ relation compared to the low-$z$ relation, although these differences are becoming increasingly ambiguous as JWST observes more high-$z$ systems. For instance, \citet{2025ApJ...981...19L} found that the high-$z$ $M_{\bullet}$–$M_{*}$ relation can be reconciled with the low-$z$ relation when accounting for selection effects and measurement uncertainties. Their sample includes 32 AGN at $z > 4$ with bolometric luminosities of $\sim10^{11-12},L_{\odot}$---comparable in brightness to most quasars analyzed in \citet{10.48550/arxiv.2211.14329} and \citet{ding2025shellqsjwstunveilshostgalaxies}, although fainter than the EIGER quasars.

Our results suggest that PSF-related measurement uncertainties are unlikely to be the only dominant source of bias in the stellar mass estimates of these host galaxies. Instead, the biases likely arise from a combination of uncertainties in the PSF, the assumed Sérsic profile and its fitted parameters, and residual errors associated with quasar subtraction. We plan to further investigate the implications of this finding for lower-luminosity AGN and future JWST detections of high-$z$ quasar hosts.

\subsection{Why set the Sérsic index = 1?}
\label{app:free_sersic}
For the purpose of isolating the bias on the recovered Sérsic magnitude introduced by quasar removal, we fix the Sérsic index to $n = 1$ throughout this work. This choice is motivated by several considerations. First, allowing $n$ to vary freely can lead to poorer convergence and increased instability in the fits. Second, introducing a free Sérsic index significantly expands the parameter space and introduces additional degeneracies (particularly between $n$, effective radius, and total magnitude as described in Appendix \ref{app:more_sersic}) that are beyond the scope of the present analysis. Finally, we find that the impact of fixing $n = 1$ on the recovered host magnitudes is minimal. 

\begin{figure*}
    \centering
    \includegraphics[width=\linewidth]{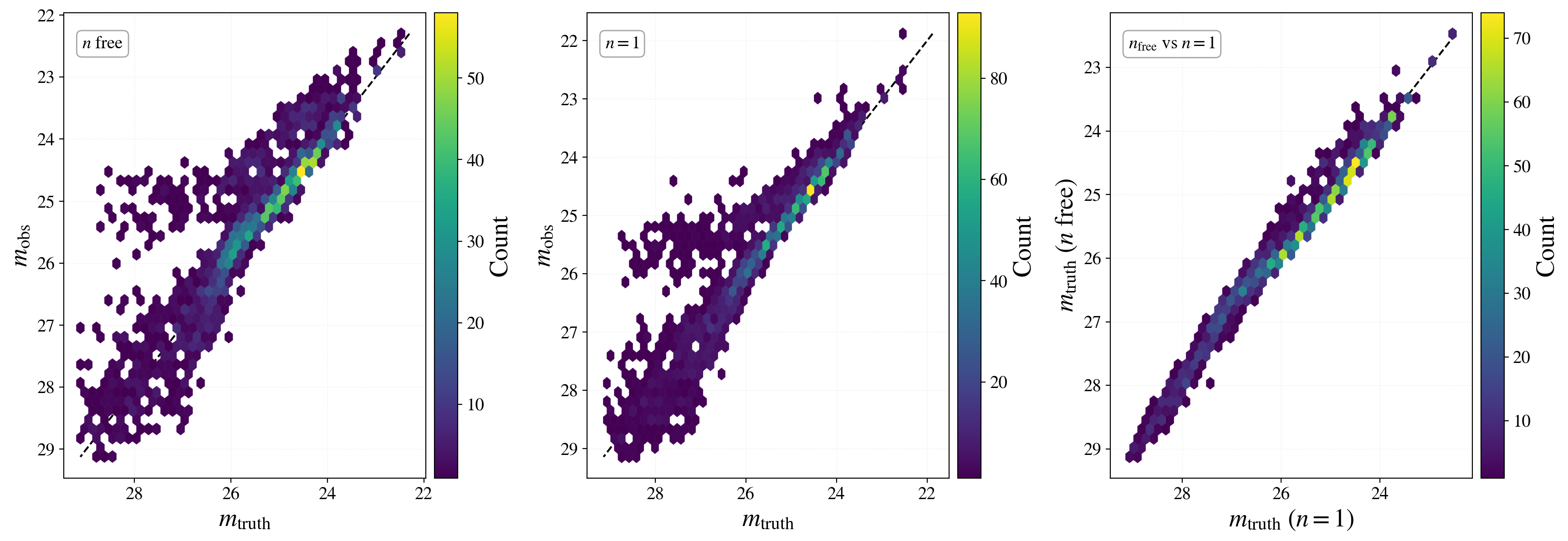}
    \caption{Comparison of inferred host magnitudes to the input truth magnitudes in F356W for Sérsic models with $n=1$ (fixed) and free $n$. From left to right: results for $n=1$, results for free $n$, and a comparison of the input host galaxy truth magnitudes only, showing that the two are recovered without significant bias. The colorbar indicates the number of hosts in that bin. We do not apply cuts based on accurate radial or quasar magnitude recovery in this figure, unlike in the main analysis of this paper.}
    \label{fig:mag_comp}
\end{figure*}

An explicit comparison for quasar J1120+0641 is shown in Figure~\ref{fig:mag_comp}, where the inferred magnitudes are generally consistent within uncertainties between the fixed-$n$ and free-$n$ models. In the F356W band, however, allowing the Sérsic index to vary produces a slight systematic shift toward brighter, more incorrect recovered magnitudes than truth magnitudes. This suggests that our fixed $n=1$ measurements are conservative and the corrections in Section \ref{subsec:galaxy_mag_biases} may be lower limits on the host magnitude errors in these filters. Future work should more comprehensively explore the structural parameter space and incorporate more realistic galaxy profile models to lessen the Sérsic profile assumption errors discussed in Appendix~\ref{appendix:sersic} and \ref{app:more_sersic}.

\subsection{\texorpdfstring{Comparison to mock observations in \citet{yue2023eiger}, \citet{Zhuang_2024}, and \citet{ding2025shellqsjwstunveilshostgalaxies}}{Comparison to mock observations in Yue et al., Zhuang et al., and Ding et al.}}

Other works have aimed to estimate the impact of PSF subtraction on measured host galaxy properties. In a similar effort, \citet{Zhuang_2024} investigated the effects of common JWST PSF mismatches in AGN–host decomposition and proposed an improved method for constructing PSFs from JWST observations, following \citet{2011ASPC..442..435B}. They performed mock PSF subtraction on thousands of simulated galaxies spanning a range of AGN-to-host flux ratios and PSF models. Their results show that systematic biases in host magnitudes arise from imperfect PSF modeling: for AGN-to-host ratios between 0.1 and 5, the true host magnitudes are on average dimmer by approximately 1.2 times the observational uncertainty. In more extreme cases—where the AGN outshines the host by a factor of 10—this bias can reach up to 100 times the uncertainty. We find more optimistic host magnitude recoveries due to PSF removal on average compared to \citet{Zhuang_2024}, although we do not account for imperfect PSF models in this analysis. 

In \citet{yue2023eiger}, mock observations are also performed to determine the accuracy on reported values such as magnitude and radius. They also find that the errors on the inferred galaxy parameters are larger for much fainter PSF stars but do not set the same limit that we do that the PSF star's magnitude must be at least as bright as the quasar. They instead set the criterion that the difference between the magnitude of the quasar and PSF star must be less than 1. As seen in Section \ref{subsec:results_psf}, we find that this criterion likely still significantly overestimates the host galaxy flux. Indeed, in Section \ref{subsec:eiger}, we find that the \citet{yue2023eiger} quasar host fluxes are overestimated. In terms of mock observations, \citet{yue2023eiger} add an exponential galaxy light profile matching their detected host galaxies to the real PSF star images. They find that the best fit magnitudes are consistent with the magnitudes they recover in their work with this method, except for J159-02 in the F115W filter. However, we find that the magnitude of the quasar host of J159-02 in the F200W and F356W filter nearly exceeds observational uncertainties. We also confirm the measurement of J1120+0641 in F200W as a detection rather than a lower limit and provide an analysis of how low the stellar mass could be in Section \ref{subsec:J1120}.

Similarly in \citet{ding2025shellqsjwstunveilshostgalaxies}, mock observations are also performed to test the reliability of their results. Our magnitude corrections shown in Figure \ref{fig:mag_biases} broadly support their image simulation tests which demonstrate that systematic biases in F356W and F150W host magnitude measurements are typically smaller than 0.3 mag in most cases. Our corrected posteriors show that the majority of quasar host magnitudes are either accurate or modestly biased (Categories A and B), with a median offset of 0.08 mag across all filters indicating quasar hosts might be slightly dimmer than measured. However \citet{ding2025shellqsjwstunveilshostgalaxies} find that  J0918+0139 in F356W is biased by $\sim$0.45 mag corresponding to a $\sim$0.1 dex offset in stellar mass. We find that the J0918+0139 quasar host is likely a robust magnitude detection and the stellar mass offset is negligible with our corrected magnitudes. 

Together, these findings validate the robustness of JWST host detections in many cases while highlighting where caution is needed, especially for lower-limit measurements and sources lacking full spectral coverage. Our approach complements existing PSF modeling efforts by providing a posterior-level reanalysis that helps distinguish true detections from uncertain or biased measurements.

As more high-$z$ quasar host galaxy detections are made, there is a critical need to ensure the properties measured for the host are robust. We can use our pipeline to correct magnitudes from future host galaxy detections. The mismatch of the PSF in the PSF removal also biases the recovered quasar host galaxy magnitude and is key to explore in future work. We can also extend our work to smaller black holes, such as LRDs, to infer the effects of AGN contamination. Future work is needed to explore the observational effects of the Sérsic index and radii of the host galaxy, which can also contribute to the host galaxy magnitude uncertainty as described in Appendices \ref{app:more_sersic} and \ref{appendix:sersic}.

\subsection{\texorpdfstring{How could the $M_{\bullet}$--$M_{*}$ relation change?}{How could the M•–M* relation change?}}
A key observable for understanding how these high-$z$ measurements constrain models of black hole and galaxy coevolution is the $M_{\bullet}$--$M_{*}$ relation. We perform consistent SED fitting using \texttt{prospector} to infer stellar masses from both the observed and bias corrected host galaxy magnitudes (Section \ref{subsec:sed}). The resulting stellar masses are shown in Table \ref{tab:stellar_mass_offsets} and plotted in Figures \ref{fig:stellar_mass} and \ref{fig:corrected}. These remain mostly consistent with those determined in  \citet{yue2023eiger}, \citet{ding2025shellqsjwstunveilshostgalaxies} and \citet{li2025dichotomynuclearhostgalaxy}.

Overall, we do not find evidence for a systematic bias in the stellar masses inferred from current JWST quasar host detections. Instead, the corrections appear to depend on quasar luminosity. For the most luminous quasars, such as J1120+0641 and J159$-$02, the host stellar masses inferred from the observed photometry may be slightly overestimated. In contrast, J2236+0032, one of the dimmer quasars from the SHELLQs sample, could have host stellar masses that are slightly larger than previously inferred once observational biases are accounted for. \citet{silverman2025shellqsjwstperspectiveintrinsicmass} finds that the high-z $M_{\bullet}$--$M_{*}$ is actually in line with what is expected when accounting for selection effects and observational uncertainties.
However, in all cases these shifts remain within the observational uncertainties. Our results are inconsistent with the ``dichotomy'' highlighted by \citet{li2025dichotomynuclearhostgalaxy}, in which more luminous sources appear less overmassive. For our subsample, we find the opposite. 

After applying our corrections, only four JWST high-$z$ quasars currently have both published black hole and stellar mass estimates. In Figure \ref{fig:corrected}, we compare these systems to the low-$z$ $M_{\bullet}$--$M_{*}$ relations from \citet{2013ARA&A..51..511K}, \citet{Reines_2015}, and \citet{2020ARA&A..58..257G}. For reference, we also show the stellar masses originally reported in the observational studies. We also compare these masses to the evolved low-z \citet{Reines_2015} relation using the high-$z$ \citet{Pacucci_2024} $M_{\bullet}$--$M_{*}$ evolutionary prescription at $z = 6$, where $M_{\bullet}$--$M_{*}$ evolves according to $(1+z)^{5/2}$ (as opposed to $(1+z)^{3/2}$ in \citealt{2003ApJ...595..614W}). \citet{Pacucci_2024} matched the new JWST black hole and stellar mass measurements (including LRDs) to an evolution proportional to $(1+z)^{5/2}$ by arguing that star formation will be inefficient when the black hole to stellar mass ratio exceeds $8\times10^{18}(n\Lambda/f_{\rm edd})[(\Omega_bM_h)/(\Omega_m M_{*}) - 1)]$, where $n$ is the gas number density, $\Lambda$ is the cooling function, and $f_{\rm edd}$ is the Eddington ratio.  With our updated stellar masses, the high-$z$ quasar hosts lie between the unevolved low-$z$ relation and the evolved relation predicted by \citet{Pacucci_2024}. The corrections are modest and sample size is small, but they reinforce the emerging picture that the relationship between black hole mass and stellar mass at high redshift may differ substantially from that observed locally.

At the same time, the current sample of luminous quasars with reliable host detections remains extremely small. In addition, black hole masses at high redshift are typically estimated using virial scaling relations calibrated at low redshift, which may introduce systematic uncertainties of up to a factor of a few or more \citep{collaboration2025spatiallyresolvedbroadline, nature_bh_dynamical}. These black hole mass uncertainties, combined with the small sample size, make it difficult to draw firm conclusions about the true evolution of the $M_{\bullet}$--$M_{*}$ relation.

\section{Conclusions}
\label{sec:conclusion}
We developed a pipeline to determine the observational bias in JWST high-$z$ quasar host detection arising from the removal of the bright quasar point source, using mock observations of BlueTides simulated galaxies. We correct the stellar masses from \citet{10.48550/arxiv.2211.14329}, \citet{yue2023eiger}, \citet{marshall2024ganifseigermerging}, and \citet{ding2025shellqsjwstunveilshostgalaxies}—representing all currently available high-$z$ quasar host detections with published magnitudes and analyzed using a Bayesian point source removal technique. We tabulate our corrected magnitude and stellar mass estimates without the effects of PSF removal in Table \ref{tab:magnitude_offsets}. Below we summarize the key findings from this work:

\begin{enumerate}[labelwidth=*, leftmargin=4mm]
\item \textbf{Quasar host decompositions are sensitive to systematic uncertainties in quasar subtraction and galaxy profile assumptions} - 
Small errors in the recovered quasar magnitude ($\sim 0.01$ mag) can correspond to flux differences comparable to or exceeding the total host galaxy flux. This leads to biased host galaxy properties. This effect is most pronounced for luminous quasars with intrinsically faint hosts, where even percent level uncertainties in quasar flux can approach the host magnitude itself. These biases arise from a combination of uncertainties in quasar subtraction, assumptions about the host galaxy Sérsic profile, and image noise. These all impact the recovered host flux and consequently the inferred stellar masses. 
\item \textbf{Extracted quasar host galaxy magnitudes are generally robust or lower limits} — Using our corrections, we find that all JWST quasar host galaxy magnitude measurements are either confirmed detections, biased detections, likely detections misclassified as lower limits, or true lower limits. Across all filters and quasars, the median offset between observed and corrected magnitudes is 0.08 mag. These differences range from a maximum positive offset of 0.14 mag (indicating slight underestimation in reported host brightness) to a minimum of $-$0.22 mag (an overestimation in J1120+0641’s F200W host detection). Overall, JWST photometry tends to either correctly recover or modestly misestimate host magnitudes, although several remain consistent with flux upper limits and should be treated as non-detections.
    
\item \textbf{Stellar mass estimates from JWST photometry are robust to quasar subtraction biases, generally being slightly overestimated} -- Systematic uncertainties from PSF removal are minimal, with a median stellar mass underestimation of just -0.07 dex (corrected - JWST rerun), a maximum of 0.22 dex, and a minimum underestimation of -0.26 dex. 
J1120+0641 has the largest difference in stellar mass (0.26 dex) but is still approximately within the quoted observational uncertainties. This has implications for the black hole to stellar mass relations at high-$z$, and may indicate that these quasar hosts are truly undermassive or overmassive compared to their black holes.

\item \textbf{The stellar mass of  J0844-0132,
J0911+0152, and J1146-0005 are not yet constrained} -- These quasar host detections include at least one key photometric band that is likely a lower limit (Category D), preventing robust stellar mass estimates. Adding deeper observations with improved magnitude estimates  on both sides of the Balmer break is essential for obtaining physically meaningful stellar mass constraints.

\item \textbf{Correcting observational biases does not remove the high-$z$ $M_{\bullet}$--$M_{*}$ tension} -- 
After accounting for quasar subtraction biases and uncertainties in host galaxy magnitude recovery, the inferred stellar masses remain broadly consistent with previous measurements within observational uncertainties. As a result, the apparent discrepancy between black hole and stellar mass at high redshift persists. Some quasar hosts remain undermassive relative to their black holes, while others appear consistent with or above expectations from local scaling relations. This supports the emerging picture that the diversity seen in JWST quasar hosts is unlikely to arise purely from methodological biases and may instead reflect genuine differences in the growth of early black holes and galaxies.

\item \textbf{Correcting future high-$z$ quasar host stellar mass estimates} -- We can extend this framework to future high-$z$ quasar host measurements with analogs in the BlueTides simulation and smaller AGN within ASTRID. As more detections of high-$z$ quasar hosts become available, this approach can test biases on estimates of the black hole–stellar mass relation. Such corrections will be increasingly important for accurately tracing galaxy and AGN co-evolution in the early universe.

\end{enumerate}

\section*{Data Availability}
The data from the BlueTides simulation is available at \url{http://bluetides.psc.edu}. To run a high-$z$ AGN host galaxy detection through the pipeline, please send an email to SB.

\section*{Acknowledgements}

SB is grateful for insightful discussions with Chris Lovell, Adam Carnall, Tong Cheunchitra, Vasily Kokorev, Fabio Pacucci, Yuxiang Qin, Emma Ryan-Weber, and Rachel Webster, which improved this work. SB also thanks Jarrett Johnson and Phoebe Upton-Sanderbeck for the valuable discussions during her visit to the Center for Theoretical Astrophysics at Los Alamos National Laboratory. SB is supported by the Melbourne Research Scholarship and N D Goldsworthy Scholarship. She also received support through the Astronomical Society of Australia Travel Grant and Alan Kenneth Head Travel Grant. MAM acknowledges support by the Laboratory Directed Research and Development program of Los Alamos National Laboratory under project number 20240752PRD1. This work was partially performed on the OzSTAR national facility at Swinburne University of Technology. OzSTAR is funded by Swinburne University of Technology and the National Collaborative Research Infrastructure Strategy (NCRIS). This research made use of Python packages mpi4py \citep{mpi4py5, mpi4py4, mpi4py3, mpi4py2, mpi4py1}, NumPy \citep{harris2020array}, Matplotlib \citep{Hunter:2007}, AstroPy \citep{astropy:2013, astropy:2018, astropy:2022}, SciPy \citep{2020SciPy-NMeth}, emcee \citep{Foreman_Mackey_2013}, and statsmodels \citep{seabold2010statsmodels}. The majority of this research was conducted on Wurundjeri, Ngunnawal (Ngunawal),  Ngambri, and Pueblo land. Sovereignty was never ceded.


\bibliographystyle{mnras}
\bibliography{main}

\appendix

\section{Quasar magnitude biases}
\label{app:quasar}
\begin{figure}
    \centering
    \includegraphics[width=\linewidth]{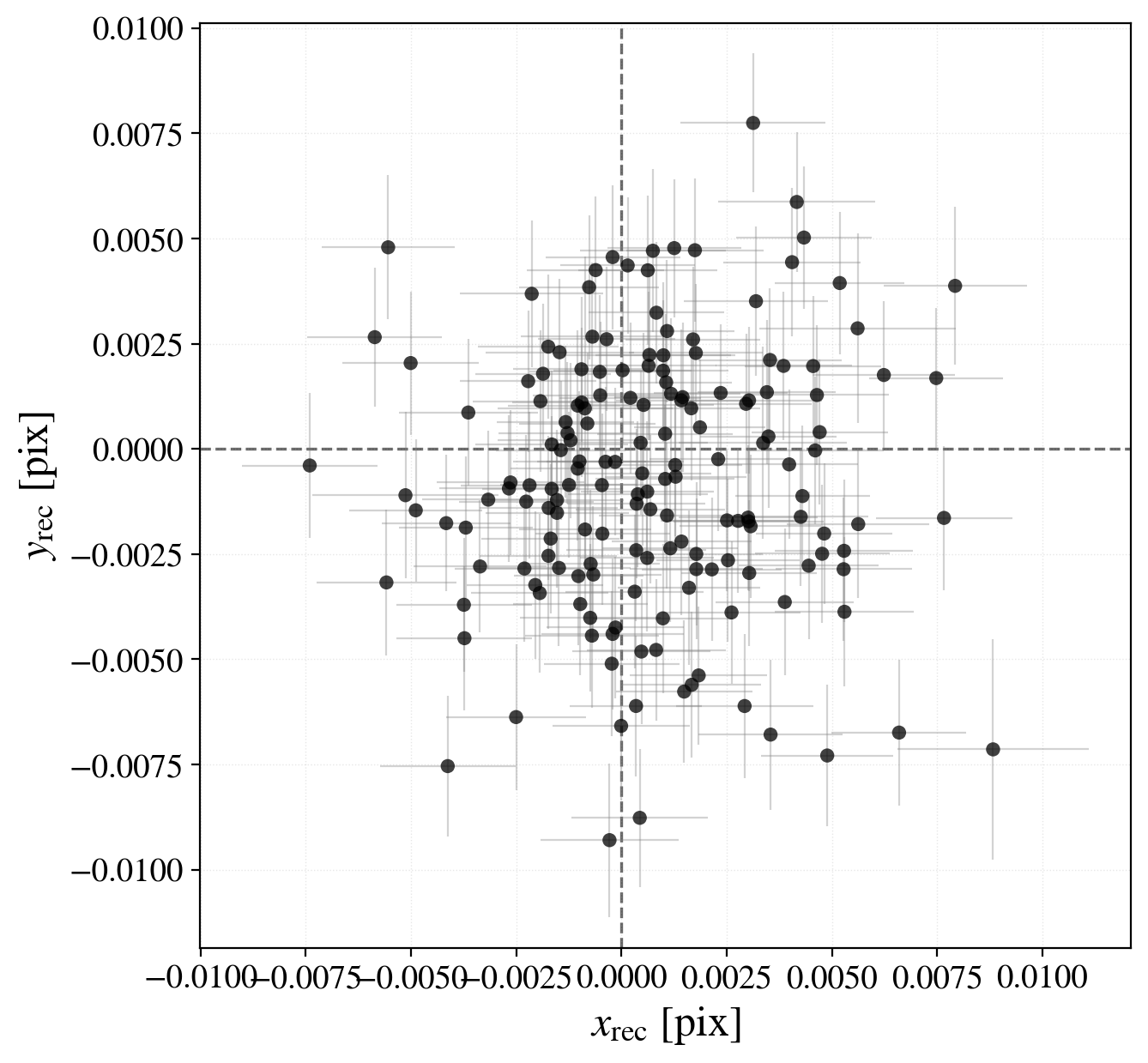}
    \caption{Example recovery of the quasar position for each BlueTides galaxies used in the J1120+0641 analysis in the F356W filter. The quasar position is always recovered within the pixel.}
    \label{fig:pos_quasar}
\end{figure}

For each quasar host decomposition, we adopt a uniform prior on the quasar magnitude spanning $\pm1$ mag around the input value. Despite this broad prior, the quasar magnitude is consistently recovered to within $\sim0.01$ mag (Table~\ref{tab:magnitude_offsets}). As illustrated in Figure~\ref{fig:pos_quasar} for J1120+0641, the inferred quasar positions are also accurate at the subpixel level. To ensure robust decompositions, we exclude systems where the recovered quasar magnitude uncertainty exceeds 0.05 mag.

\begin{figure}
    \centering
    \includegraphics[width=\linewidth]{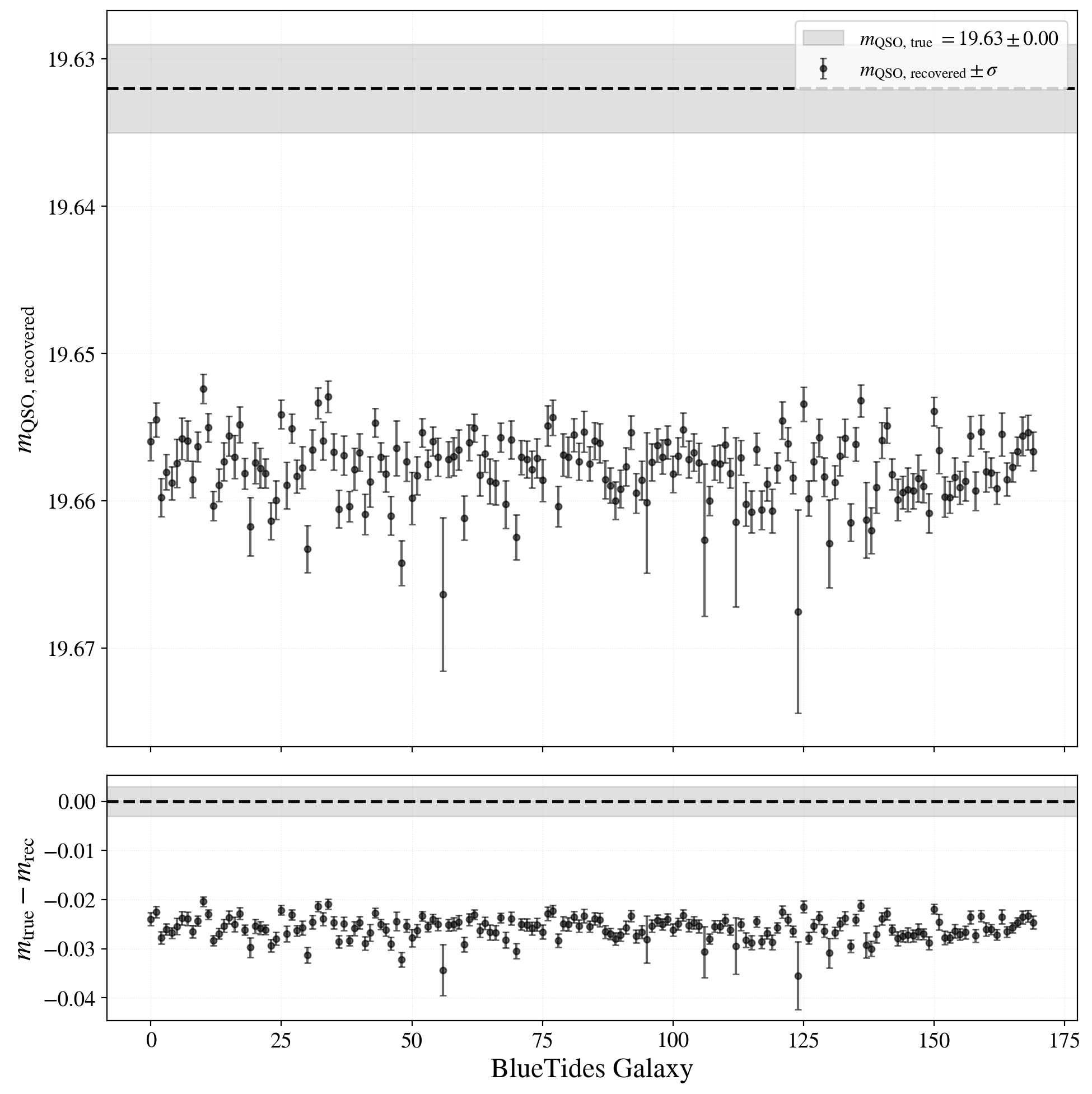}
    \caption{Example recovery of the quasar magnitude for each BlueTides galaxies used in the J1120+0641 analysis in the F356W filter. The quasar flux residual corresponds to approximately 2\% change in absolute flux on average, which is comparable to or larger than the total host galaxy flux.}
    \label{fig:mag_error_bar_quasar}
\end{figure}

Although the recovered magnitudes track the true values closely, even small uncertainties correspond to large absolute flux differences. Figure~\ref{fig:mag_error_bar_quasar} shows the quasar magnitude residuals for the BlueTides galaxies used in the F356W analysis of J1120+0641. A 0.01 mag uncertainty corresponds to a $\sim$2\% change in quasar flux. This is on order of the total host flux in this quasar. Consequently, even minor mismatches in the quasar model can lead to significant misattribution of quasar light to the host component. As described in this work, this effect is further exacerbated by PSF uncertainties, noise, and radial profile degeneracies (Sections~\ref{app:free_sersic}–\ref{appendix:psf}).

\section{Bin Width Tolerance for Each Updated Magnitudes}
Due to the discretization introduced by the binning in our SBI pipeline (see Section \ref{sec:stat_methods}), we adopt a bin width tolerance for each updated magnitude in each filter. This tolerance represents the precision to which we can reliably infer the updated magnitudes. In Table \ref{tab:bin_sizes}, we show the bin sizes for each quasar host magnitude updated posterior.

\begin{figure}
    \centering
    \includegraphics[width=\linewidth]{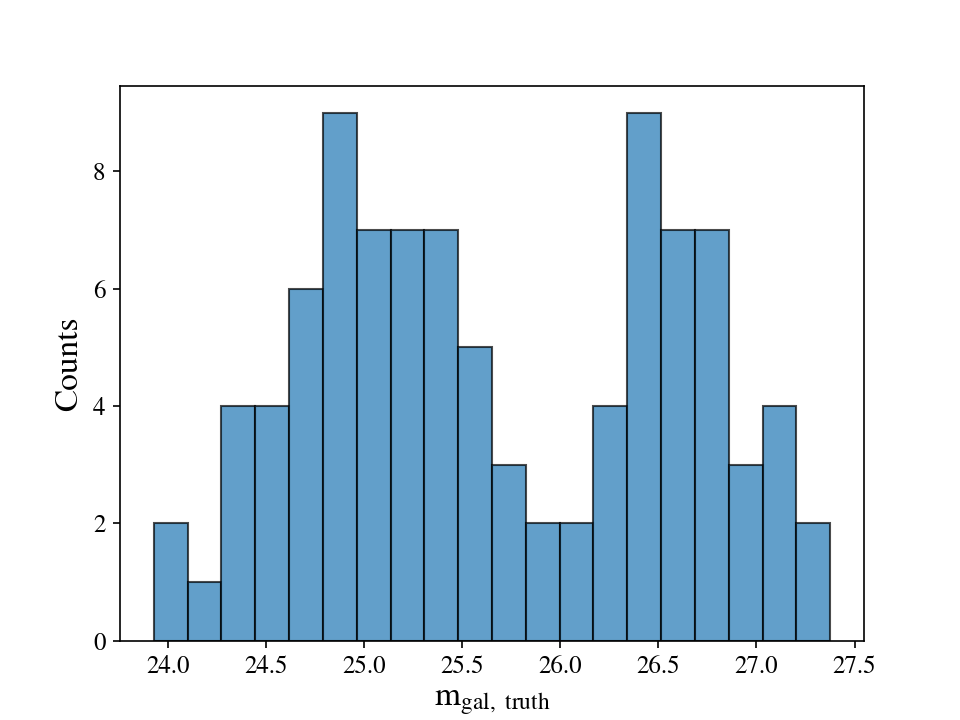}
    \caption{Histogram of the truth magnitude bins used in the J1120+0641 F200W quasar host magnitude inference.}
    \label{fig:hist_mags_example}
\end{figure}
\begin{table}
\centering

\begin{tabular}{l c c}
\textbf{Quasar} & \textbf{Filter} & \textbf{Bin Width (mag)} \\
\hline
J0844-0132      & F150W & 0.27 \\
                & F356W & 0.24 \\
J0911+0152      & F150W & 0.23 \\
                & F356W & 0.22 \\
J0918+0139      & F150W & 0.24 \\
                & F356W & 0.26 \\
J1425-0015      & F150W & 0.20 \\
                & F356W & 0.26 \\
J1525+4303      & F150W & 0.17 \\
                & F356W & 0.22 \\
J1146-0005      & F150W & 0.21 \\
                & F356W & 0.23 \\
J1146+0124      & F150W & 0.17 \\
                & F356W & 0.20 \\
J0217-0208      & F150W & 0.11 \\
                & F356W & 0.13 \\
J0844-0052      & F150W & 0.16 \\
                & F356W & 0.27 \\
J2236+0032      & F150W & 0.15 \\
                & F356W & 0.13 \\
J2255+0251      & F150W & 0.26 \\
                & F356W & 0.22 \\
J159-02         & F115W & 0.12 \\
                & F200W & 0.13 \\
                & F356W & 0.18 \\
J1120+0641      & F115W & 0.20 \\
                & F200W & 0.17 \\
                & F356W & 0.23 \\
\hline
\end{tabular}

\caption{An overview of all the bin widths used in the generation of each quasar host magnitude posterior. Each bin width is the limit on our truth magnitude recovery precision.\label{tab:bin_sizes}}
\end{table}

\section{Example Forming Final Weighted Posteriors from Individual Sample Magnitudes}

\begin{figure*}
\centering
\includegraphics[width=\linewidth]{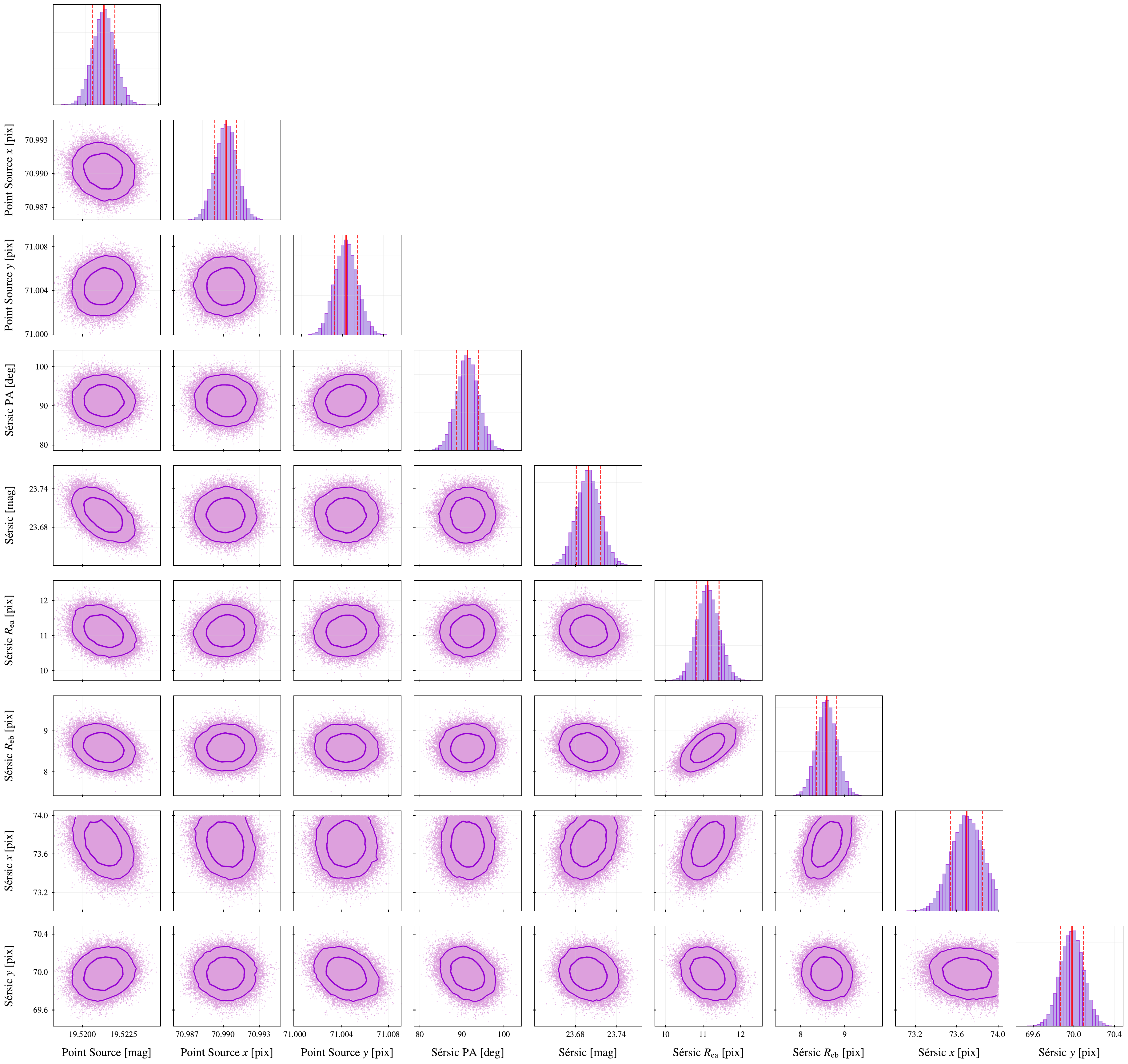}
\caption{An example corner plot for a quasar point source removal and host detection. We extract the median from the posterior and standard deviation values (using the 16th and 84th percentile marked in dashed red lines on the histograms and not assuming a Gaussian distribution) from these sampled values to construct our updated Sérsic magnitude posteriors. The dark purple contours indicate the $68\%$ and $95\%$ confidence intervals. PA refers to Sérsic proper angle. \label{fig:corner_sample}}
\end{figure*}

\begin{figure*}
    \centering
    \includegraphics[width=0.9\linewidth]{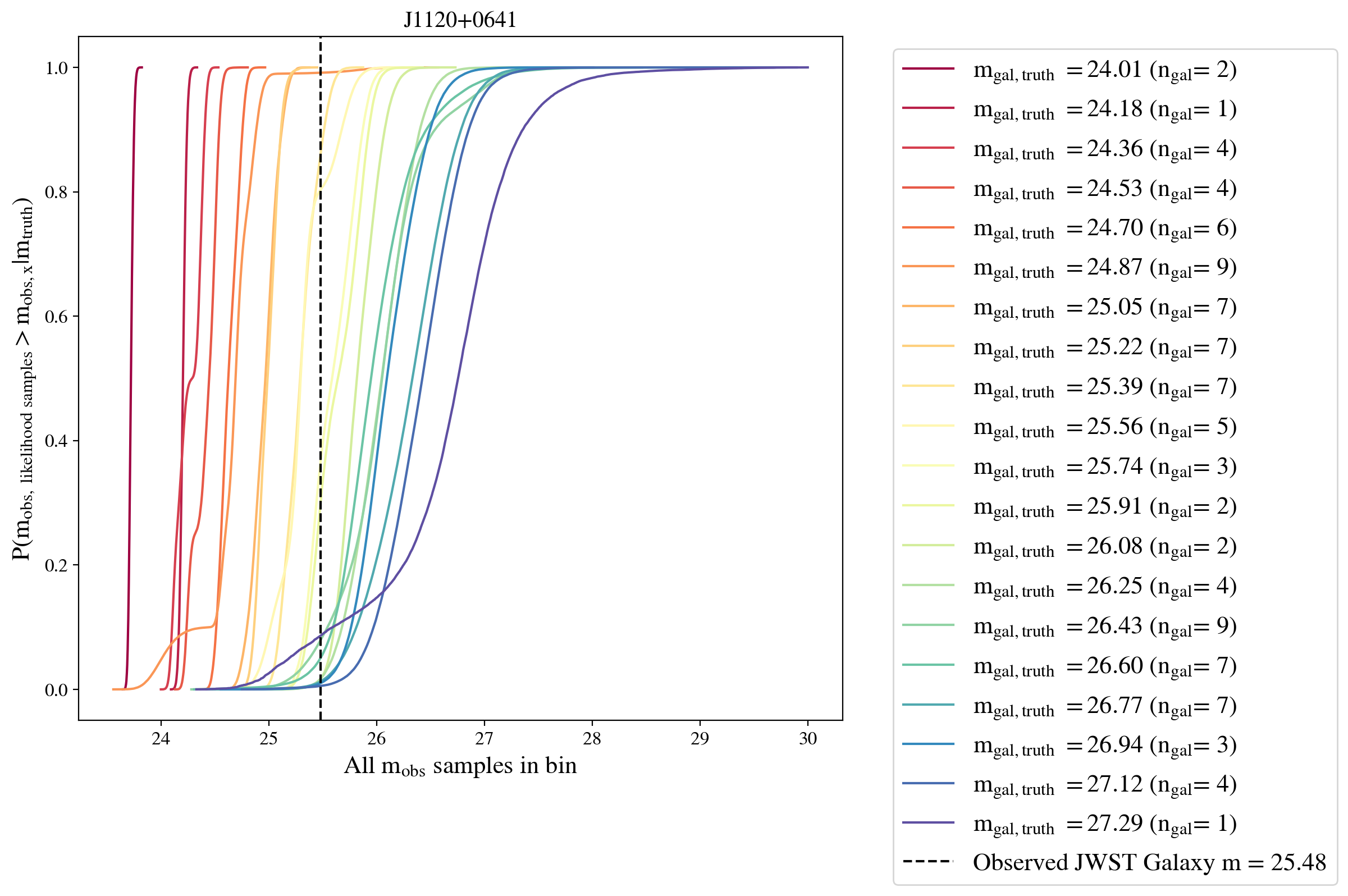}
    \caption{For J1120+0641 F200W, we show example CDFs of all mock observed likelihood samples for each truth magnitude bin. Each color CDF is the same color as the point used to generate the posterior in Figure \ref{fig:pdf_one}.}
    \label{fig:cdfs_one}
\end{figure*}

\begin{figure*}
    \centering
    \includegraphics[width=0.6\linewidth]{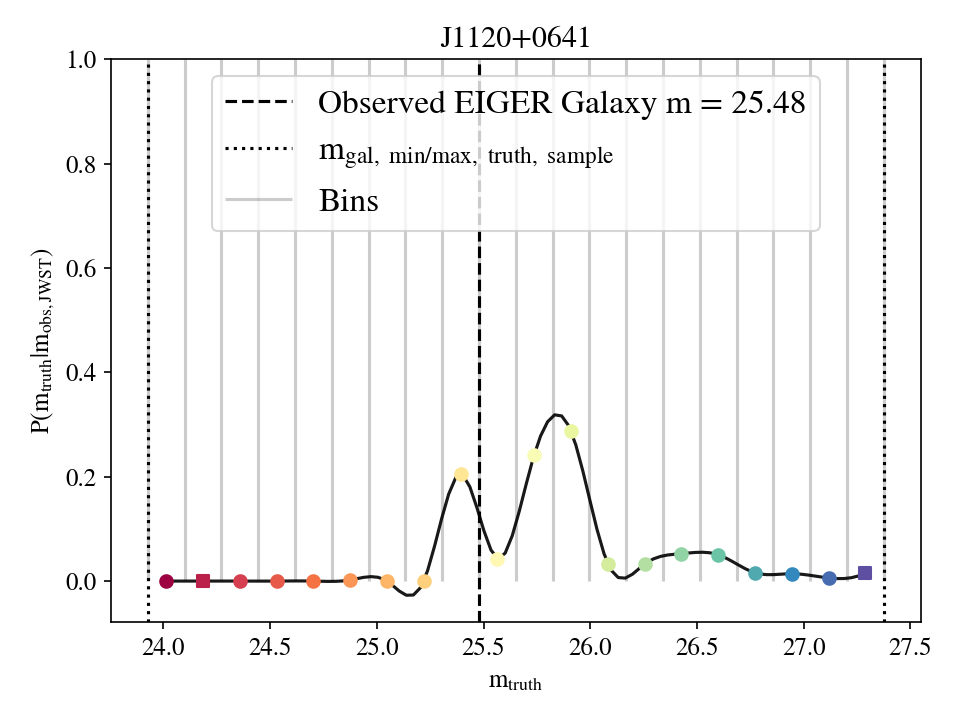}
    \caption{For J1120+0641 F200W, the PDF of the BlueTides truth magnitude posterior inferred from the derivative of the CDF along the black dashed line.}
    \label{fig:pdf_one}
\end{figure*}

\begin{figure*}
    \centering
    \includegraphics[width=\textwidth,height=0.9\textheight,keepaspectratio]{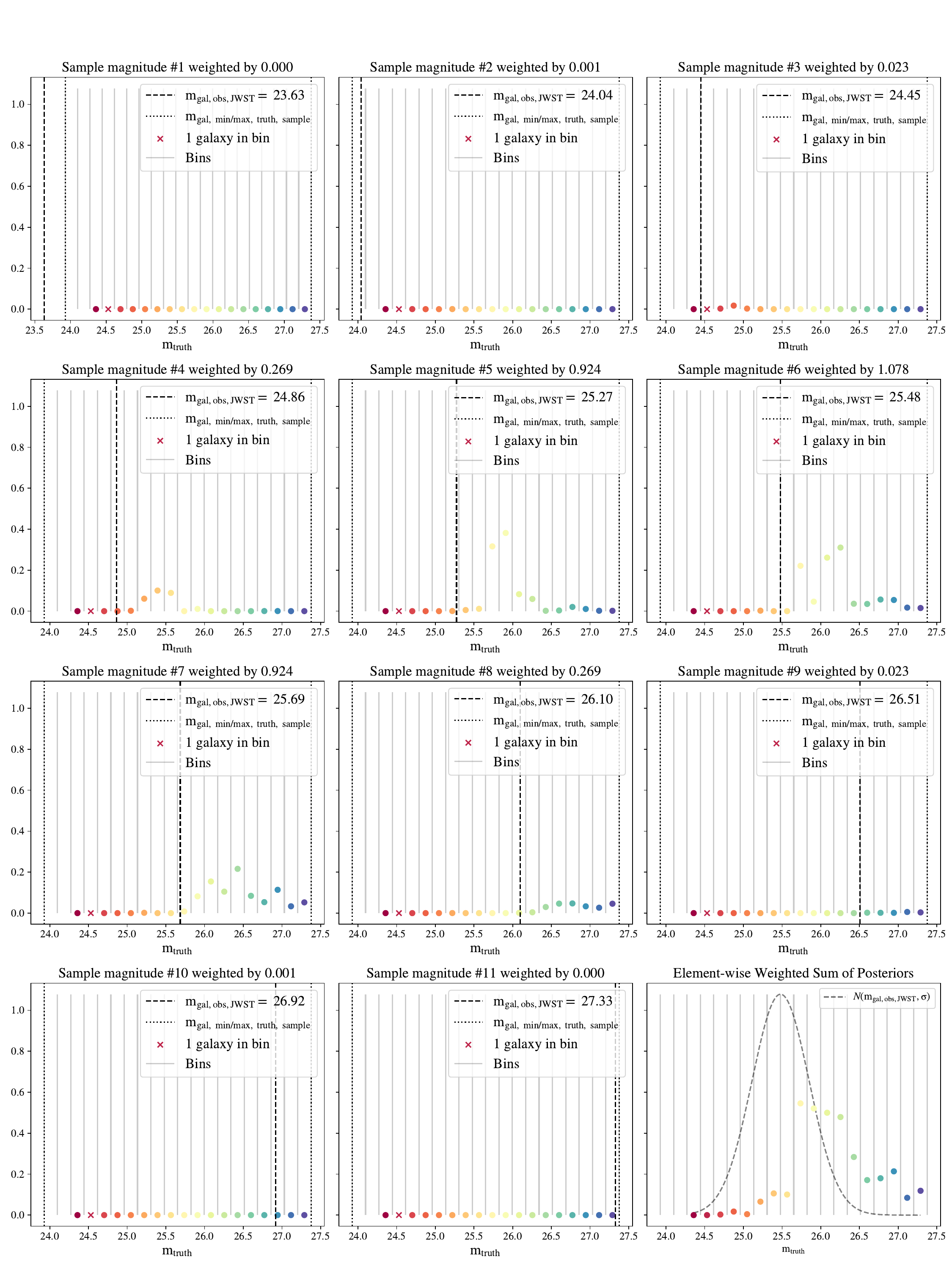}
    \caption{An example of the individual posteriors that make up the final weighted posterior plots in Figure \ref{fig:weighted_images_1} through Figure \ref{fig:weighted_images_2}. Each individual posterior is weighted by the measured (assumed Gaussian) posterior evaluated at the observed magnitude, corresponding to the point at which the CDFs in Figure \ref{fig:cdfs_one} are differentiated. This particular example is for J1120+0641 in F200W. \label{fig:sum_figure}}
\end{figure*}

In Figure \ref{fig:corner_sample}, we show an example corner plot for a PSF removal from a mock image. We use all posterior samples from the Sérsic magnitude posterior shown in the fifth row from the top in the corner plot in our analysis. The CDF of these posteriors for J1120+0641 in F200W is shown in Figure \ref{fig:cdfs_one}. An individual PDF from Figure \ref{fig:cdfs_one} (differentiated along one dashed line) is shown in Figure \ref{fig:pdf_one}. Figure \ref{fig:sum_figure} shows the individual posteriors corresponding to each sample magnitude used to construct the weighted posteriors in Appendix \ref{appendix:post}.

\section{All Corrected Posteriors for High-$z$ Quasar Host Detections}
\label{appendix:post}

\begin{figure*}
\centering
\setlength{\tabcolsep}{0pt}
\renewcommand{\arraystretch}{0}
\begin{tabular}{cc}
\includegraphics[width=0.48\textwidth]{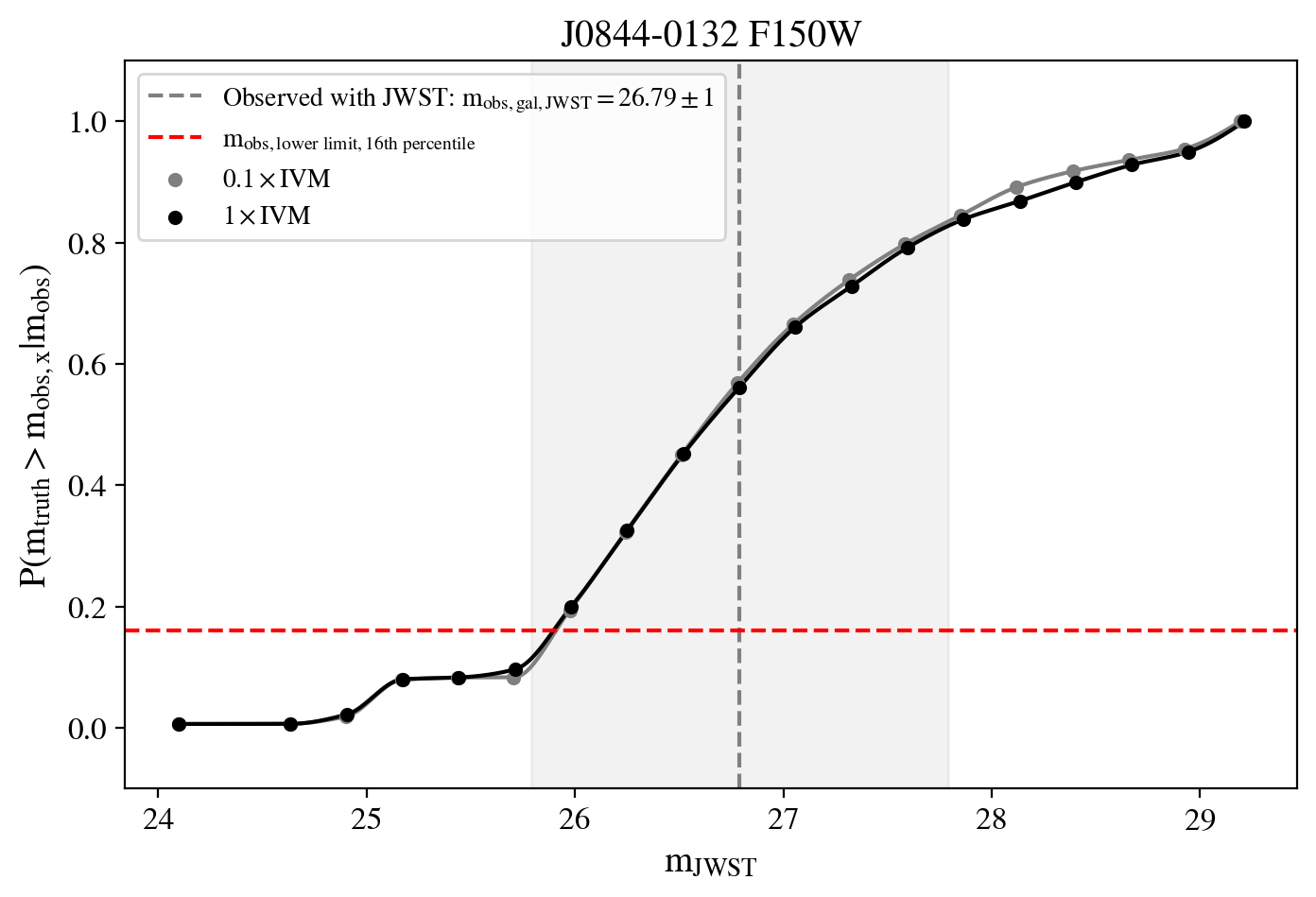} &
\includegraphics[width=0.48\textwidth]{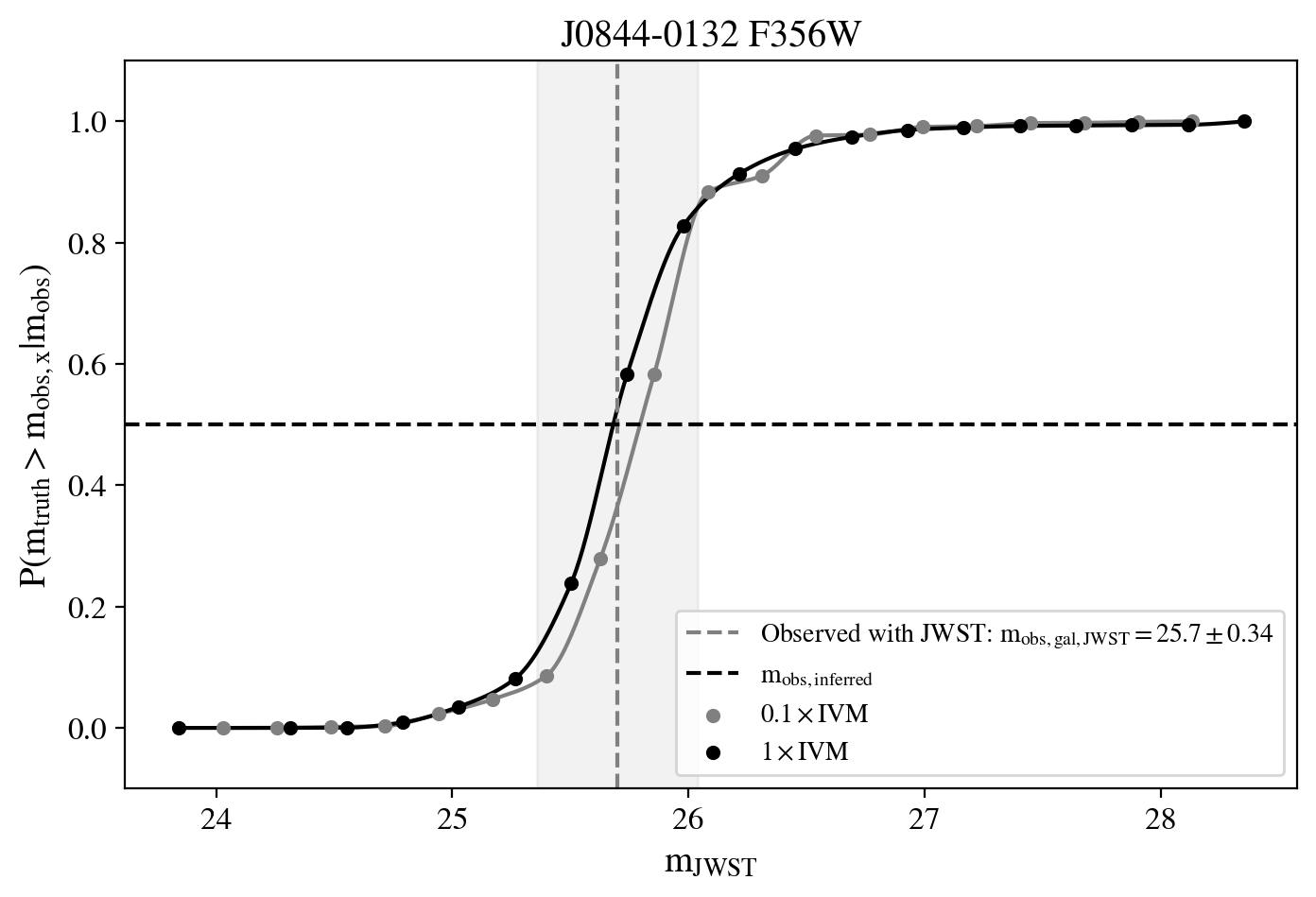} \\

\includegraphics[width=0.48\textwidth]{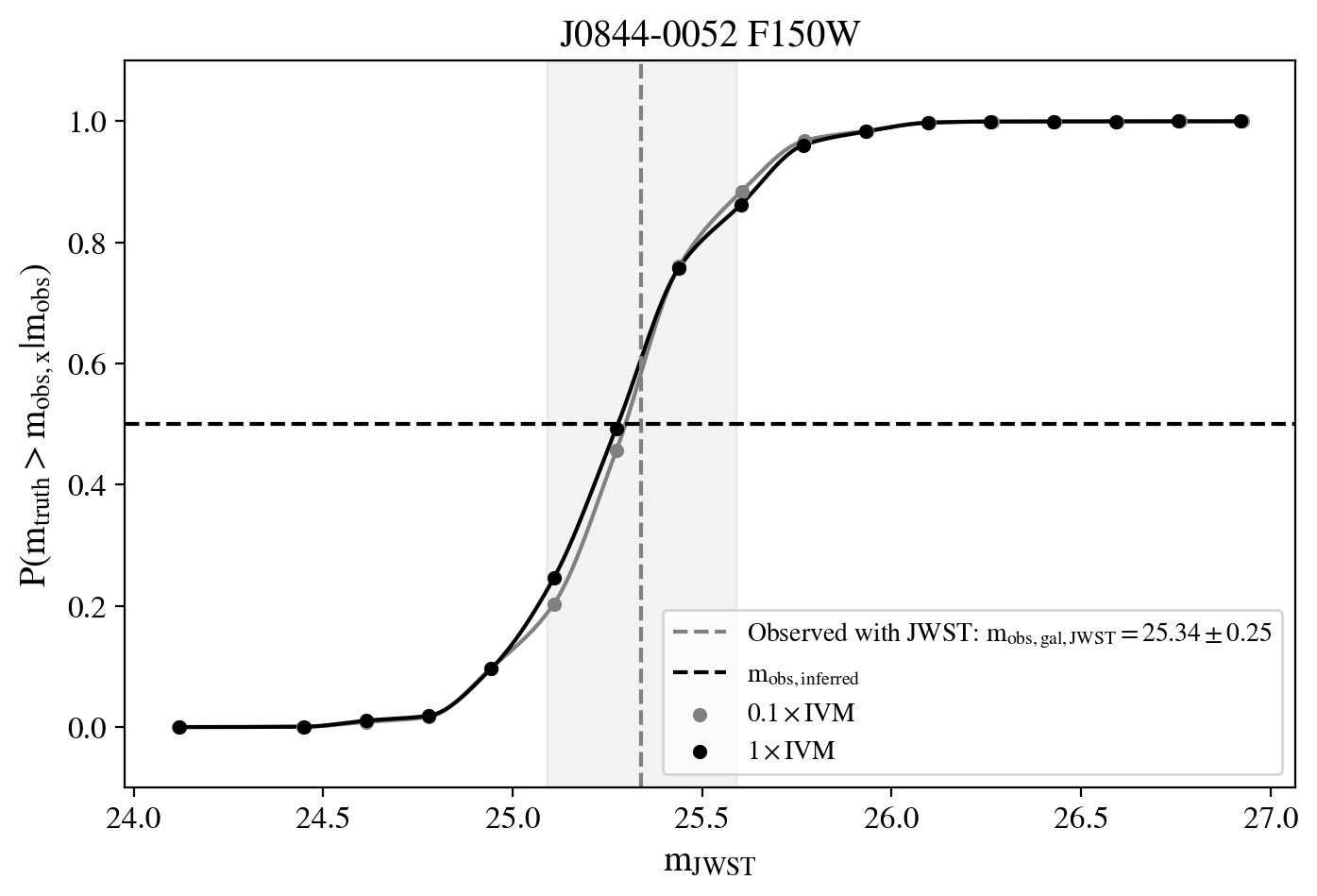} &
\includegraphics[width=0.48\textwidth]{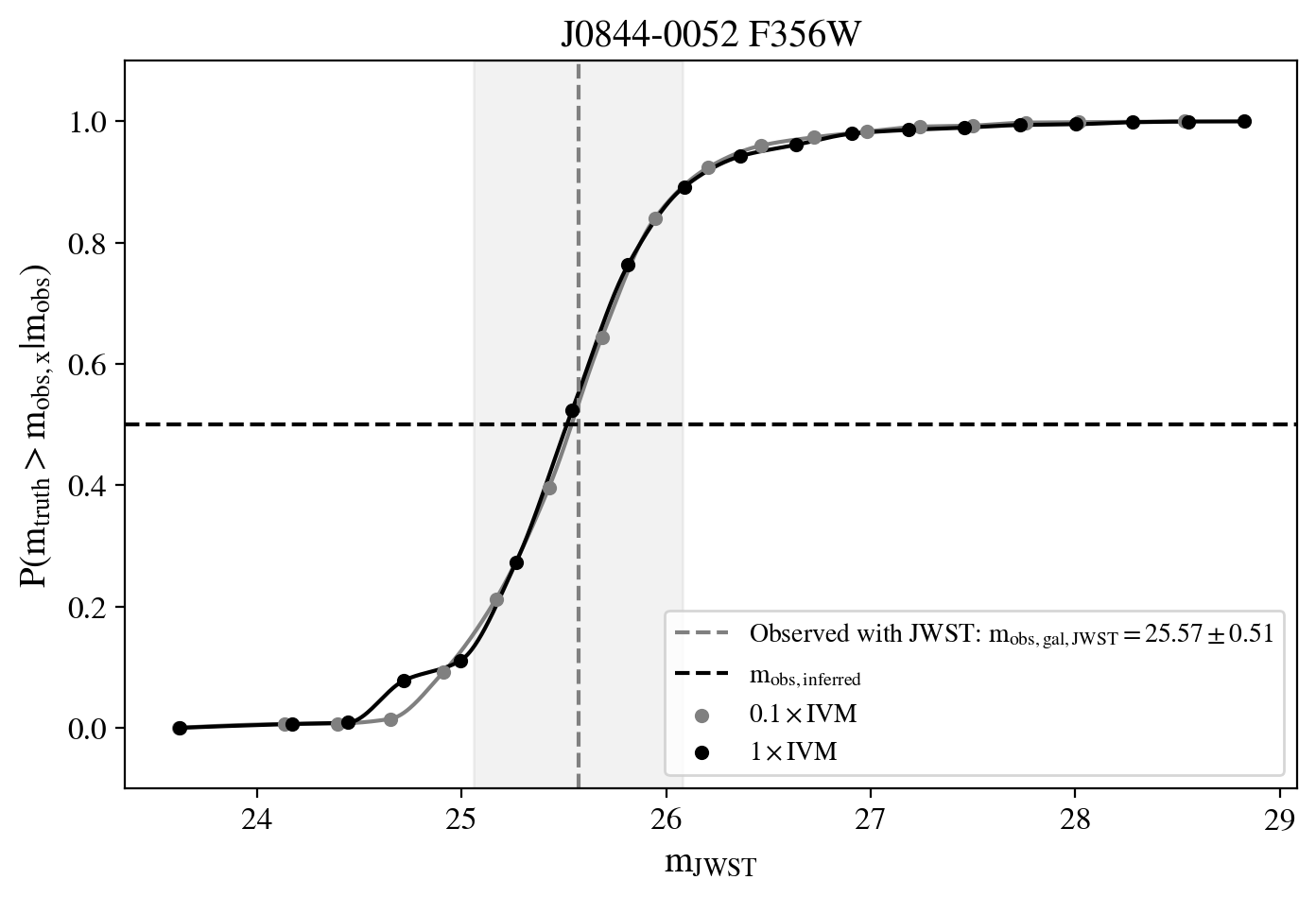} \\

\includegraphics[width=0.48\textwidth]{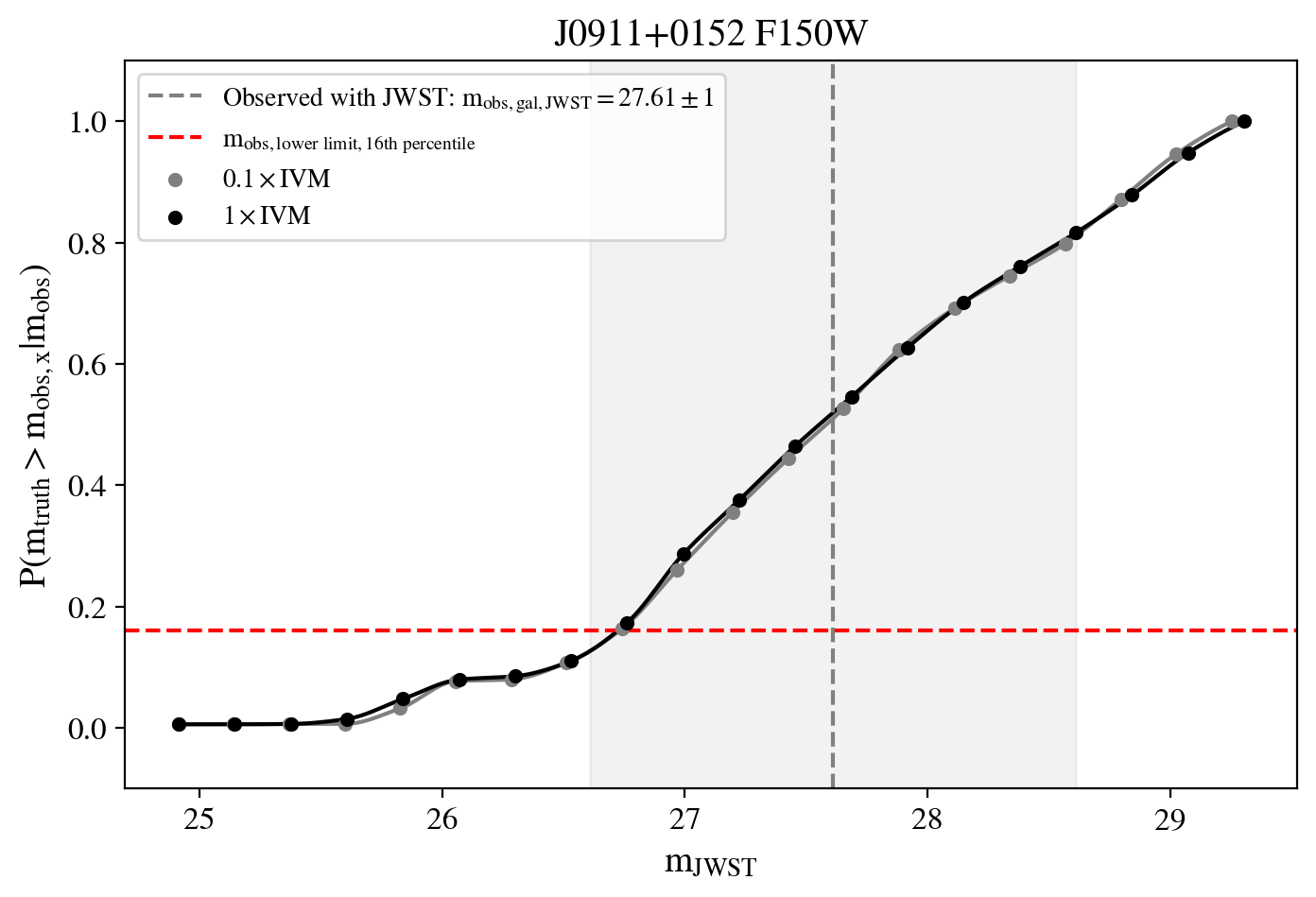} &
\includegraphics[width=0.48\textwidth]{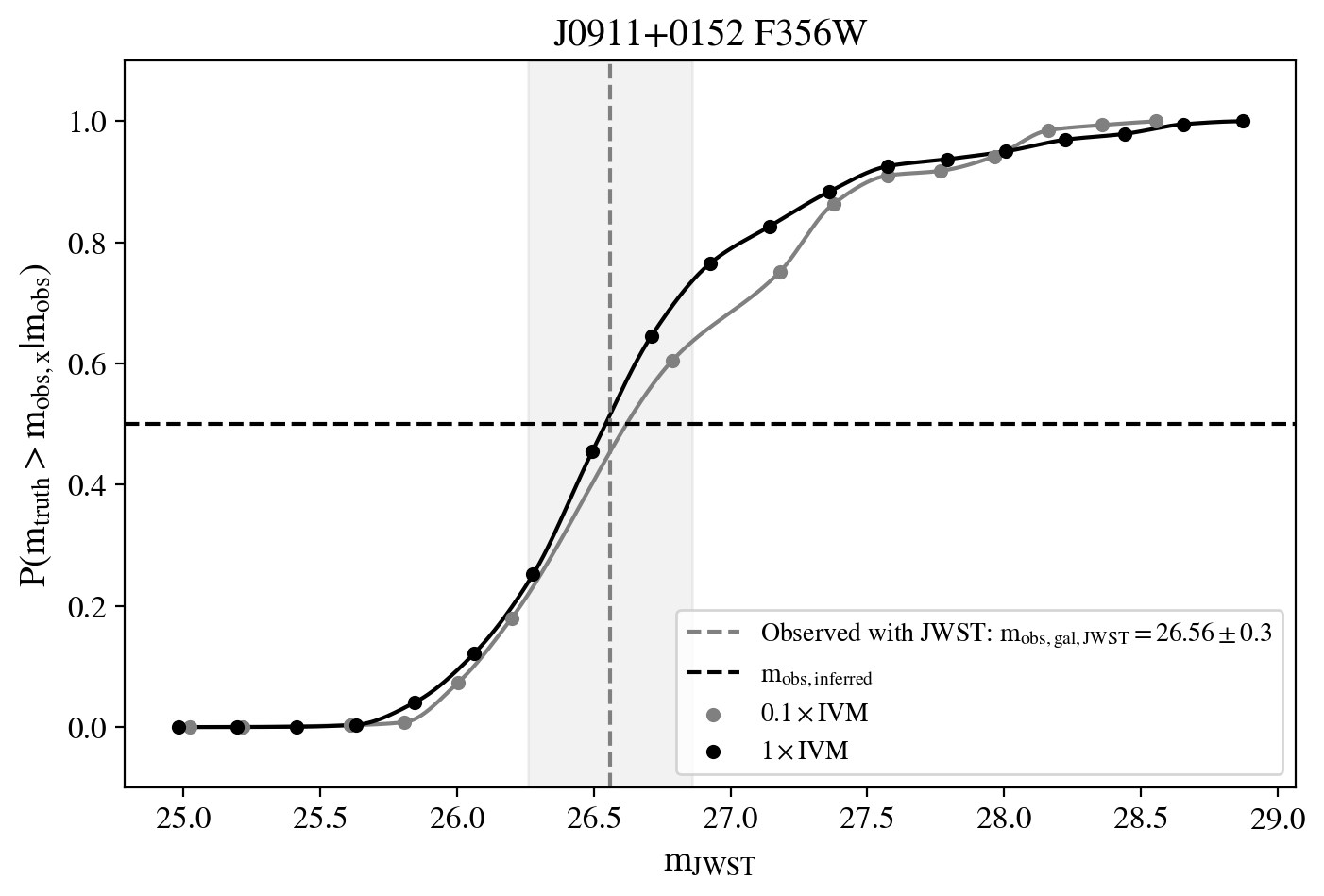} \\
\end{tabular}
\caption{The final inferred weighted posterior and observed posterior for all quasar hosts with detections from \citet{10.48550/arxiv.2211.14329} and \citet{ding2025shellqsjwstunveilshostgalaxies}. Each subfigure is labeled with the quasar name and NIRCam filter used. We show the weighted posterior described in Section \ref{sec:stat_methods}. The grey area shows 1$\sigma$ from the nominal value (shown as a vertical dashed line) of each JWST host magnitude measurement in each filter. For Category D quasar hosts, the BlueTides updated lower limit magnitude is taken to be the red horizontal dashed line at 0.16. All inferred median values and errors are shown in Table \ref{tab:magnitude_offsets}.}
\label{fig:weighted_images_1}
\end{figure*}

\begin{figure*}
\centering
\setlength{\tabcolsep}{0pt}
\renewcommand{\arraystretch}{0}
\begin{tabular}{cc}
\includegraphics[width=0.48\textwidth]{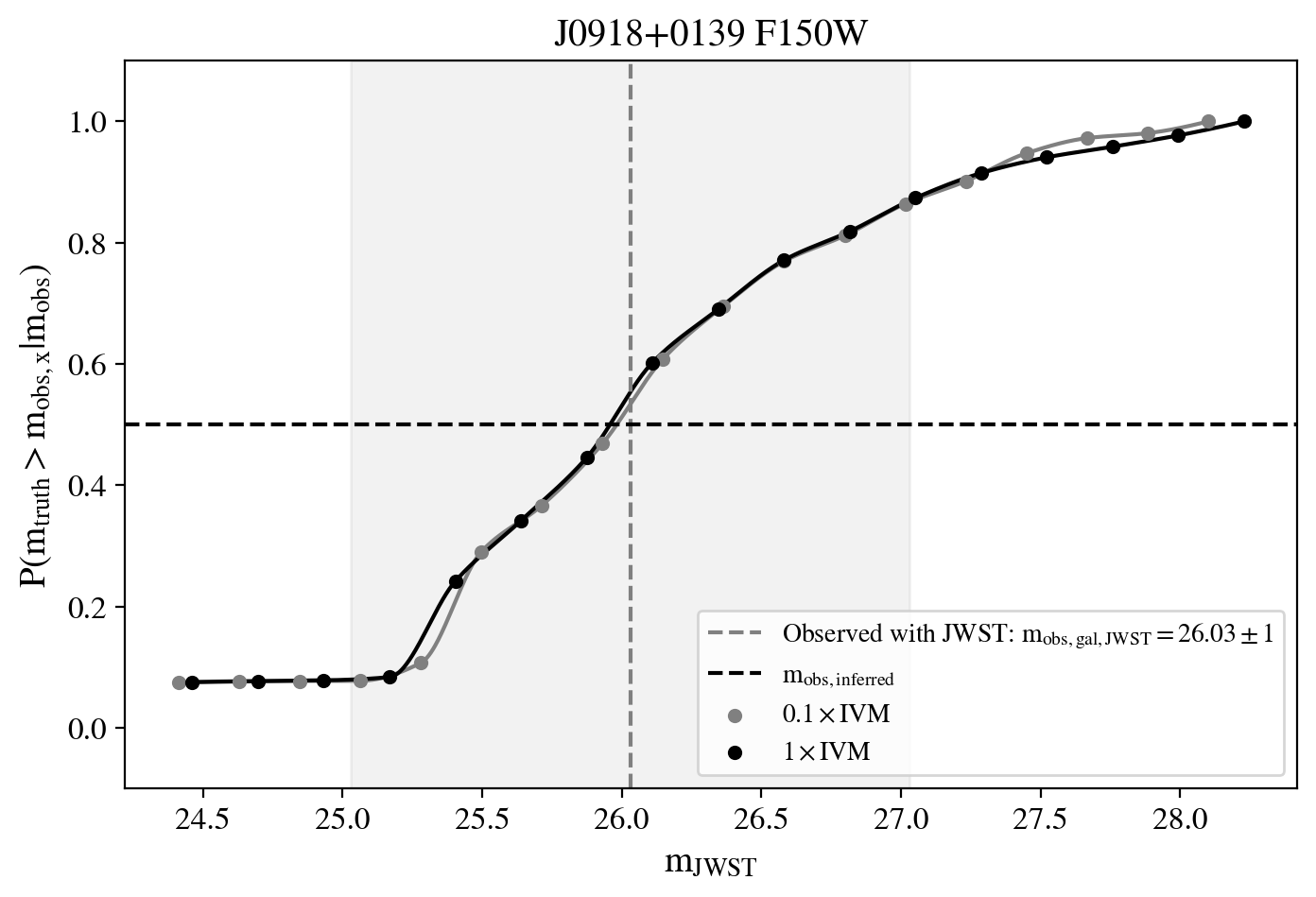} &
\includegraphics[width=0.48\textwidth]{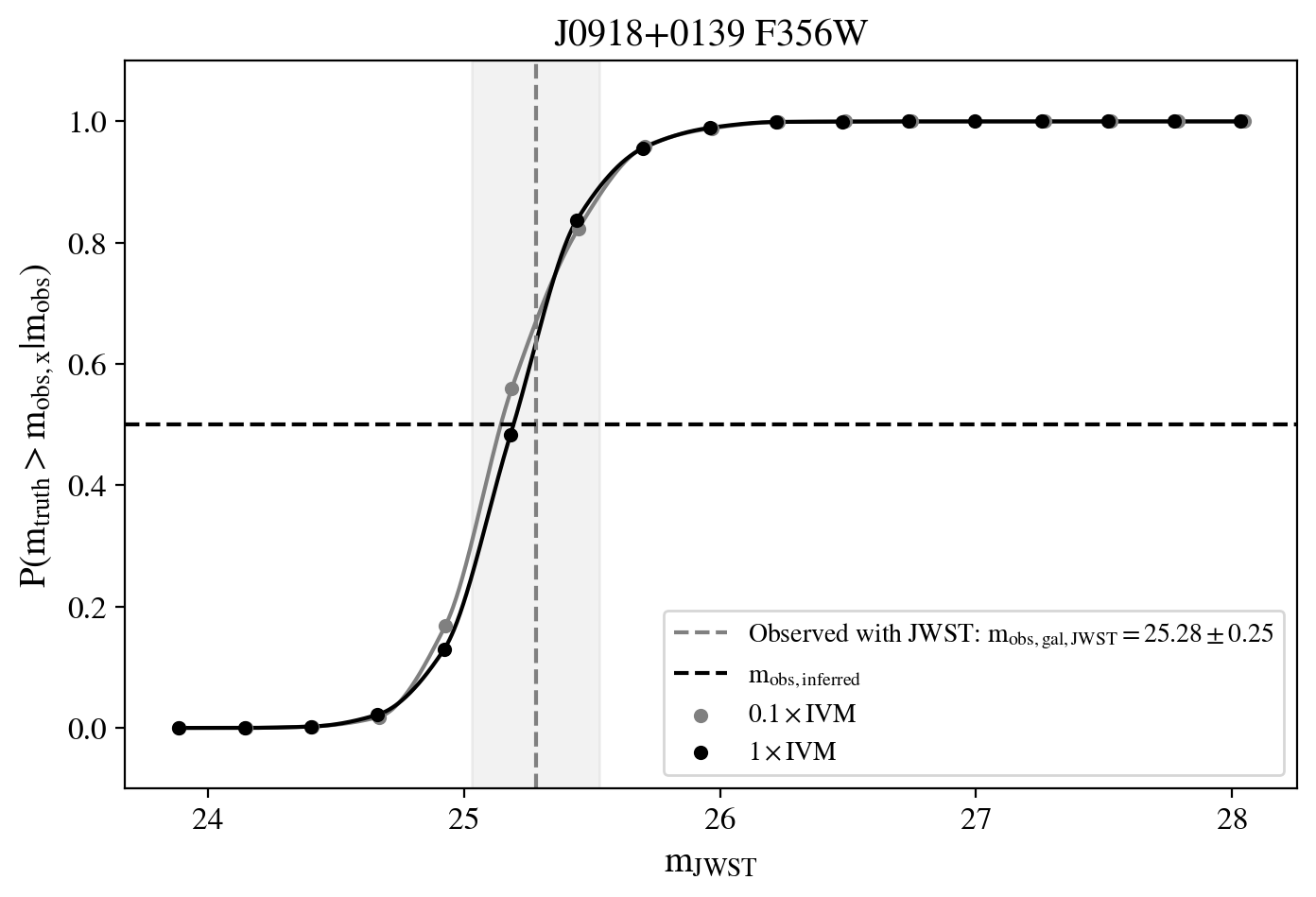} \\

\includegraphics[width=0.48\textwidth]{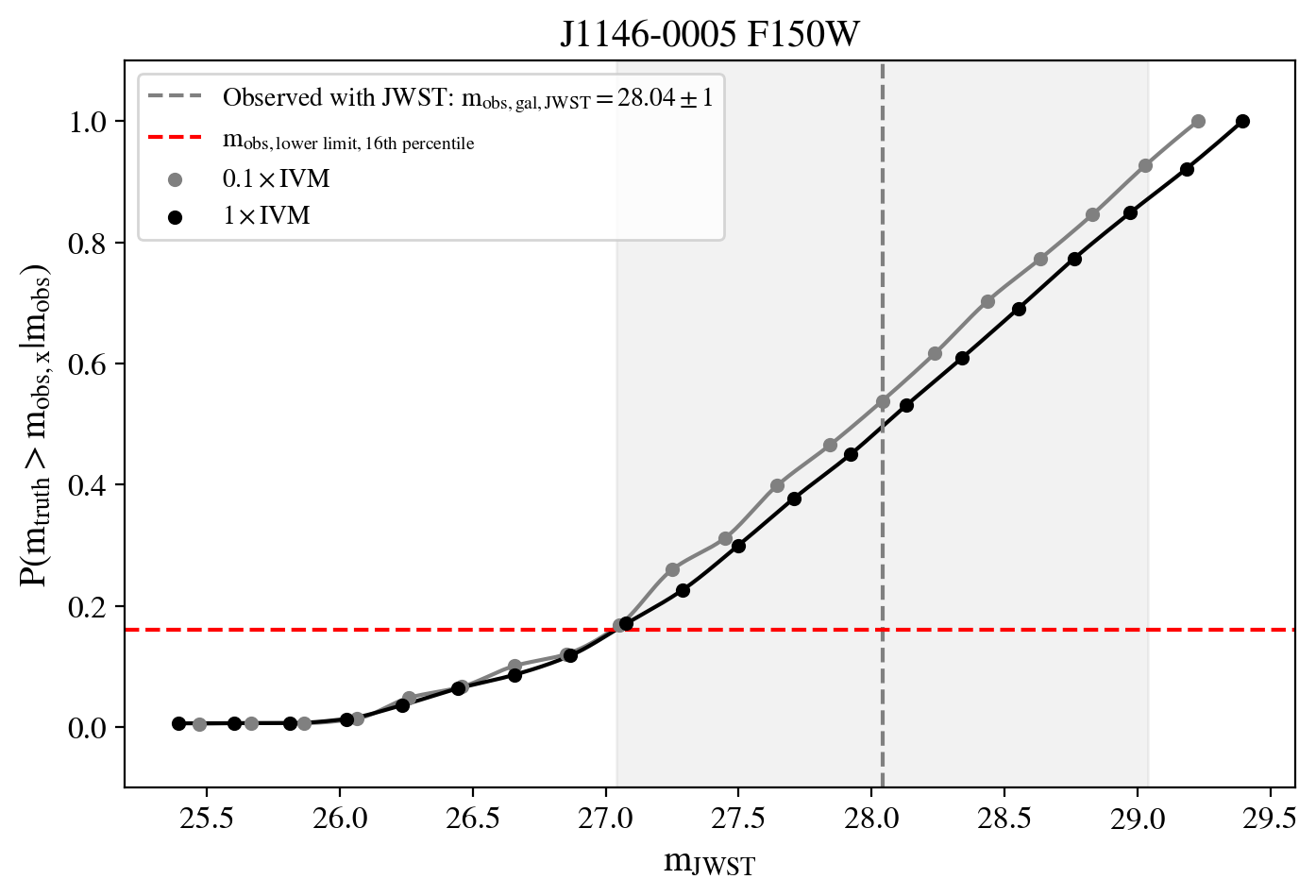} &
\includegraphics[width=0.48\textwidth]{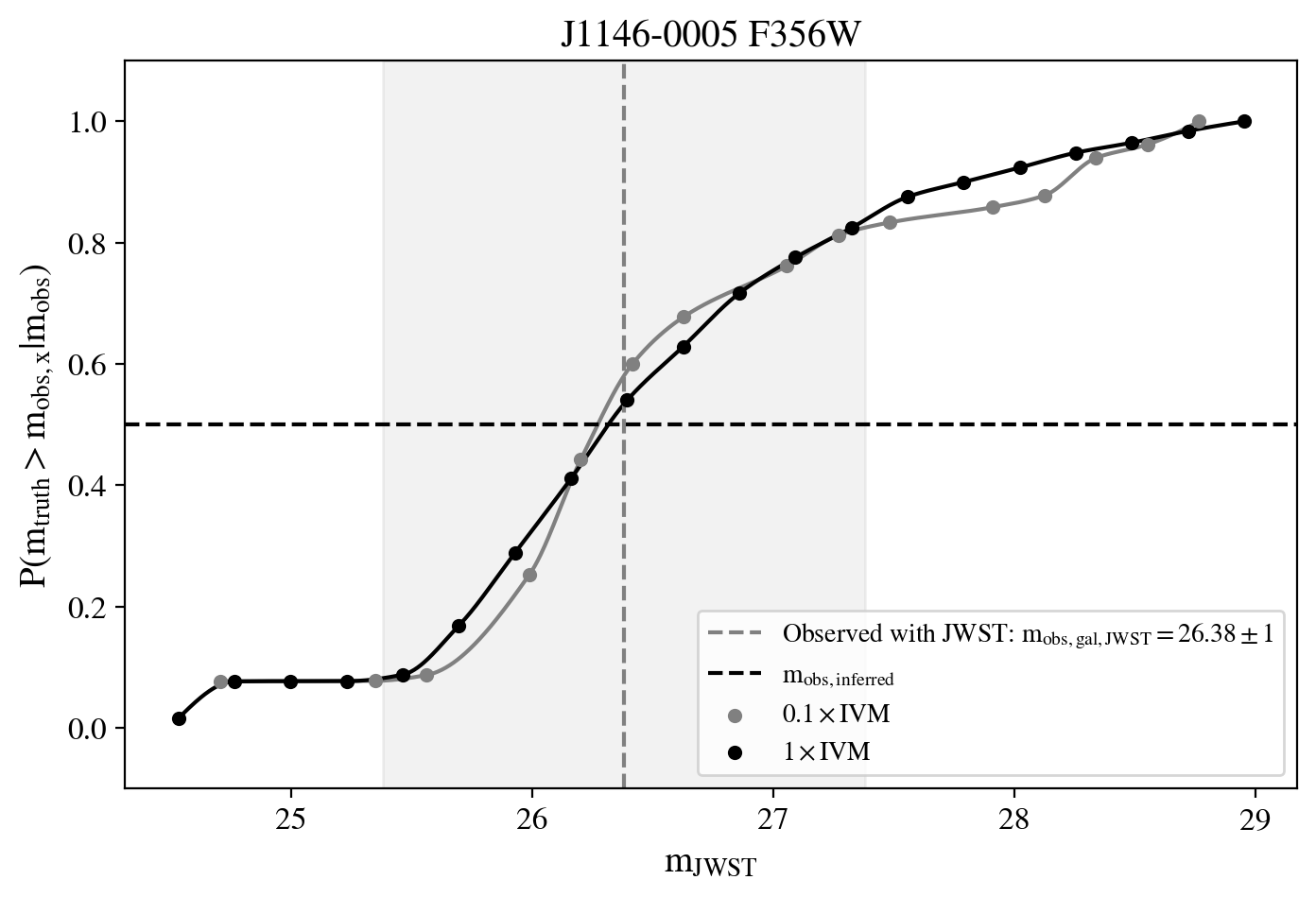} \\

\includegraphics[width=0.48\textwidth]{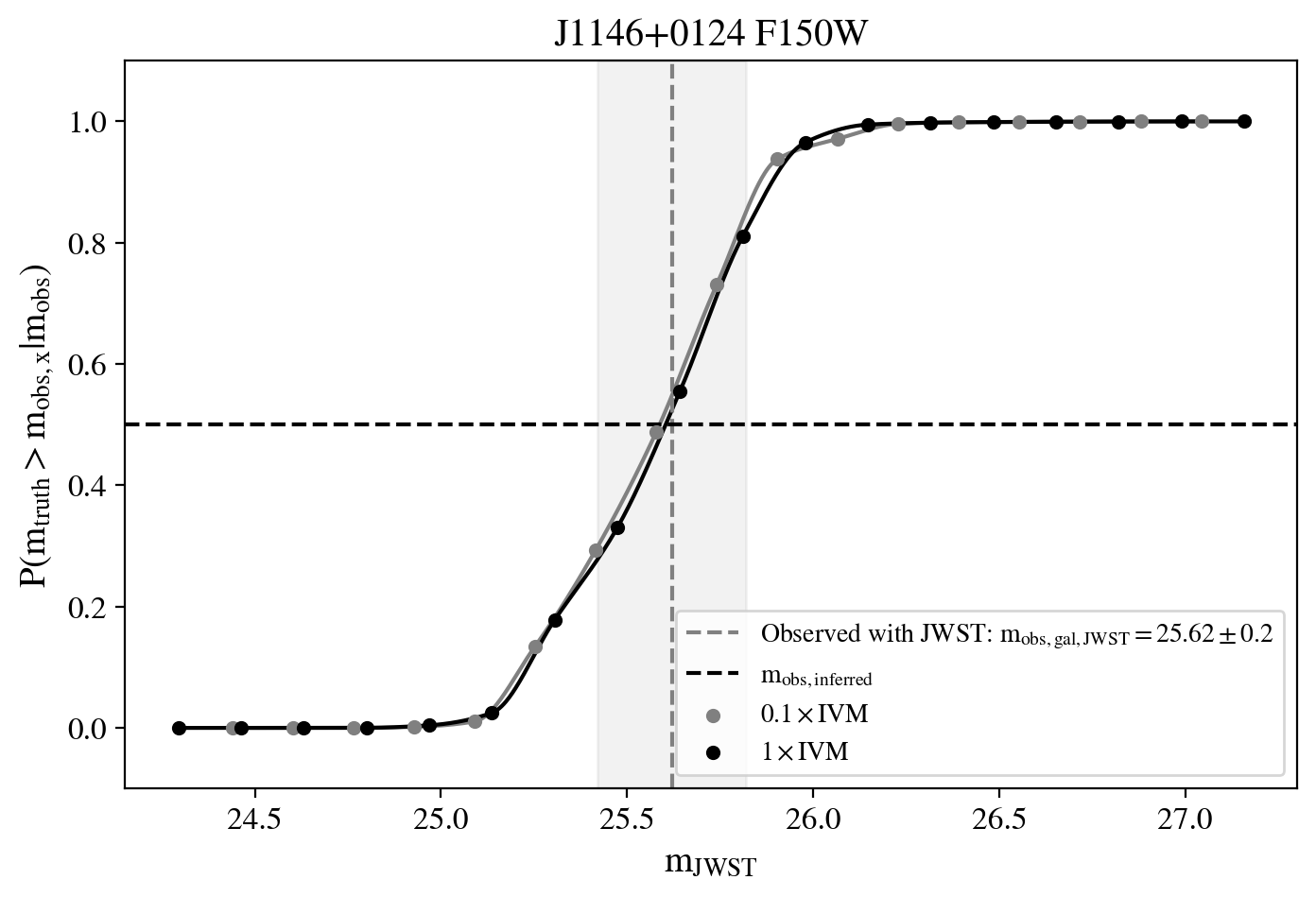} &
\includegraphics[width=0.48\textwidth]{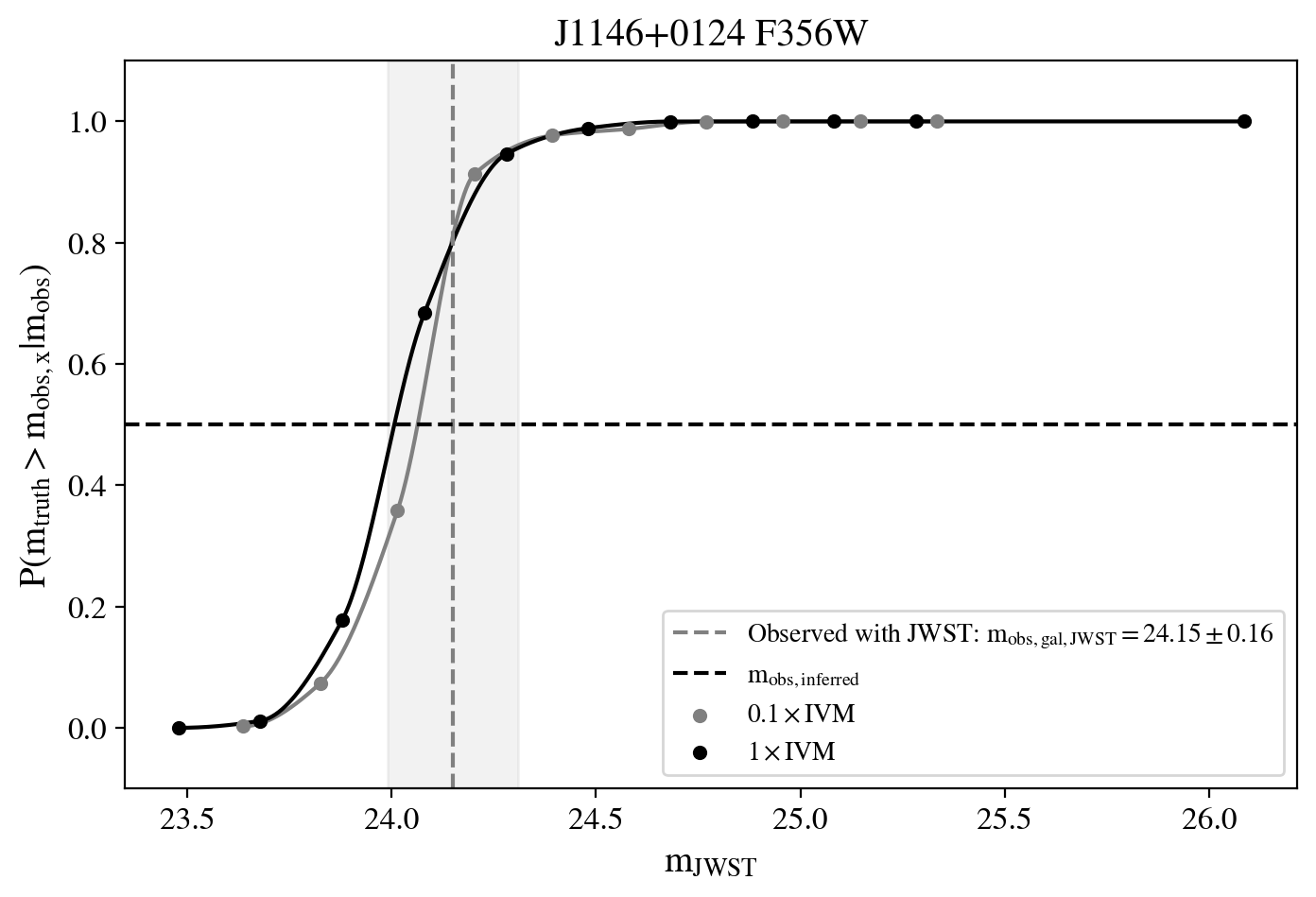} \\
\end{tabular}
\caption{Inferred weighted and observed posterior host magnitudes (part 2). See Figure~\ref{fig:weighted_images_1} for details.}
\label{fig:weighted_images_2}
\end{figure*}

\begin{figure*}
\centering
\setlength{\tabcolsep}{0pt}
\renewcommand{\arraystretch}{0}
\begin{tabular}{cc}
\includegraphics[width=0.48\textwidth]{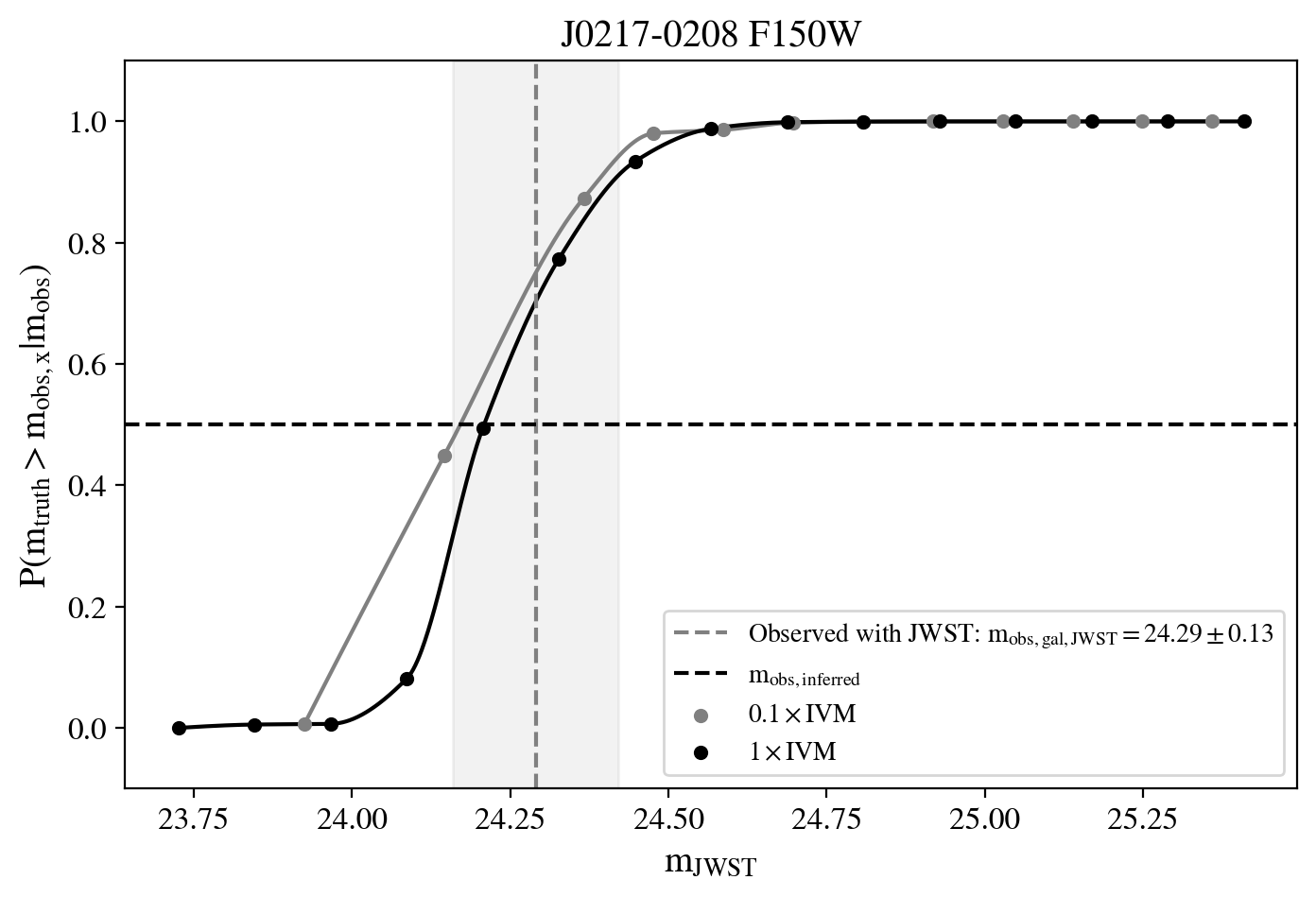} &
\includegraphics[width=0.48\textwidth]{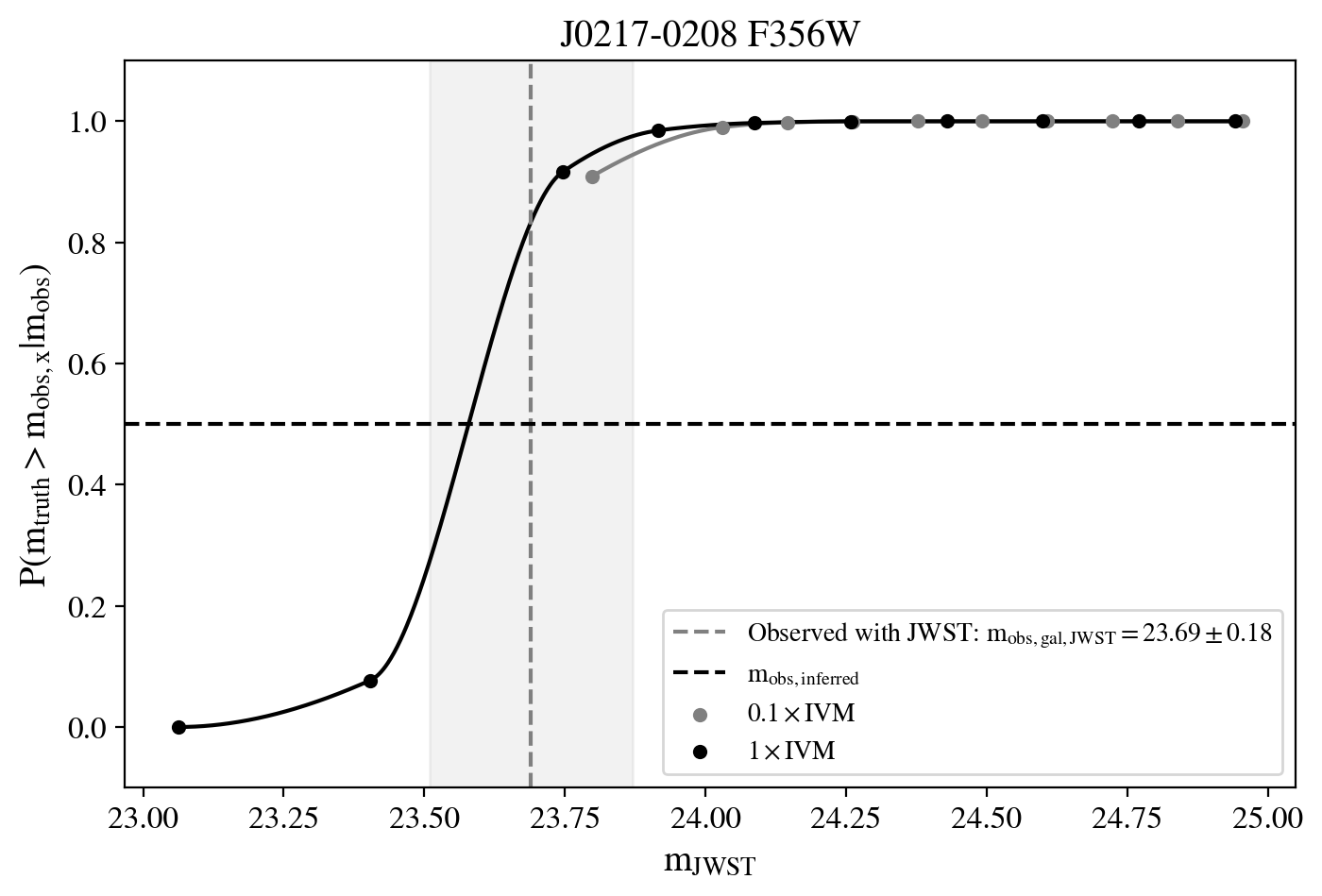} \\

\includegraphics[width=0.48\textwidth]{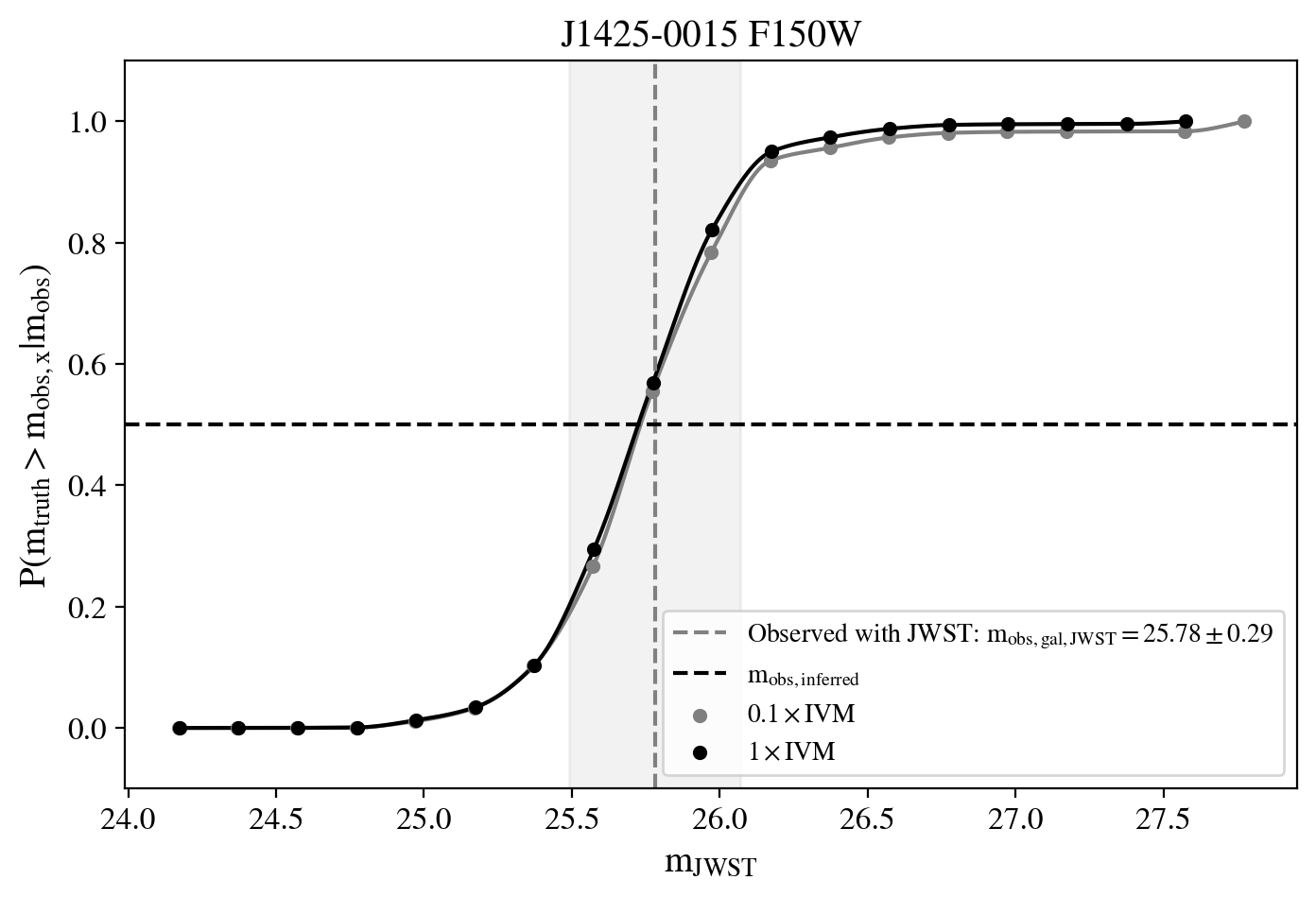} &
\includegraphics[width=0.48\textwidth]{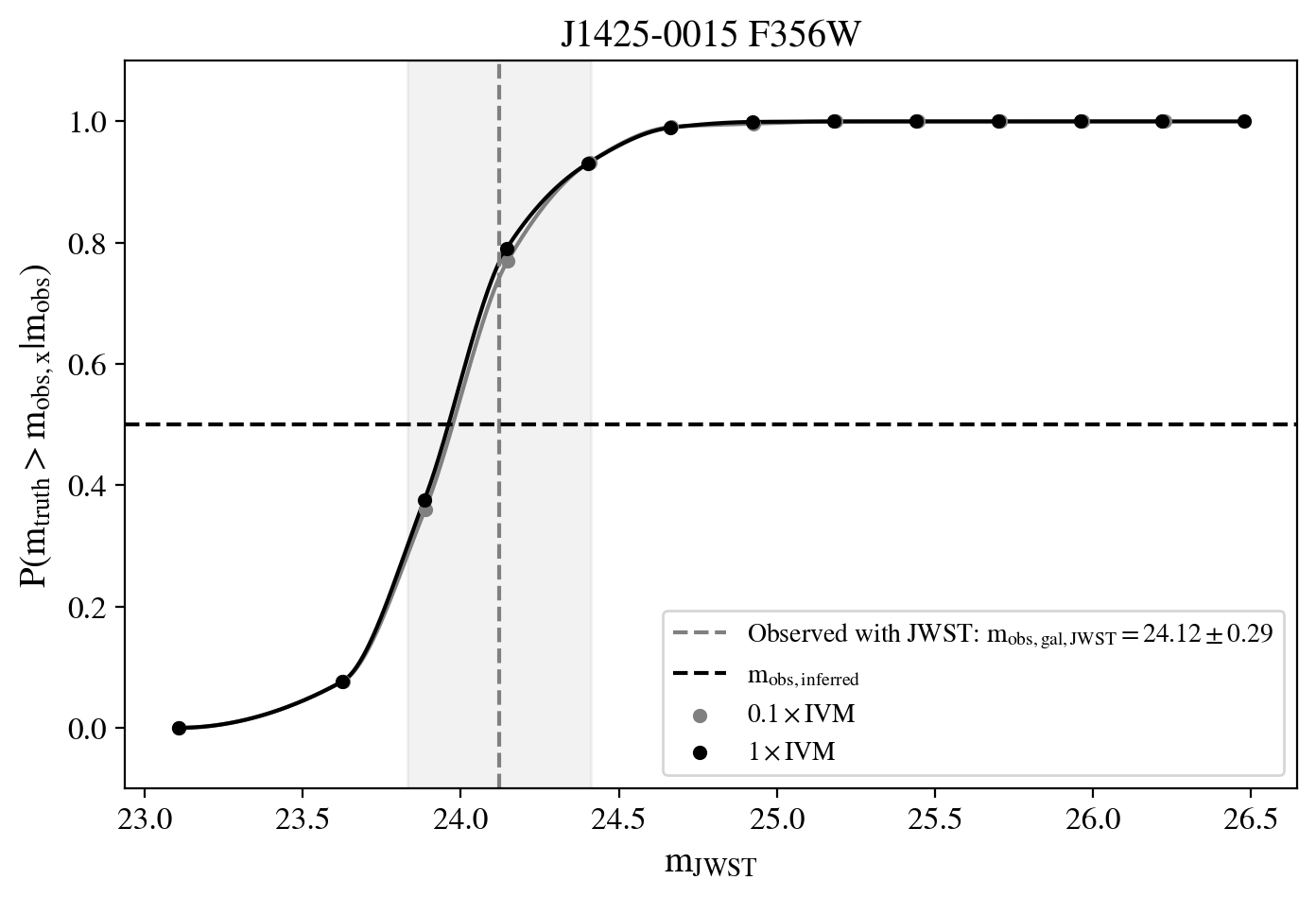} \\
\includegraphics[width=0.48\textwidth]{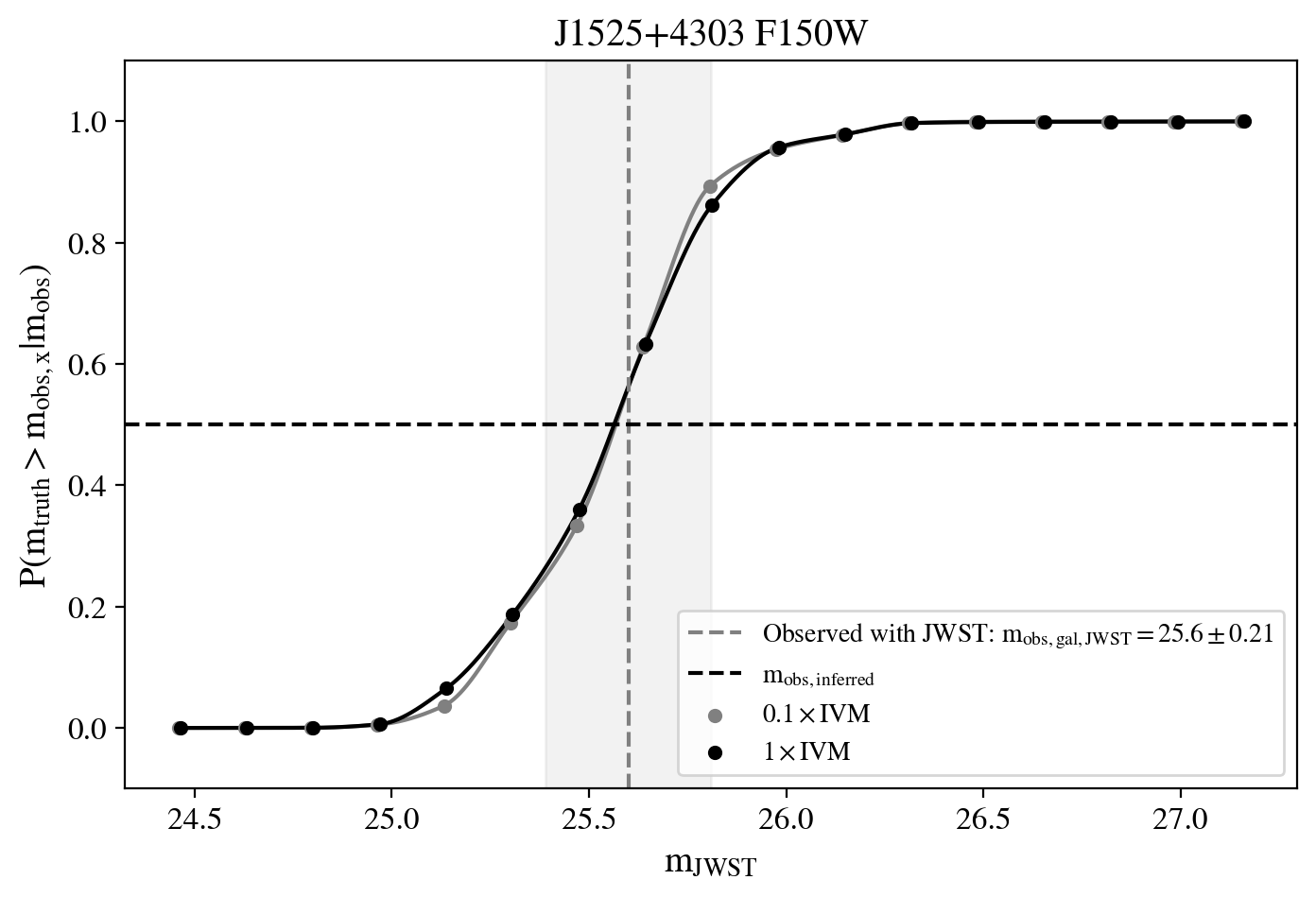} &
\includegraphics[width=0.48\textwidth]{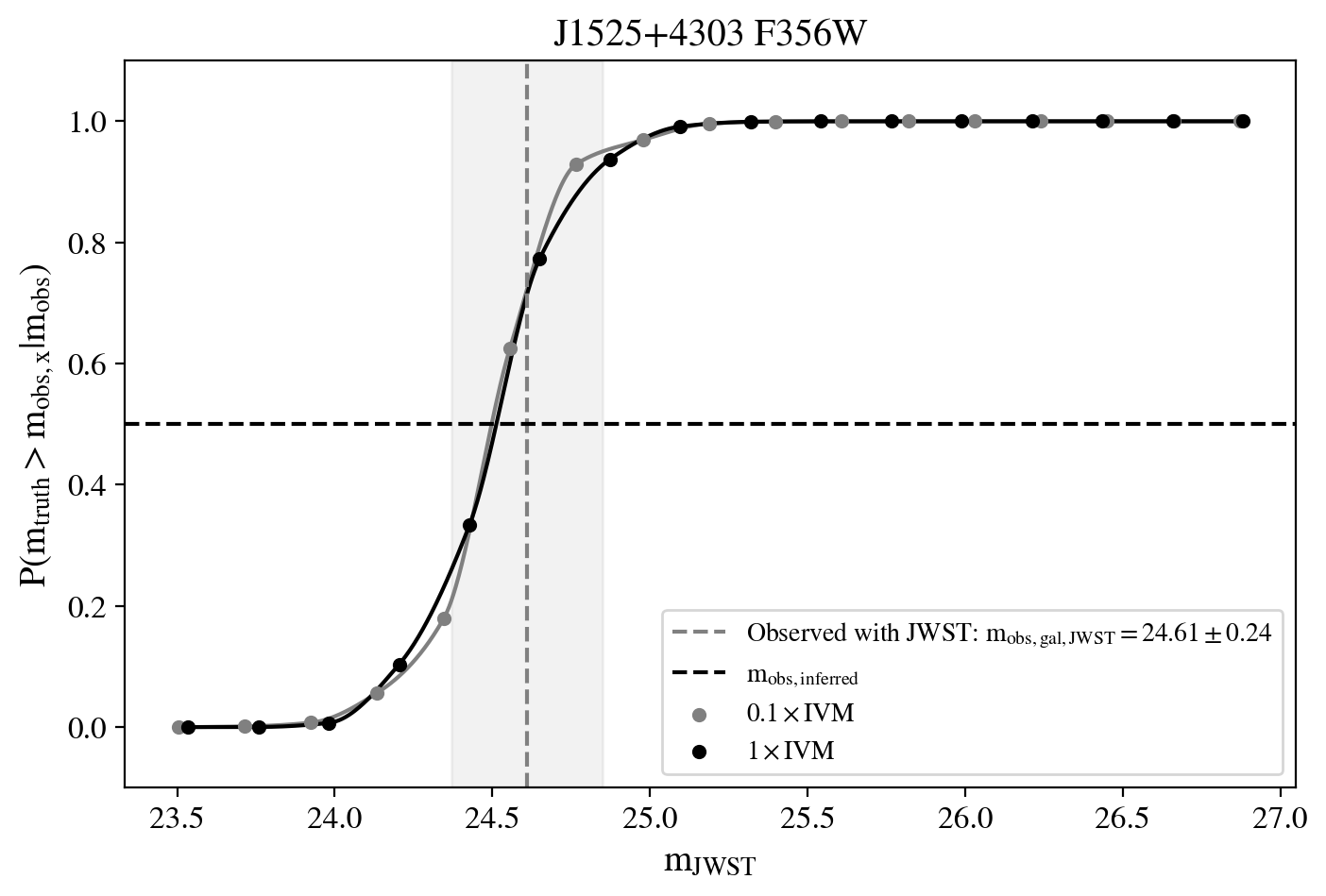} \\
\end{tabular}
\caption{Inferred weighted and observed posterior host magnitudes (part 3). See Figure~\ref{fig:weighted_images_1} for details.}
\label{fig:weighted_images_3}
\end{figure*}

\begin{figure*}
\centering
\setlength{\tabcolsep}{0pt}
\renewcommand{\arraystretch}{0}
\begin{tabular}{cc}

\includegraphics[width=0.48\textwidth]{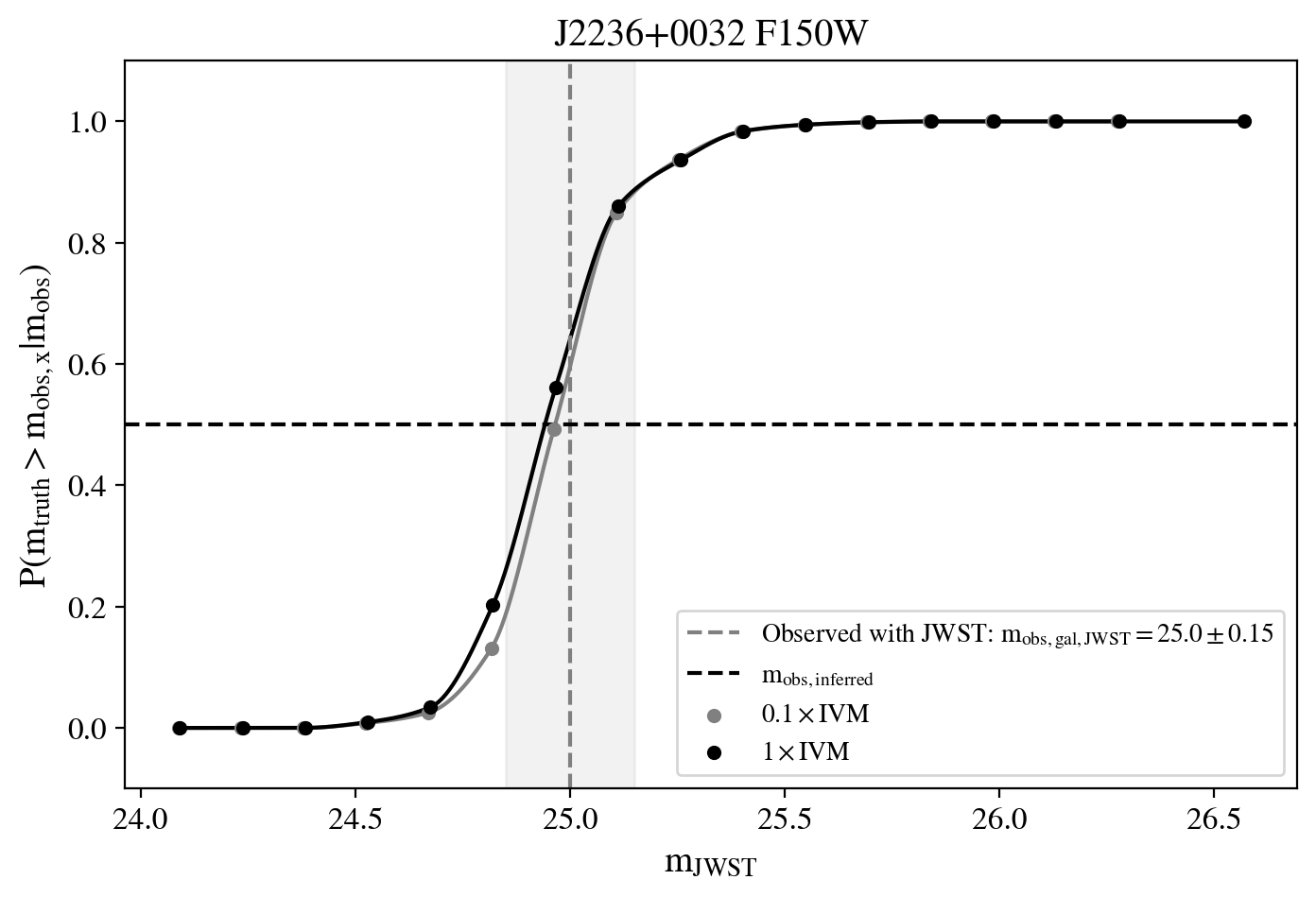} &
\includegraphics[width=0.48\textwidth]{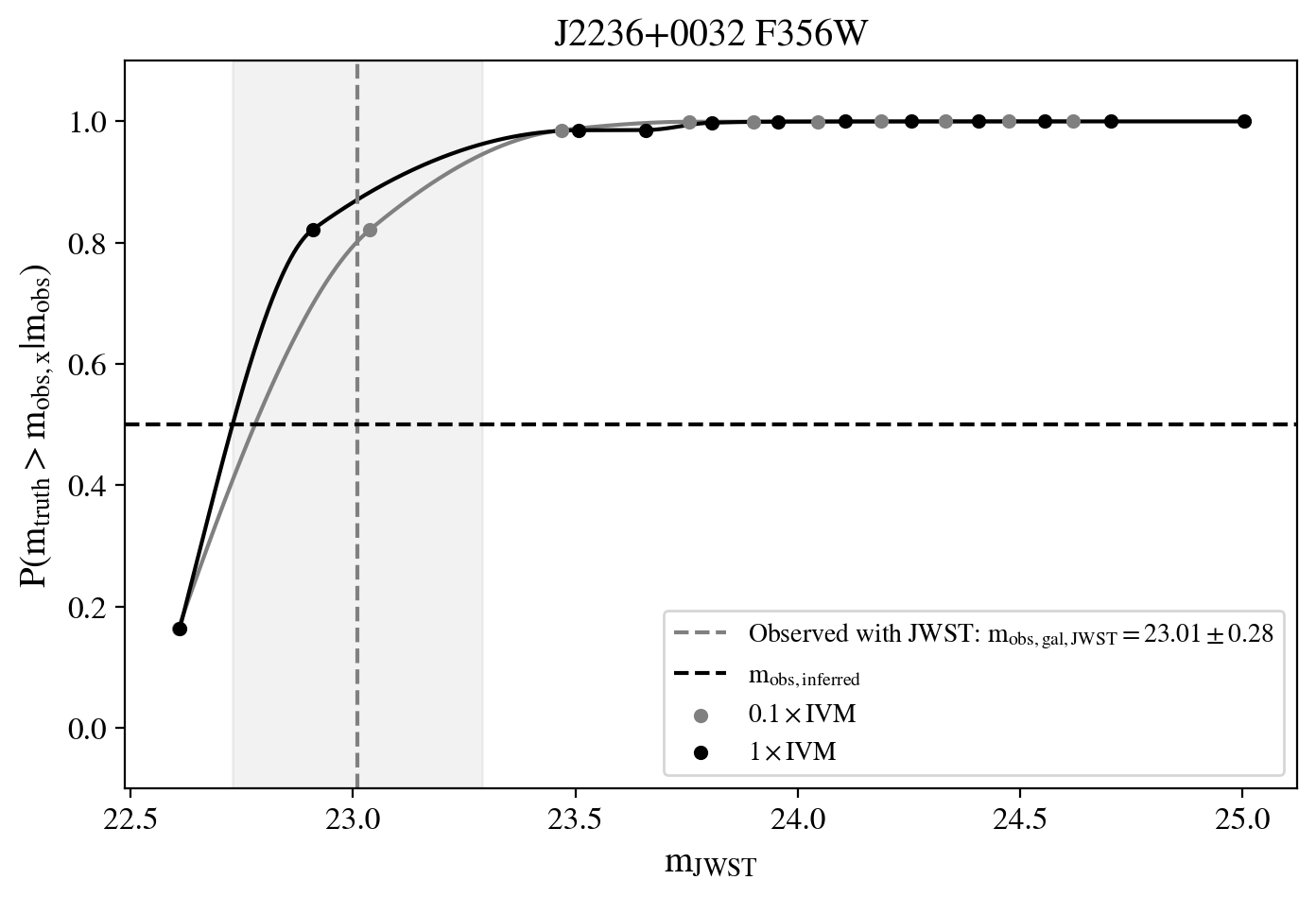} \\

\includegraphics[width=0.48\textwidth]{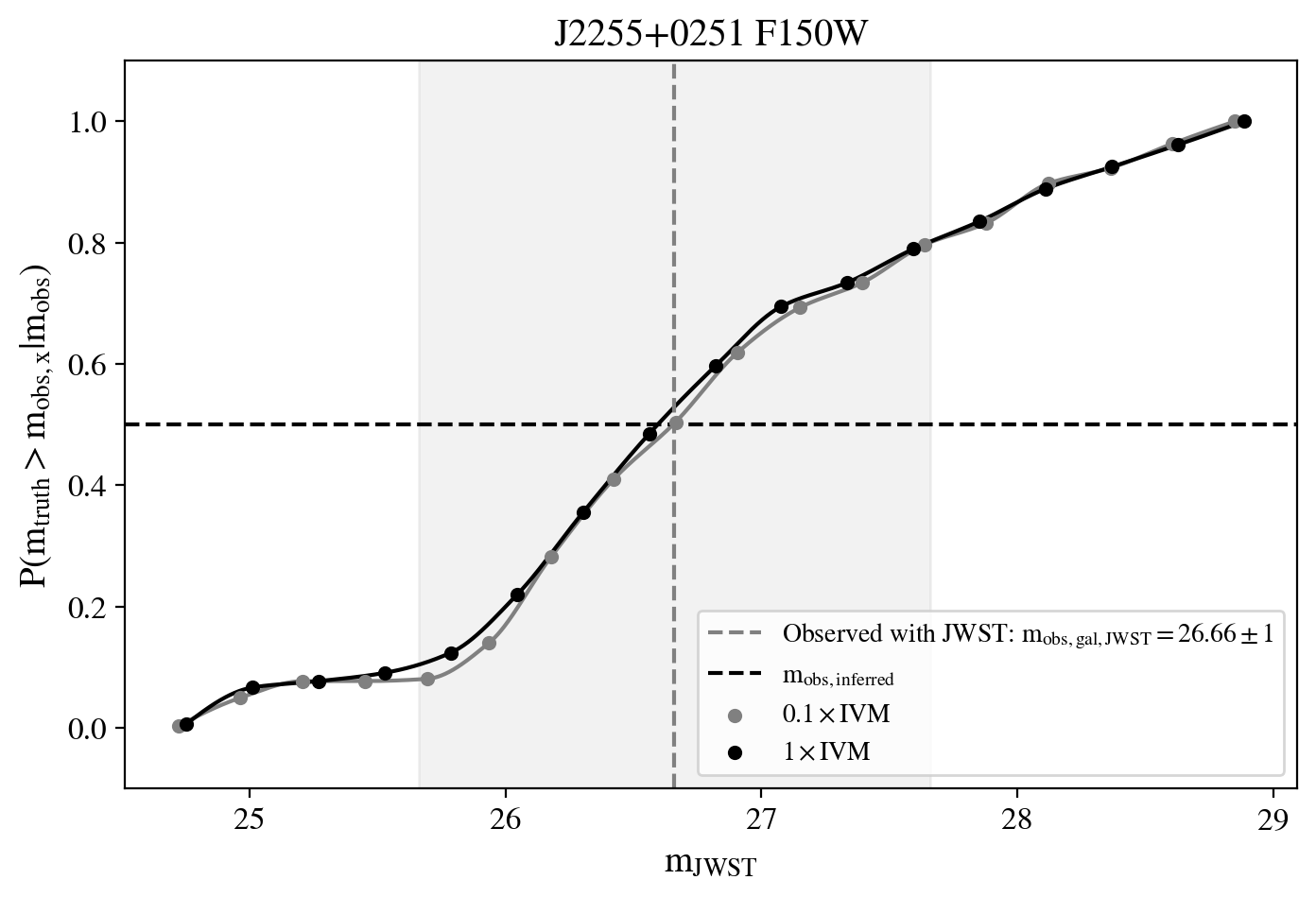} &
\includegraphics[width=0.48\textwidth]{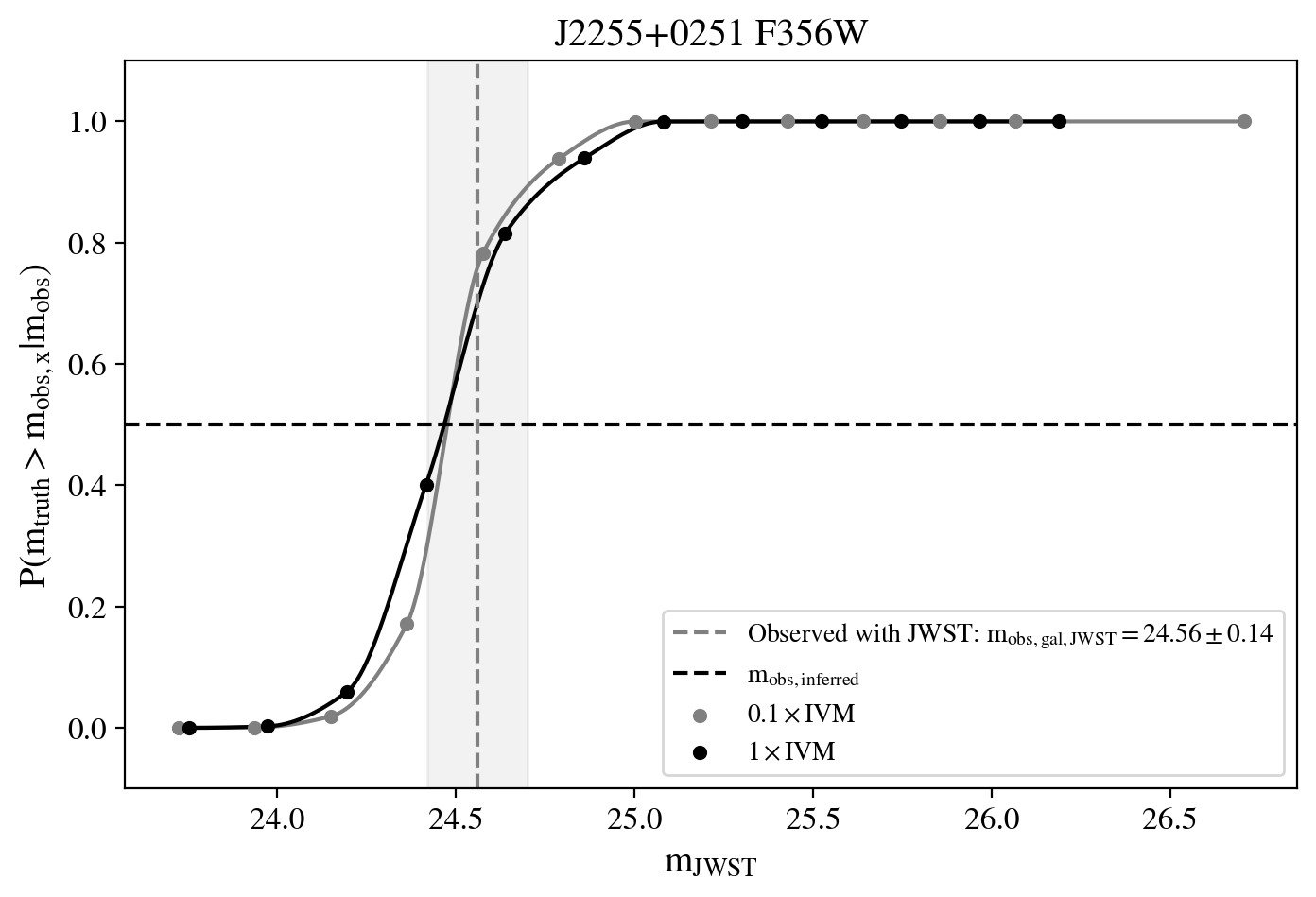} \\
\end{tabular}

\caption{Inferred weighted and observed posterior host magnitudes (part 4). See Figure~\ref{fig:weighted_images_1} for details.}
\label{fig:weighted_images_4}
\end{figure*}

\begin{figure*}
\centering
\setlength{\tabcolsep}{0pt}
\renewcommand{\arraystretch}{0}
\begin{tabular}{cc}
\includegraphics[width=0.45\textwidth]{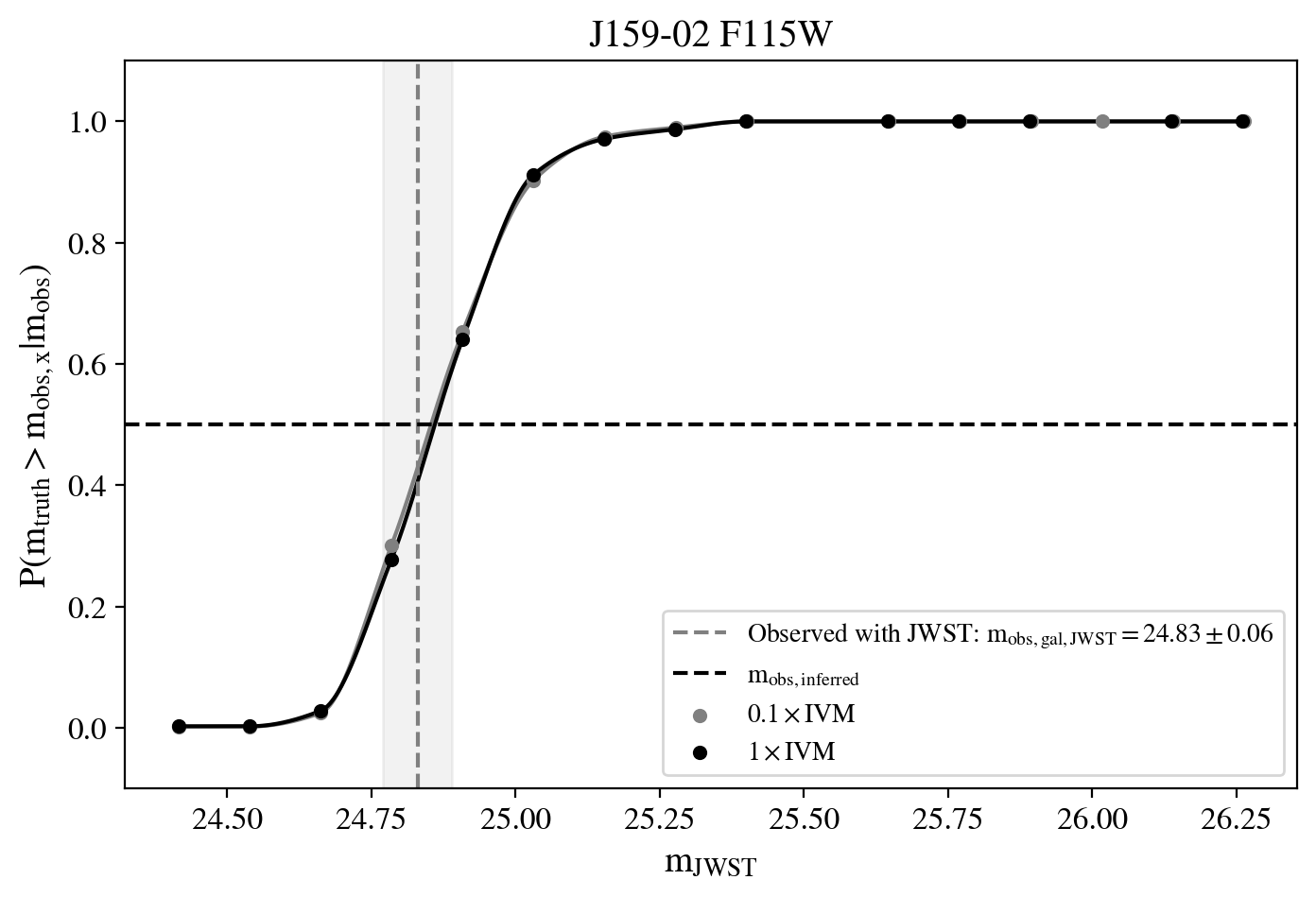} &
\includegraphics[width=0.45\textwidth]{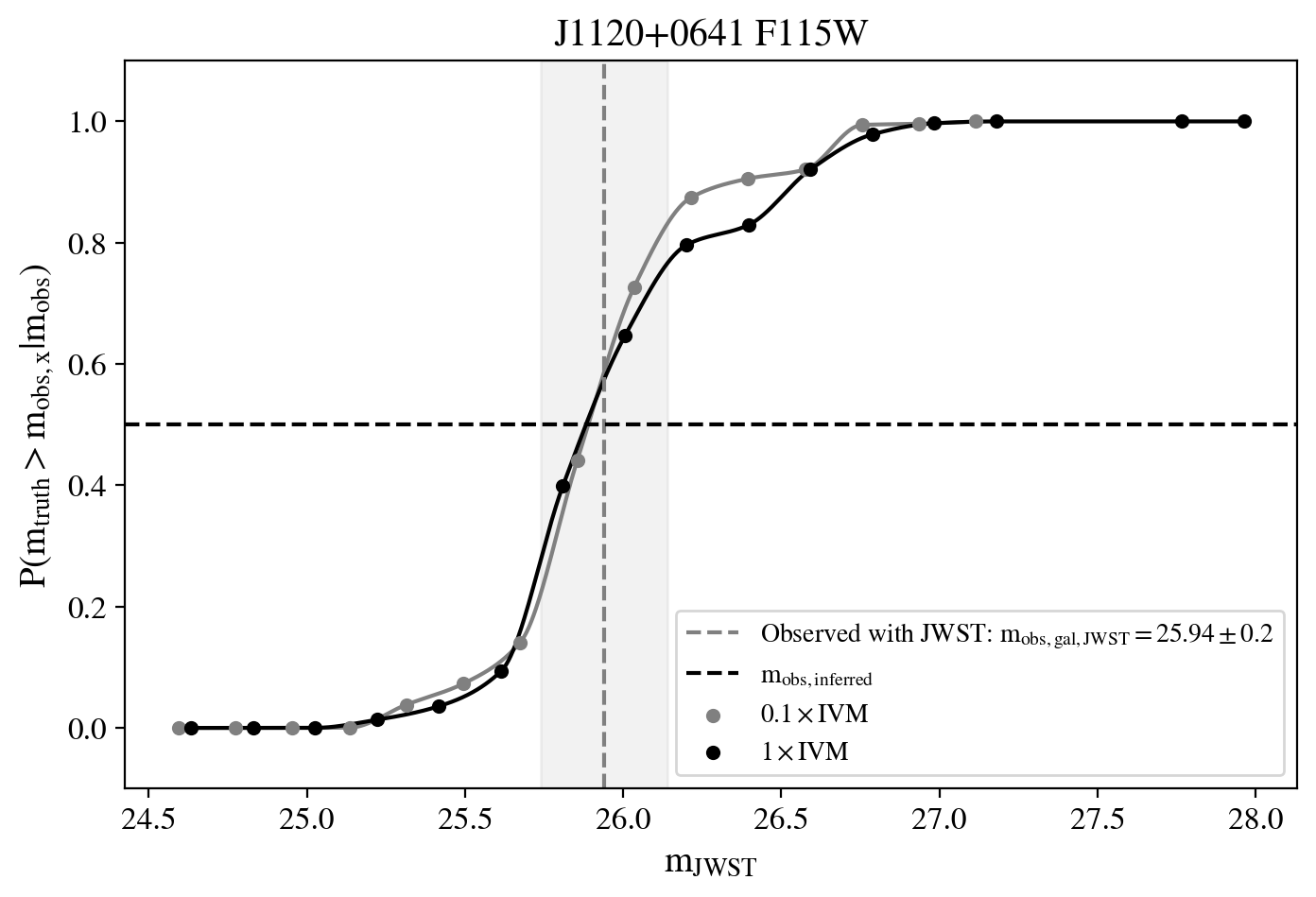} \\[1em]
\includegraphics[width=0.45\textwidth]{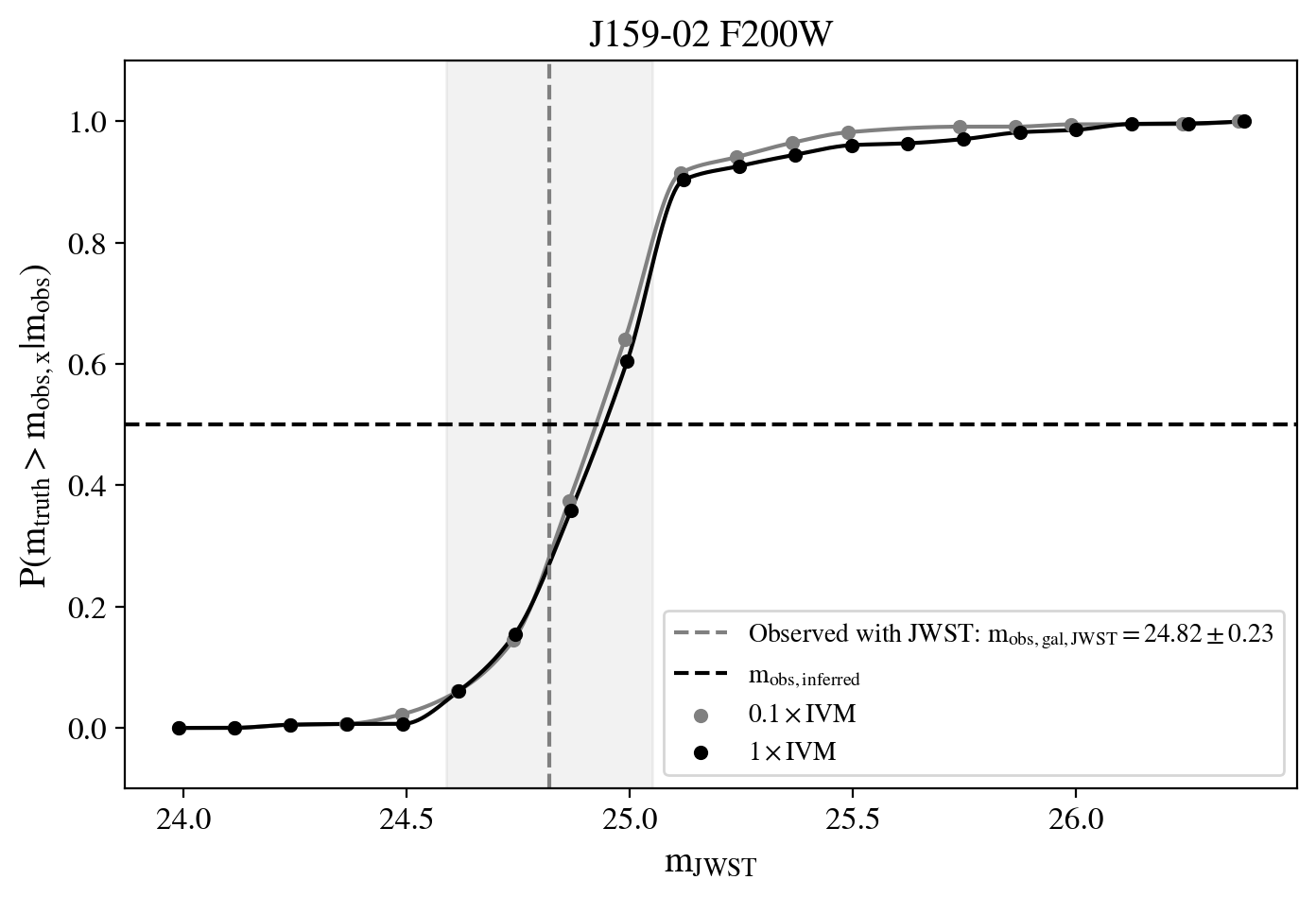} &
\includegraphics[width=0.45\textwidth]{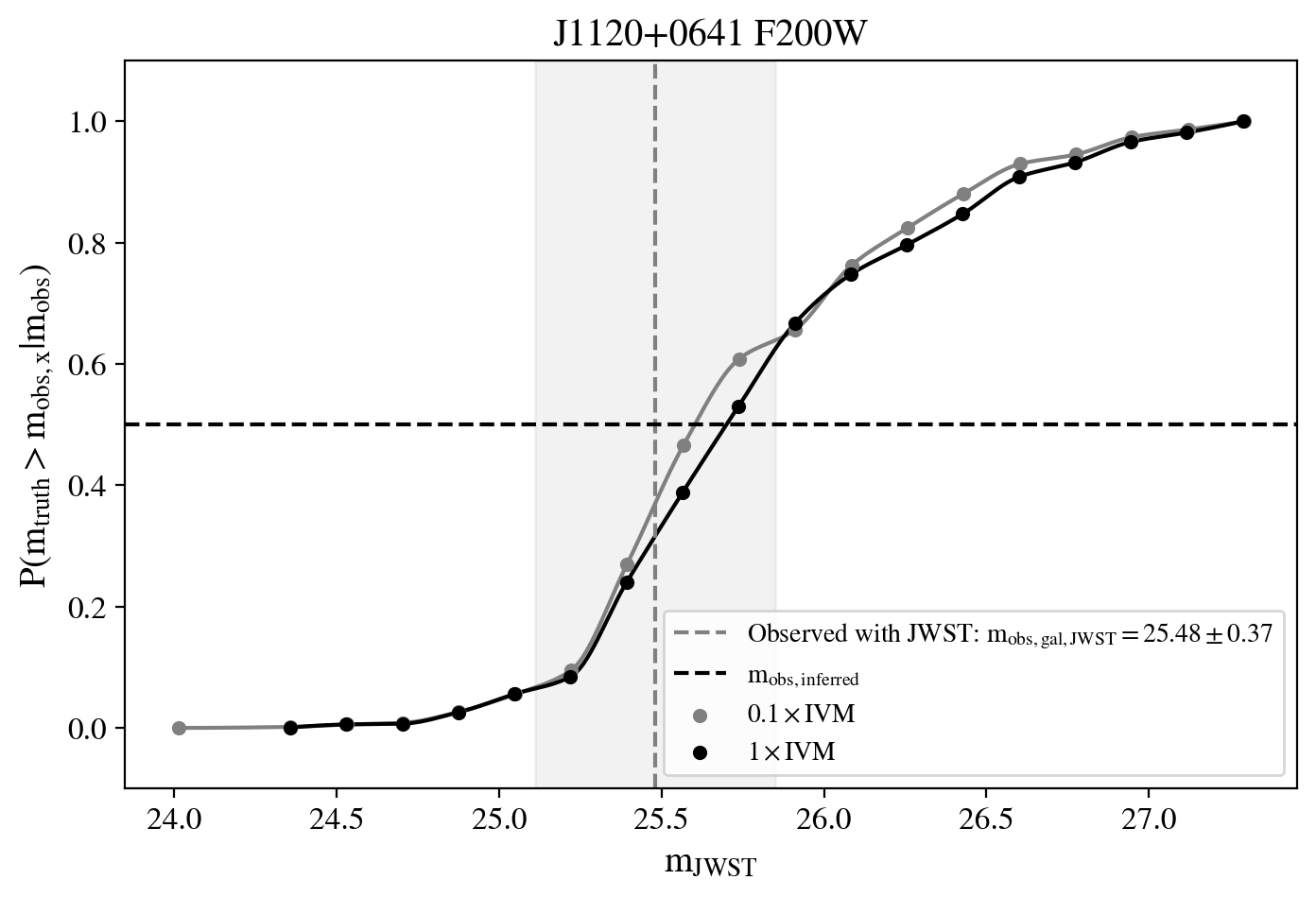} \\[1em]

\includegraphics[width=0.45\textwidth]{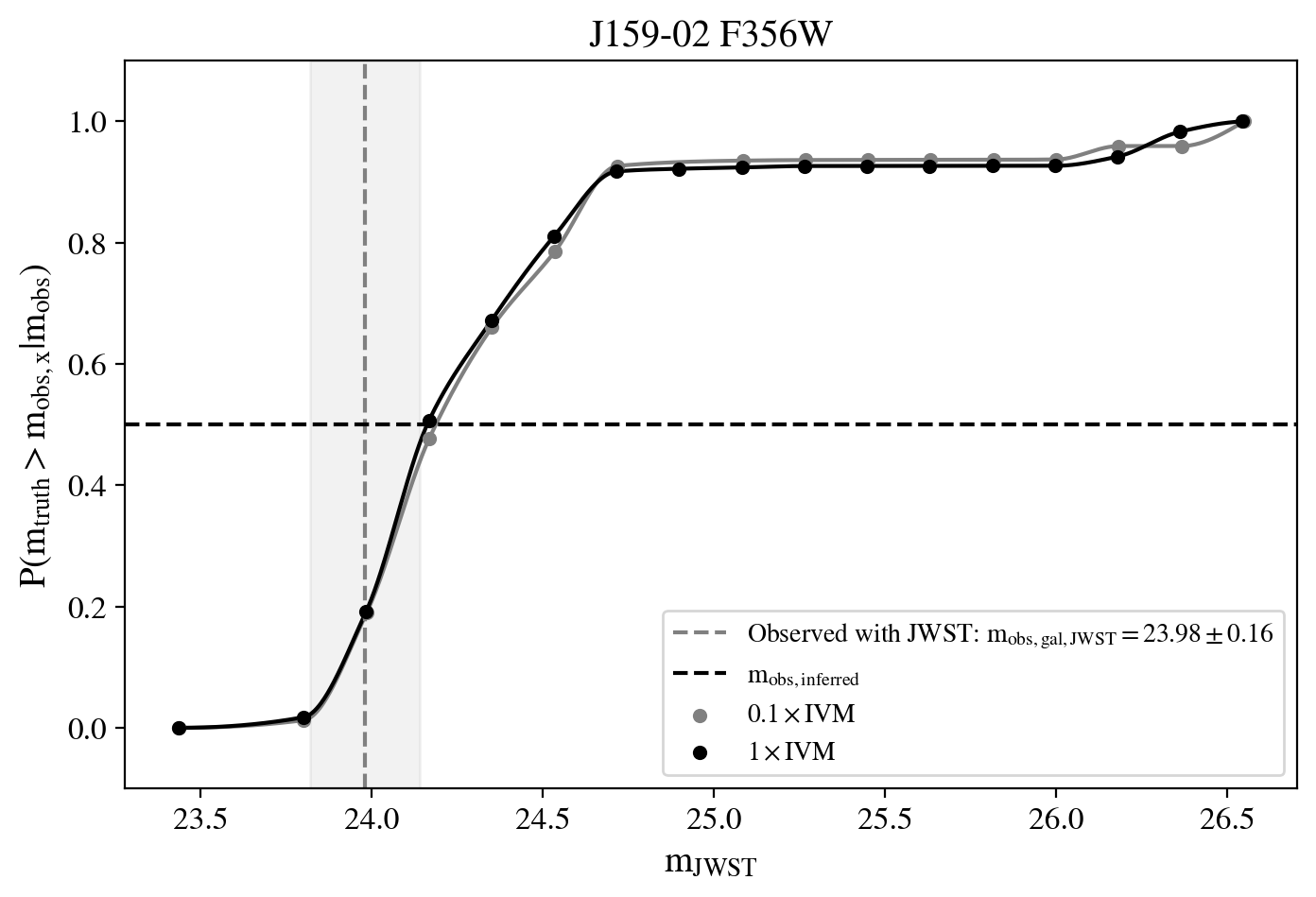} &
\includegraphics[width=0.45\textwidth]{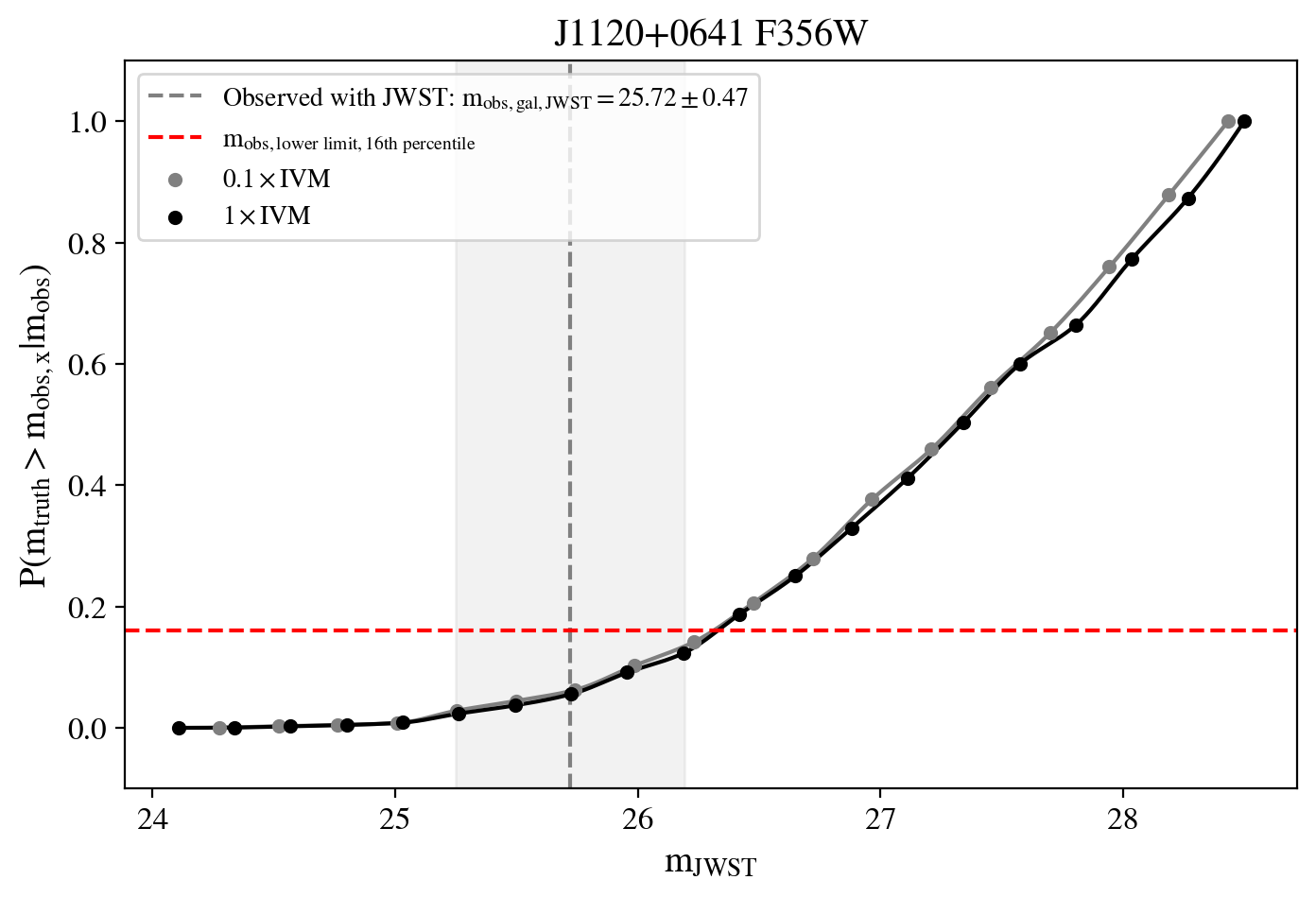} \\
\end{tabular}
\caption{Same as Figure \ref{fig:weighted_images_1}, but showing the final inferred weighted CDF for all quasar hosts with detections from \citet{yue2023eiger} and \citet{marshall2024ganifseigermerging}. Subfigures are now organized by NIRCam filter (rows) and quasar name (columns).}
\label{fig:weighted_images_5}
\end{figure*}

We use BlueTides to update the posteriors of all high-$z$ quasar host galaxy magnitudes from \citet{10.48550/arxiv.2211.14329}, \citet{ding2025shellqsjwstunveilshostgalaxies}, \citet{yue2023eiger}, and \citet{marshall2024ganifseigermerging}. The median and errors on the magnitudes are shown in Table \ref{tab:magnitude_offsets}. We display the CDFs of all updated posteriors in Figures \ref{fig:weighted_images_1} through \ref{fig:weighted_images_5}.

\section{Freeing the Sérsic Index}
\label{app:more_sersic}

We briefly explore the effects of allowing the Sérsic index to vary during the quasar removal process. In Section~\ref{sec:results}, we show that biases in the recovered host magnitudes are primarily driven by the presence of the quasar rather than the choice of Sérsic index. Figure~\ref{fig:mag_comp} illustrates this for J1120+0641 in F356W, demonstrating that the magnitude biases introduced by quasar contamination are significantly larger than those arising from fixing or freeing the Sérsic index.

\begin{figure*}
    \centering
    \includegraphics[width=1\linewidth]{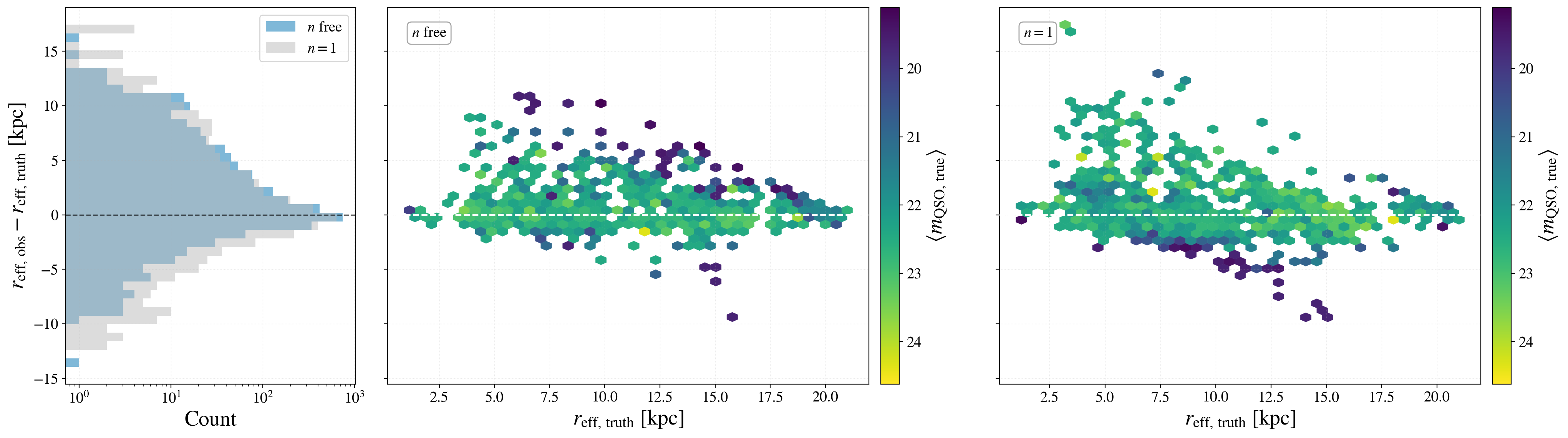}
\caption{Comparison of inferred radii and input truth radii for galaxies in the F356W filter. The colorbar indicates the quasar magnitude used in each galaxy. Biases in the recovered radii persist regardless of whether the Sérsic index is fixed or allowed to vary. We do not apply cuts based on accurate radial or quasar magnitude recovery in this figure, unlike in the main analysis of this paper.}
\label{fig:r_comp}
\end{figure*}

We also examine the impact on recovered effective radii. As shown in Figure~\ref{fig:r_comp}, radial differences remain biased regardless of whether the Sérsic index is fixed or allowed to vary. This indicates that allowing $n$ to vary does not mitigate the dominant bias introduced by quasar contamination. For a free Sérsic index, we also do not find a clear trend that incorporating a quasar with a brighter flux under or overbiases the recovered Sérsic radius. For a free Sérsic index and high quasar flux, we find a slight bias toward overestimating the host galaxy radius, whereas fixing the profile to $n = 1$ results in an underestimation. This effect is particularly relevant for the compact galaxies commonly observed at high redshift.

\begin{figure*}
    \centering
    \includegraphics[width=0.7\linewidth]{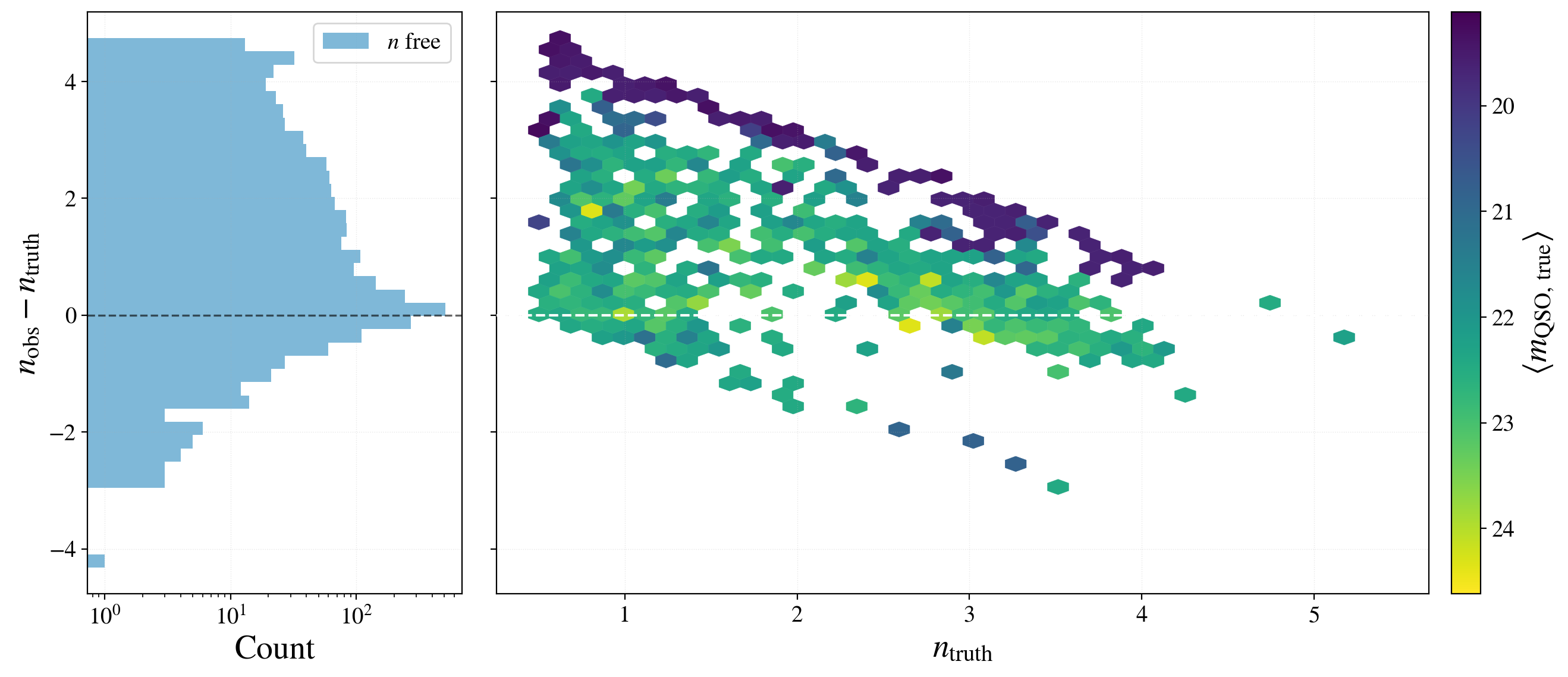}
\caption{Comparison of recovered Sérsic indices and truth values for F356W across the host galaxy and quasar magnitude sample used in this analysis. The colorbar indicates the count per hexbin. The recovered Sérsic indices can deviate significantly from the input values, demonstrating the additional degeneracy introduced when allowing $n$ to vary. We do not apply cuts based on accurate radial or quasar magnitude recovery in this figure, unlike in the main analysis of this paper.}
\label{fig:n_comp}
\end{figure*}

Allowing the Sérsic index to vary also introduces additional degeneracies between the Sérsic index, effective radius, and host magnitude during the quasar and host decomposition. In Figure~\ref{fig:n_comp}, we show that the recovered Sérsic indices can vary substantially relative to the input values. Interestingly, a high quasar flux tends to bias the inferred Sérsic index toward larger values. \textit{psfmc} is likely compensating for quasar and host degeneracy by concentrating more of the host light toward the center. We also observe that the spread in the residuals decreases toward higher Sérsic index, suggesting that more centrally concentrated profiles are less sensitive to this degeneracy.

This additional degree of freedom can therefore lead to unstable fits without improving the recovery of host galaxy properties. For the purposes of this study, we therefore restrict our analysis to galaxies modeled with $n=1$, selecting cases where the input and recovered (with quasar) radii agree within $1\sigma$.

\clearpage


\clearpage

\section{Photometric and Sérsic Magnitude Discrepancies}
\label{appendix:sersic}
\begin{figure}
    \centering
    \includegraphics[width=1.1\linewidth]{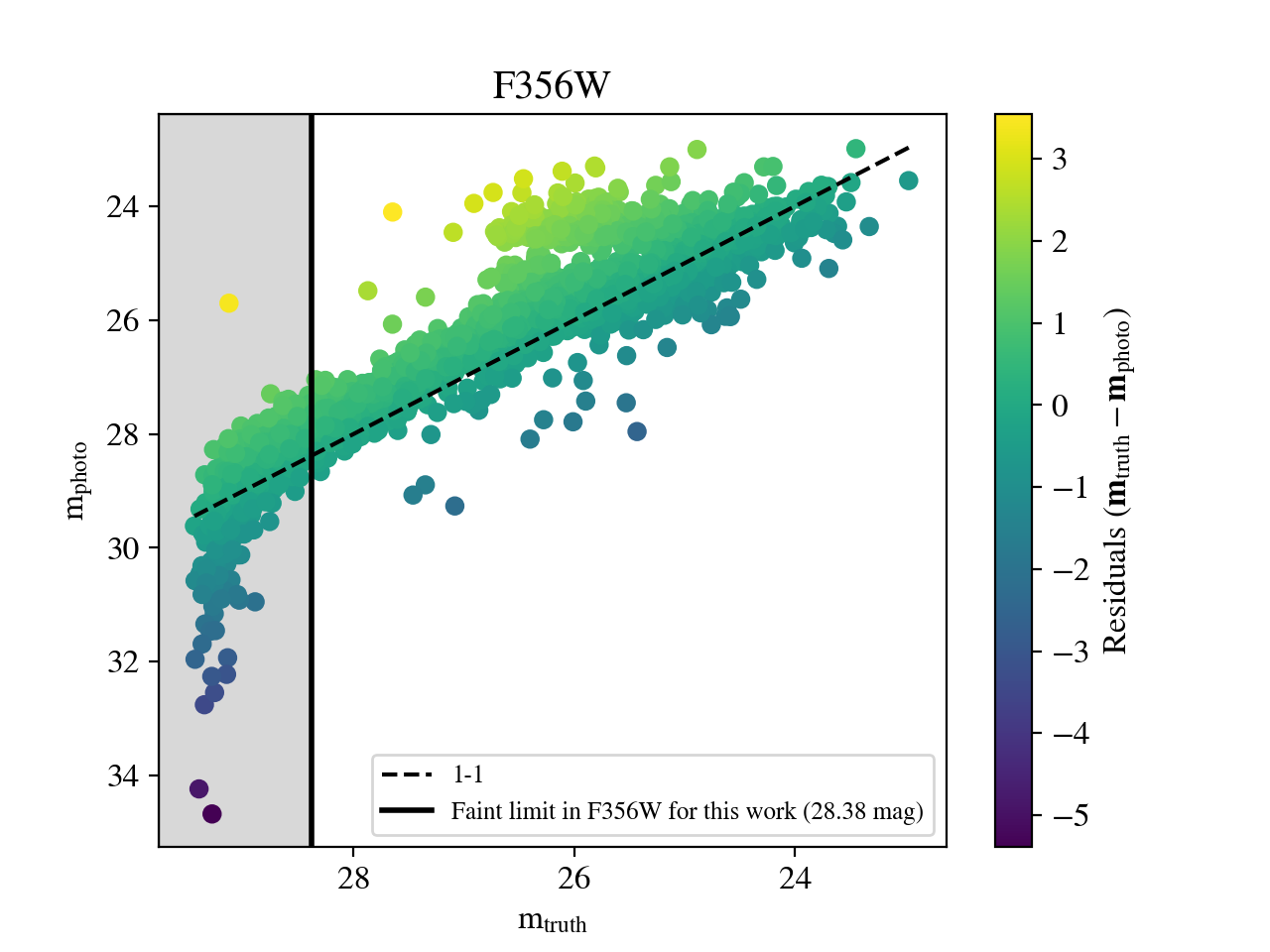}
    \caption{A sample of 2,000+ BlueTides galaxies with magnitudes estimated (without a quasar) using a \textit{psfmc}’s Sérsic profile fit ($\rm m_{truth}$) and a
photometric magnitude calculated using the radius of the galaxy ($\rm m_{photo}$).
As described in Appendix \ref{appendix:sersic}, using Sérsic profile magnitude values for galaxies
must be done carefully for dimmer galaxies.}
    \label{fig:photometric}
\end{figure}

In Figure \ref{fig:photometric}, we show a large sample of BlueTides galaxies with F356W magnitudes determined through a Sérsic profile fit and a photometric aperture. In Figure \ref{fig:photometric_residuals_panel_page1} through Figure \ref{fig:photometric_residuals_panel_page2}, we show the photometric versus Sérsic magnitudes and residuals for each galaxy analyzed in this work. The photometric aperture is six times the half light radius found in \citet{marshall_2022_dust}. The half-light radii are calculated by performing aperture photometry with 20 circular apertures logarithmically spaced between 0.05 to 2 kpc to determine the radius containing half the flux. We find that the Sérsic profile predicted magnitude and the photometric magnitude calculated can vary significantly, especially for extremely dim galaxies (m < 29 mag) where we see a difference of up to 5 magnitudes. 

High-$z$ galaxy magnitude estimates using alternative galaxy light fitting methods or aperture photometry (if possible) could provide better estimates for dimmer galaxies. We anticipate that the majority of the difference between the two magnitude estimates is due to including more than one galaxy in the photometric aperture magnitude estimate, using too large of an aperture, or alternatively the Sérsic model providing a poor fit to the 2D light profile. Either could lead to a significant flux difference. Since this work relies exclusively on the Sérsic magnitude—except during the initial galaxy selection, where the photometric magnitude is used—we do not explore this discrepancy further. We also rely on a 2D light profile to update each posterior such that using the Sérsic profile is necessary in this work. However, future studies should carefully account for differences between Sérsic fit and photometric magnitudes or explore other more realistic galaxy light profiles such as those found in \citet{stoneAstrophotFittingEverything2023}.

\begin{figure*}
    \centering
    \includegraphics[width=0.28\textwidth]{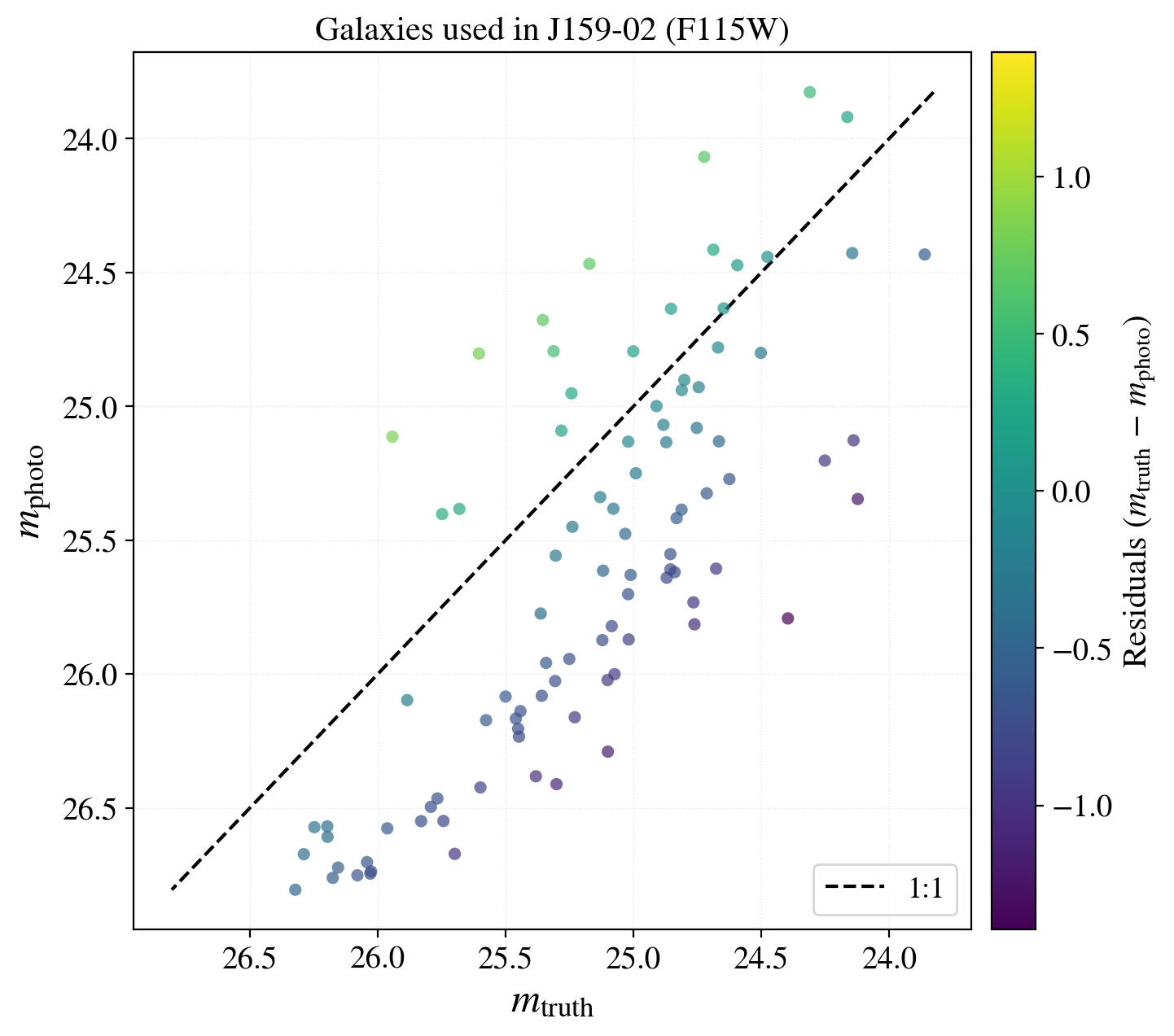}
    \includegraphics[width=0.28\textwidth]{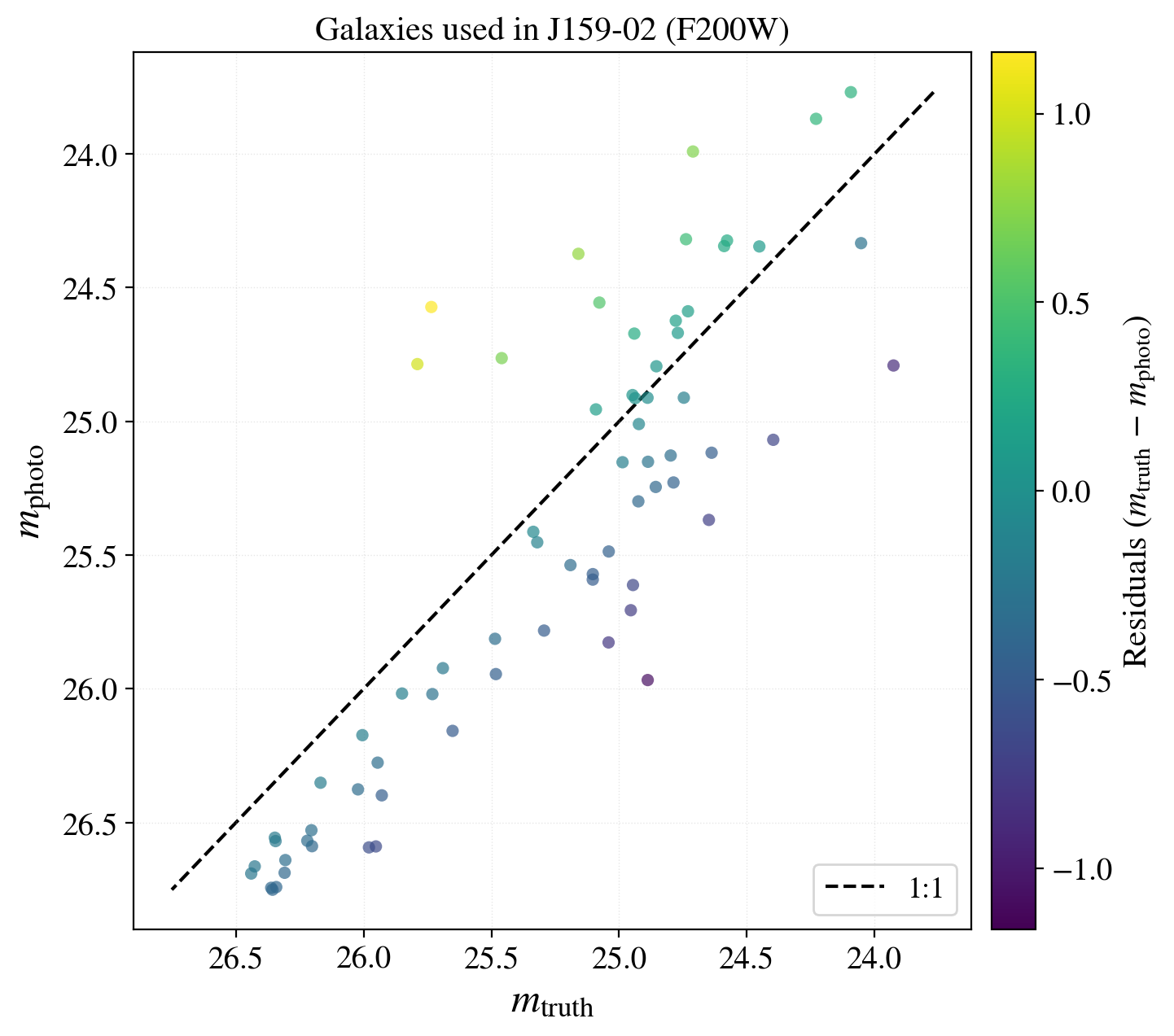}
    \includegraphics[width=0.28\textwidth]{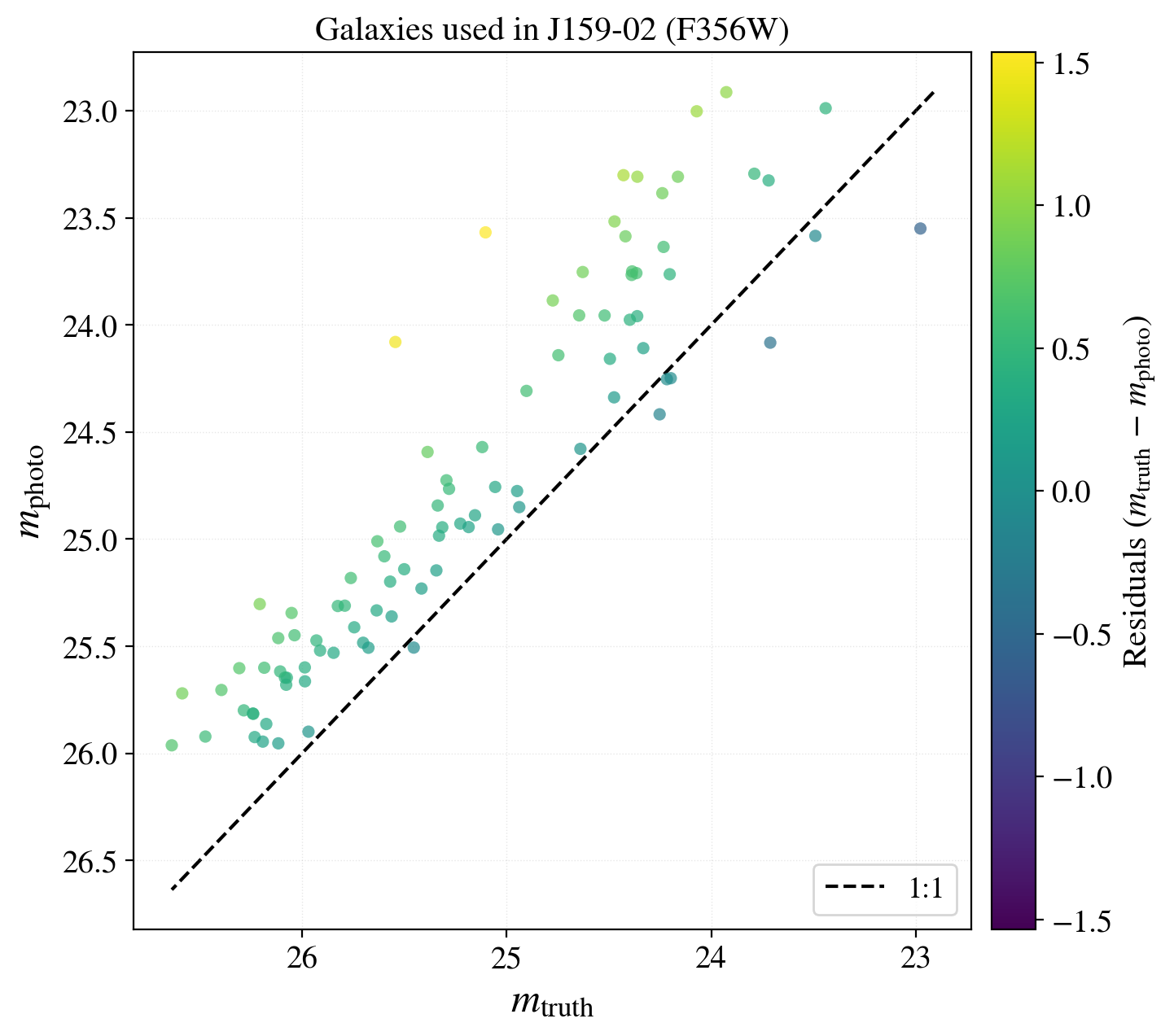}
    \\[0.2cm]
    \includegraphics[width=0.28\textwidth]{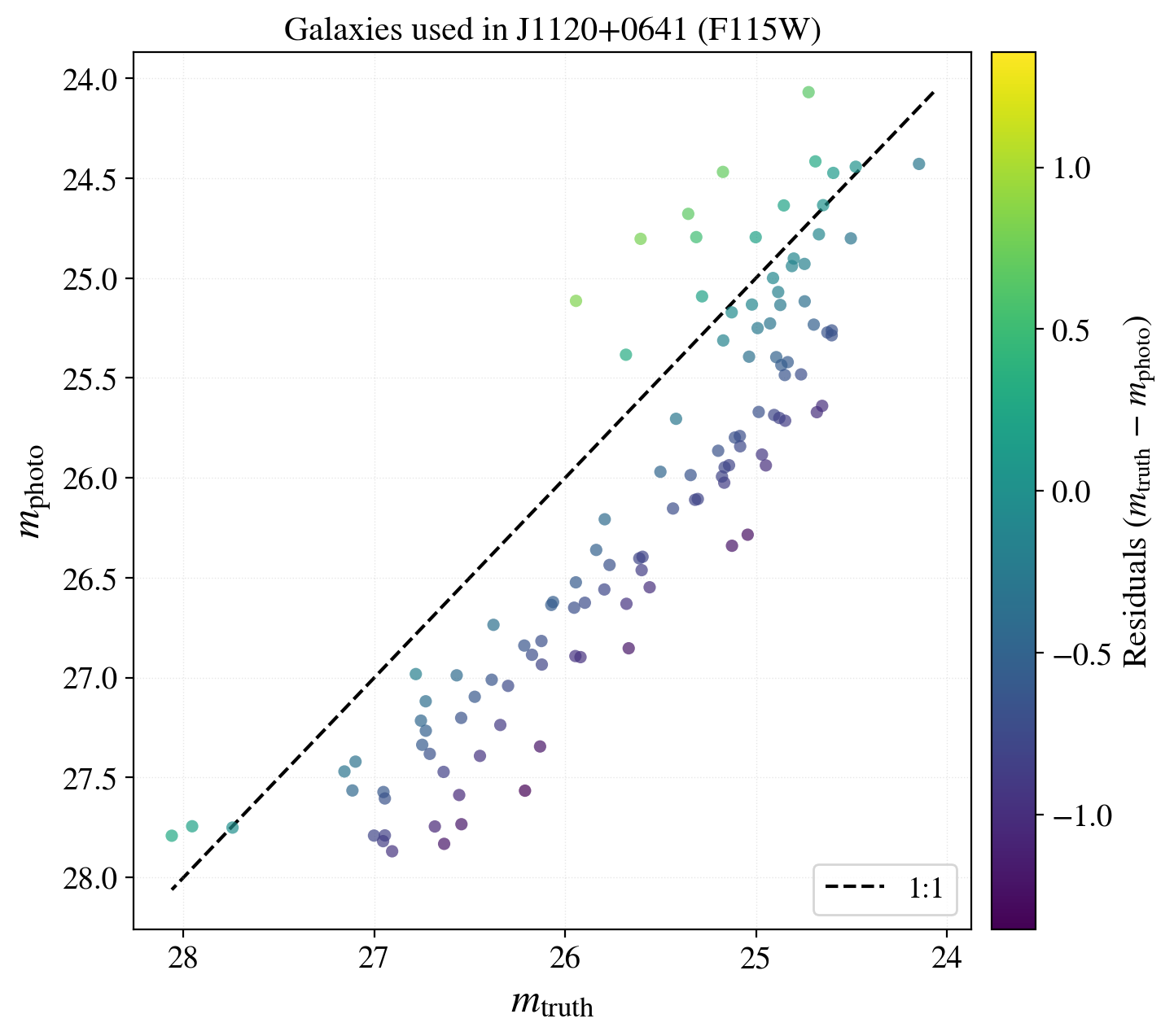}
    \includegraphics[width=0.28\textwidth]{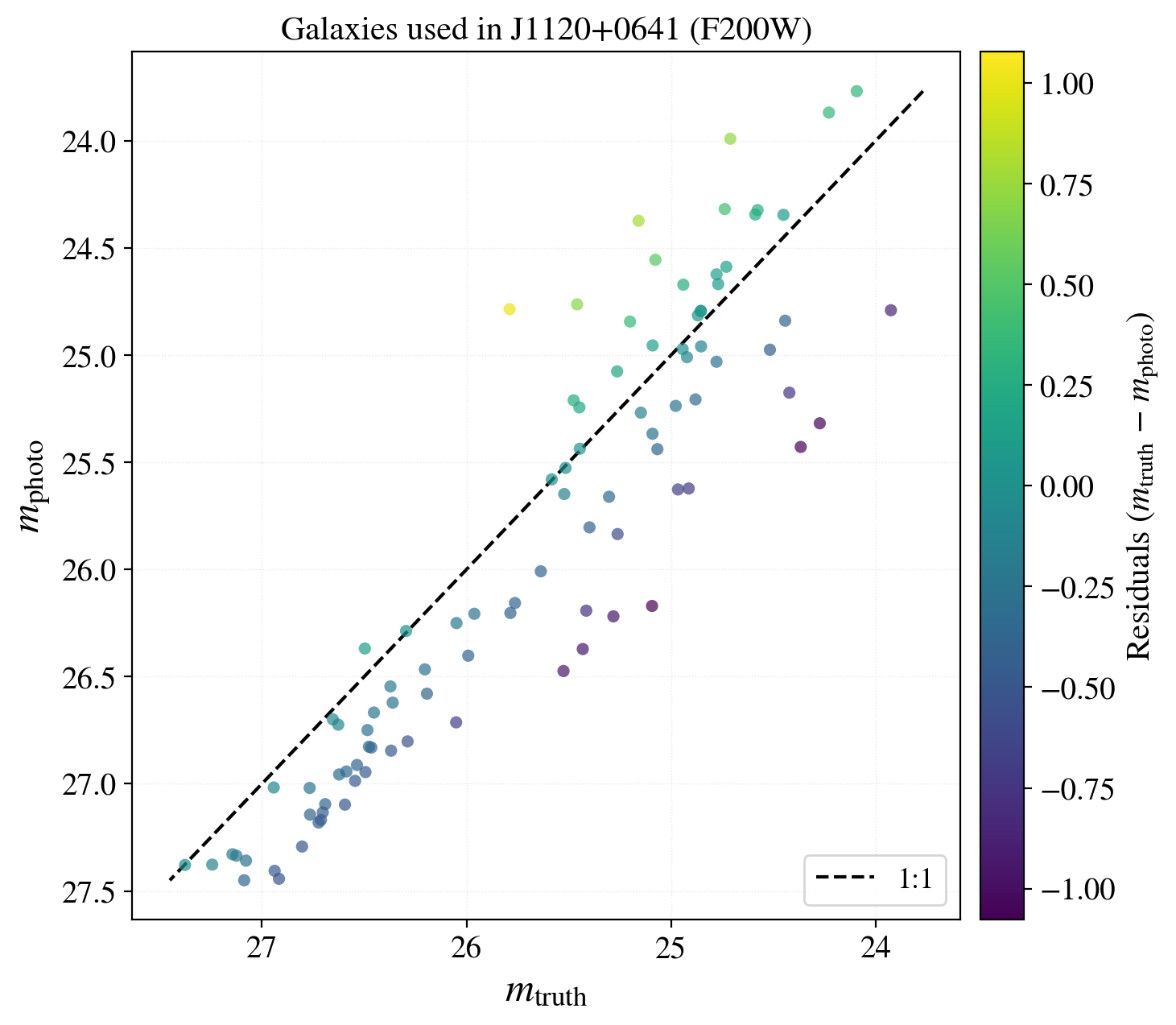}
    \includegraphics[width=0.28\textwidth]{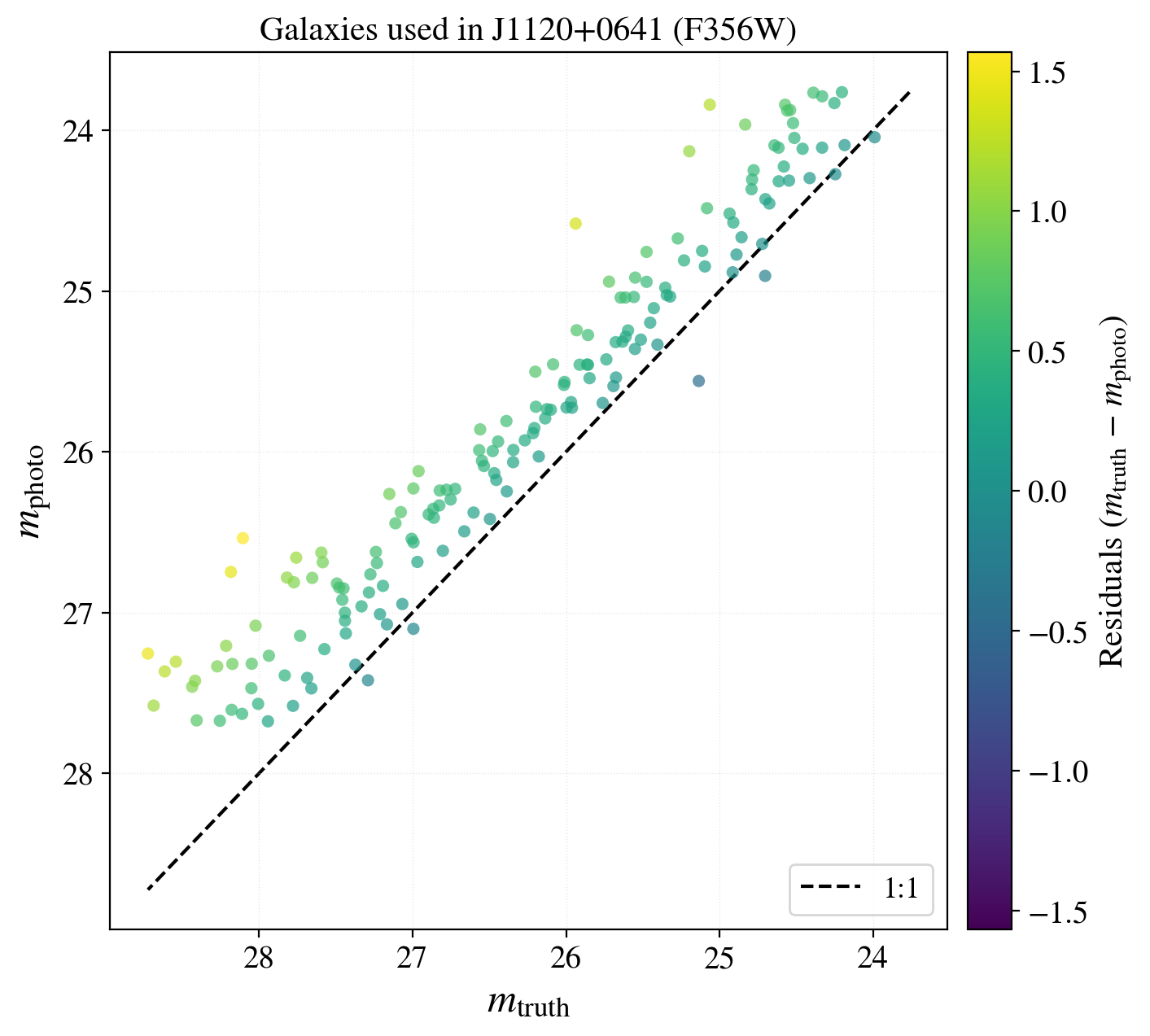}
    \\[0.2cm]
    \includegraphics[width=0.28\textwidth]{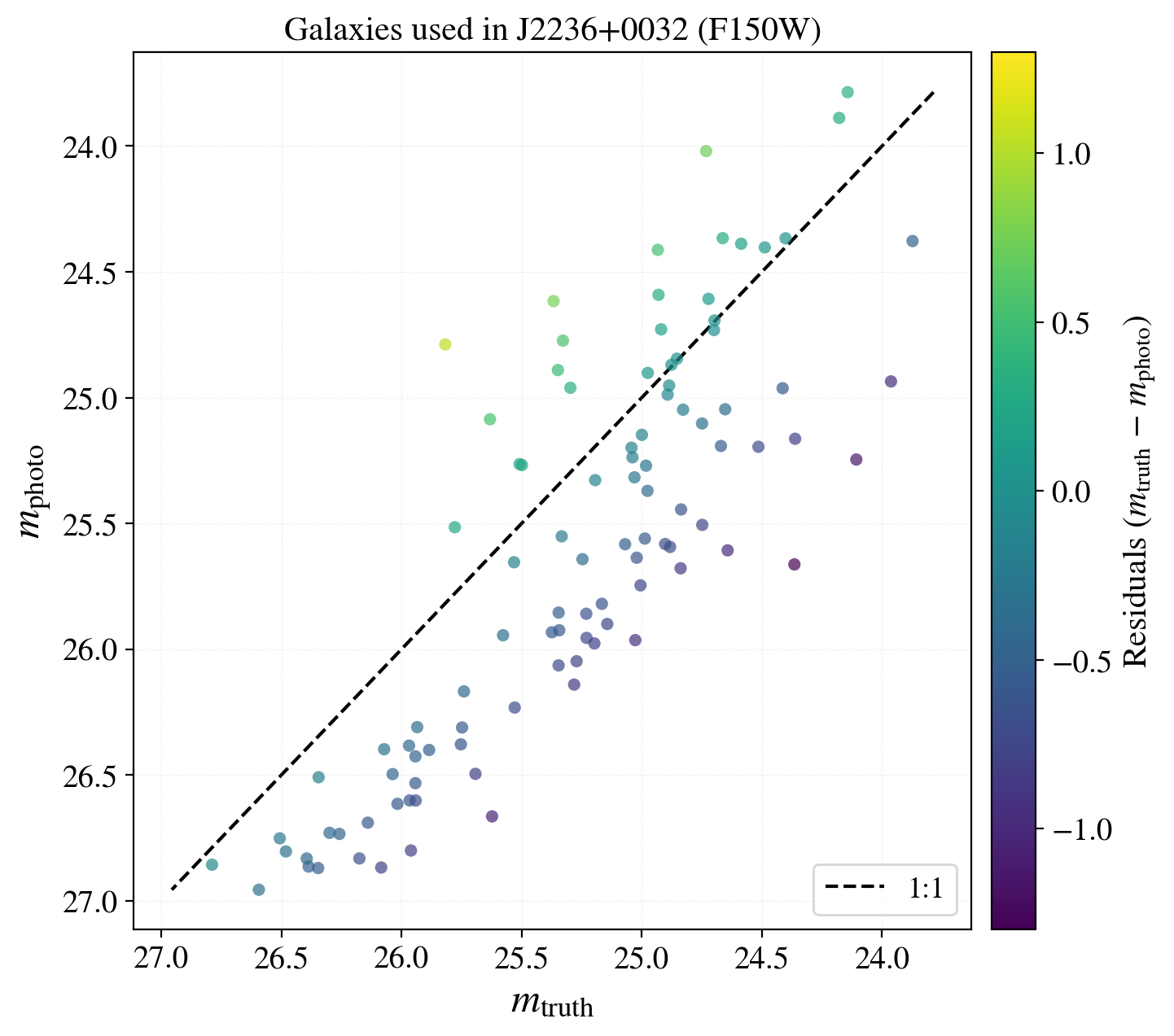}
    \includegraphics[width=0.28\textwidth]{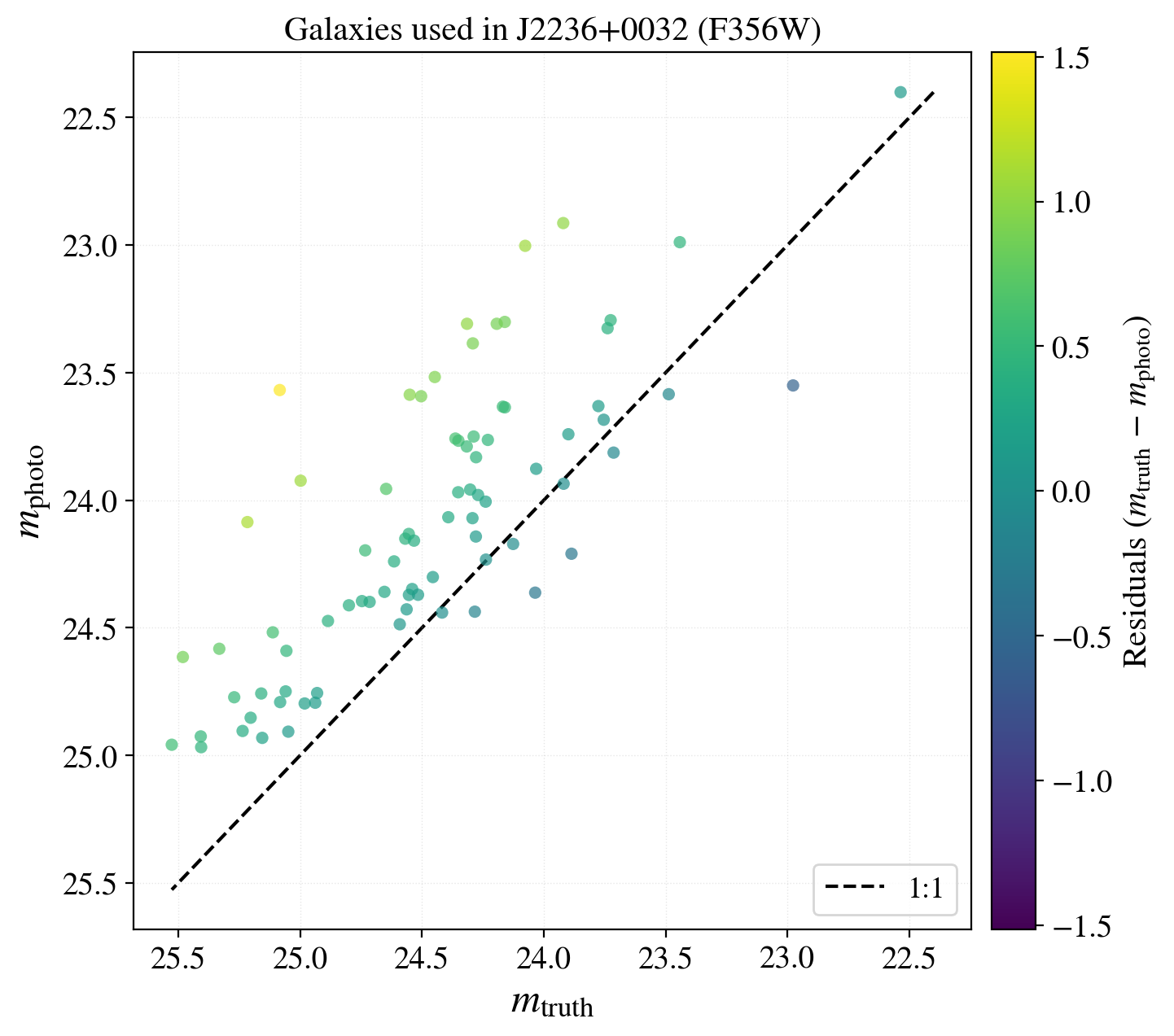}
    \includegraphics[width=0.28\textwidth]{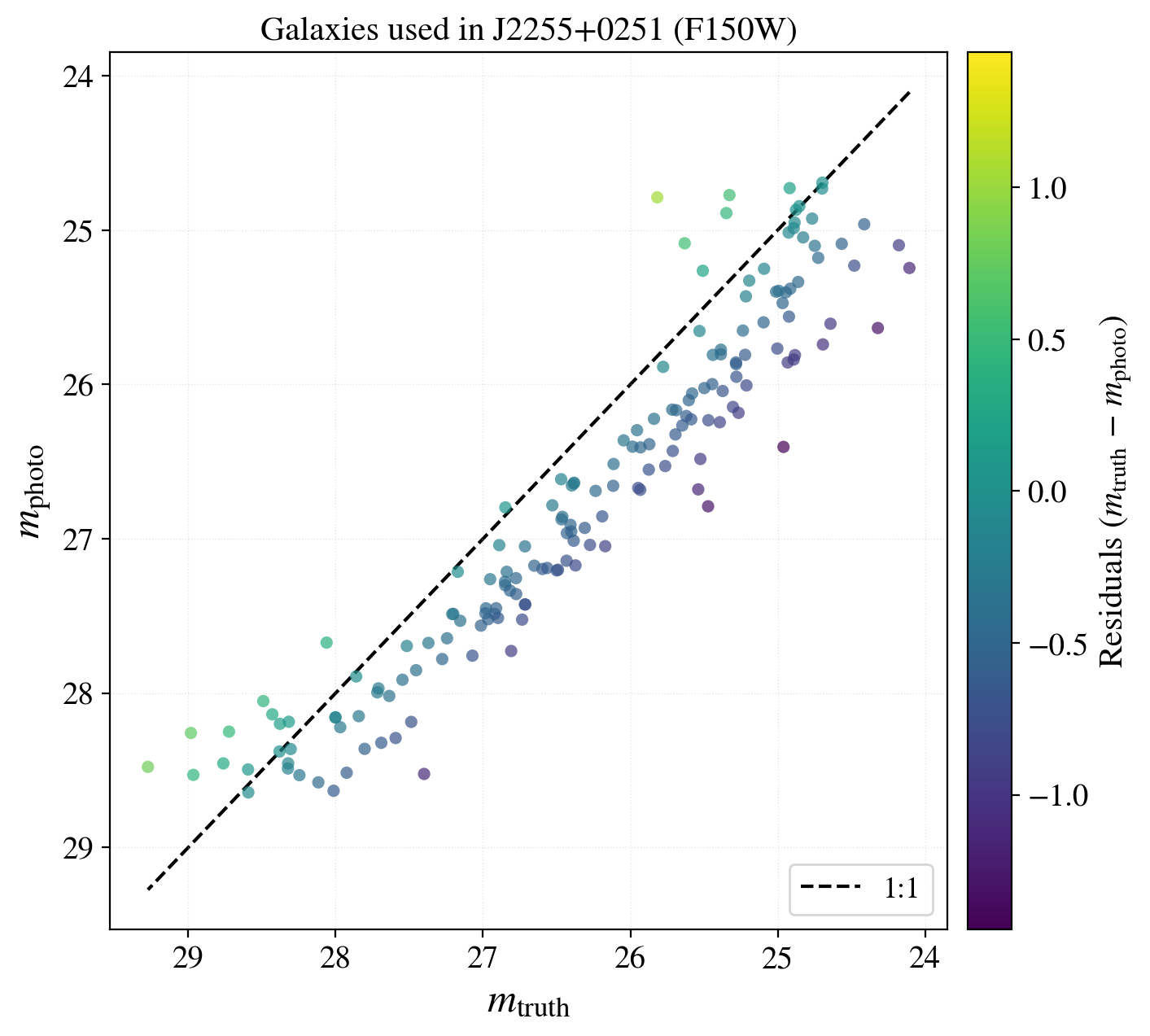}
    \\[0.2cm]
    \includegraphics[width=0.28\textwidth]{residuals_photo/photometric_residuals_J2255+0251_F150W.png}
    \includegraphics[width=0.28\textwidth]{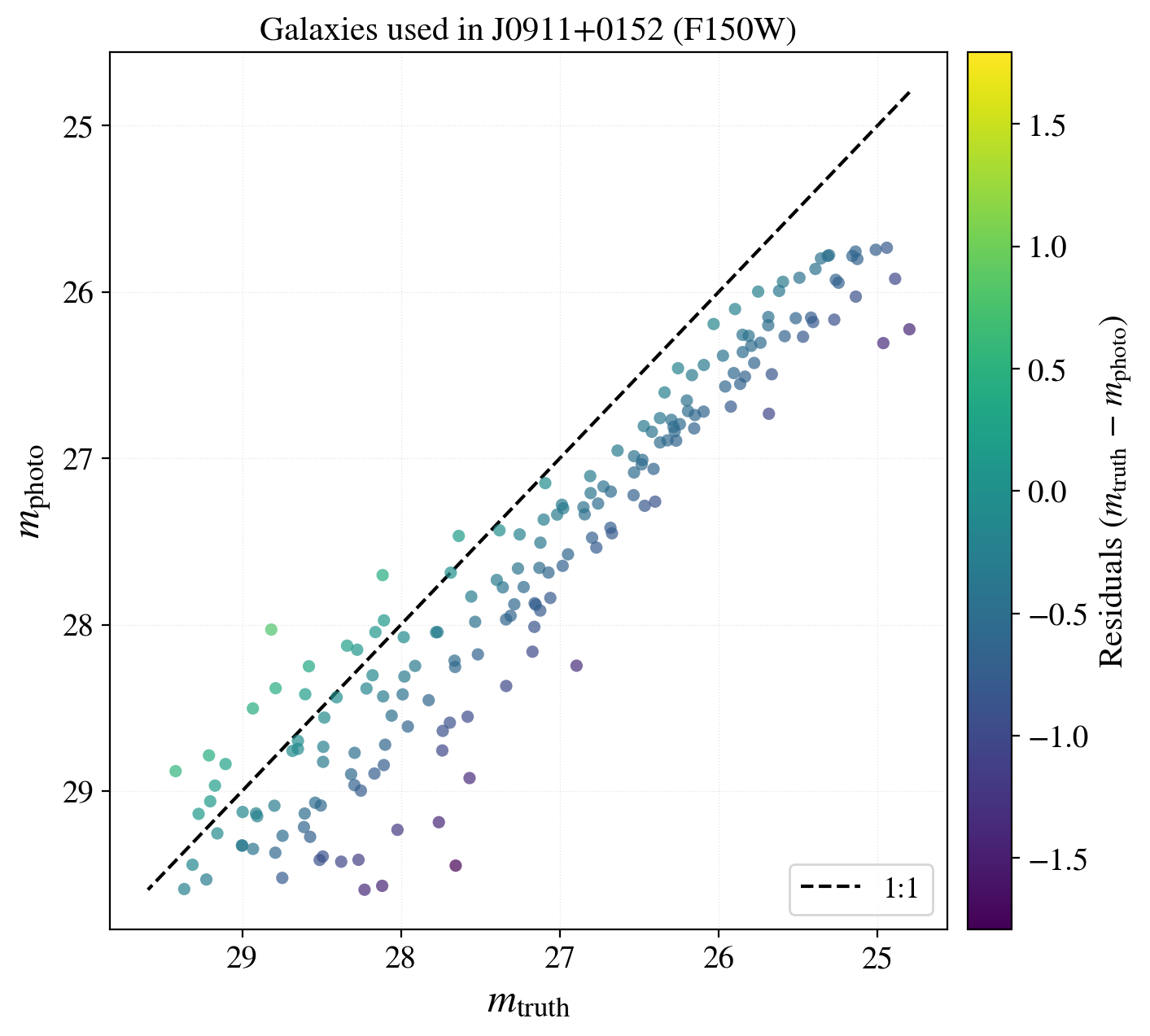}
    \includegraphics[width=0.28\textwidth]{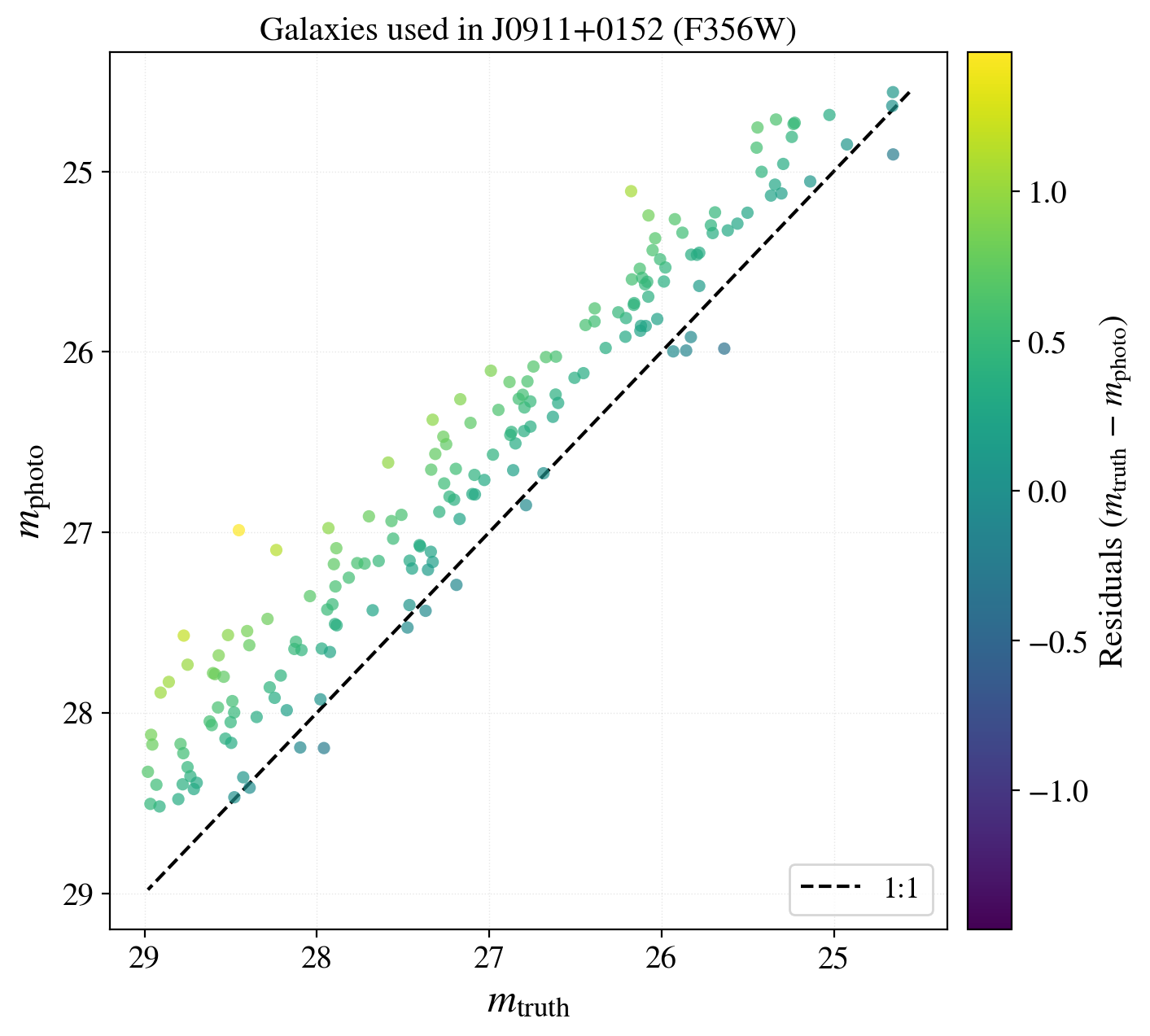}
    \\[0.2cm]
    \includegraphics[width=0.28\textwidth]{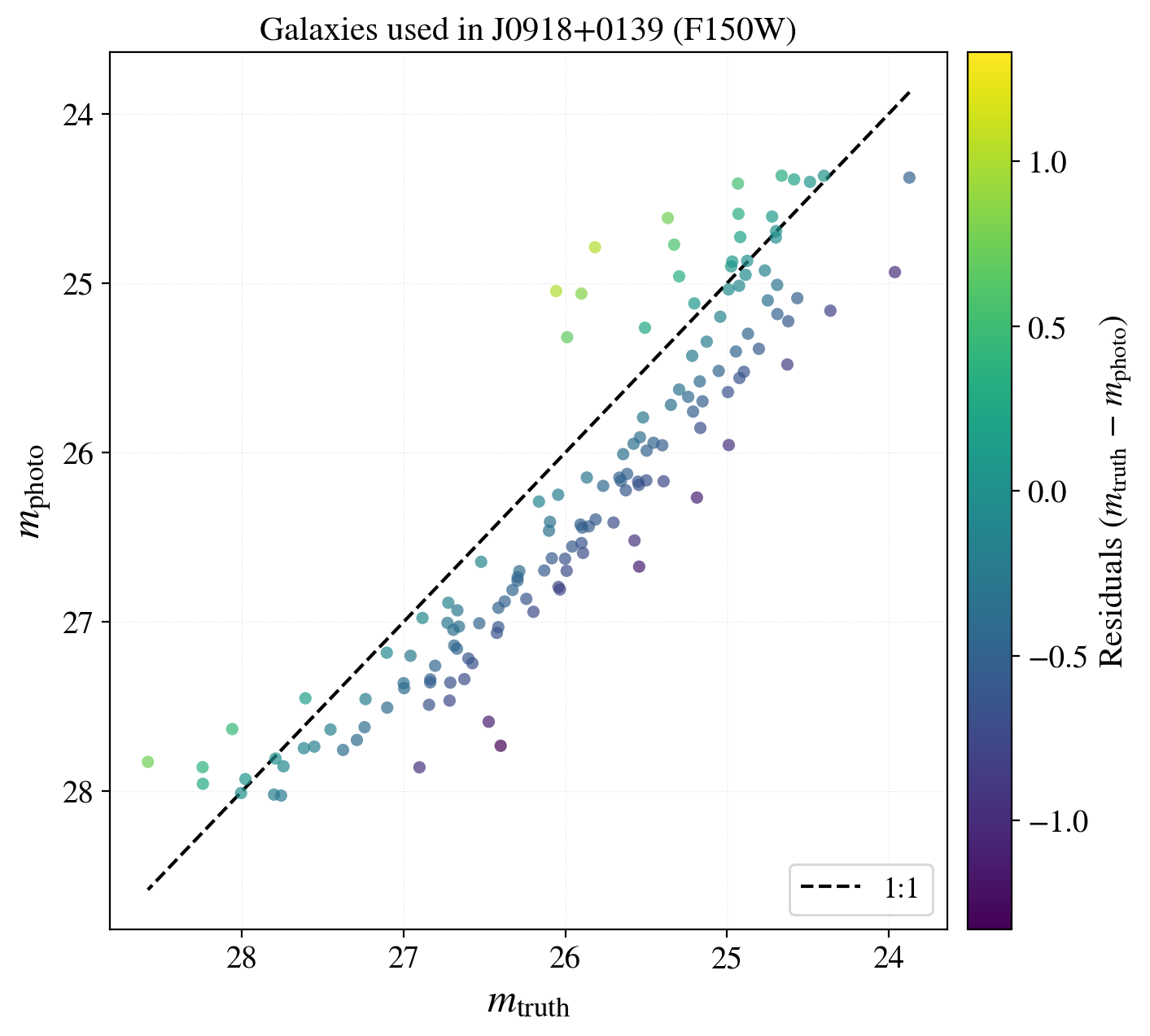}
    \includegraphics[width=0.28\textwidth]{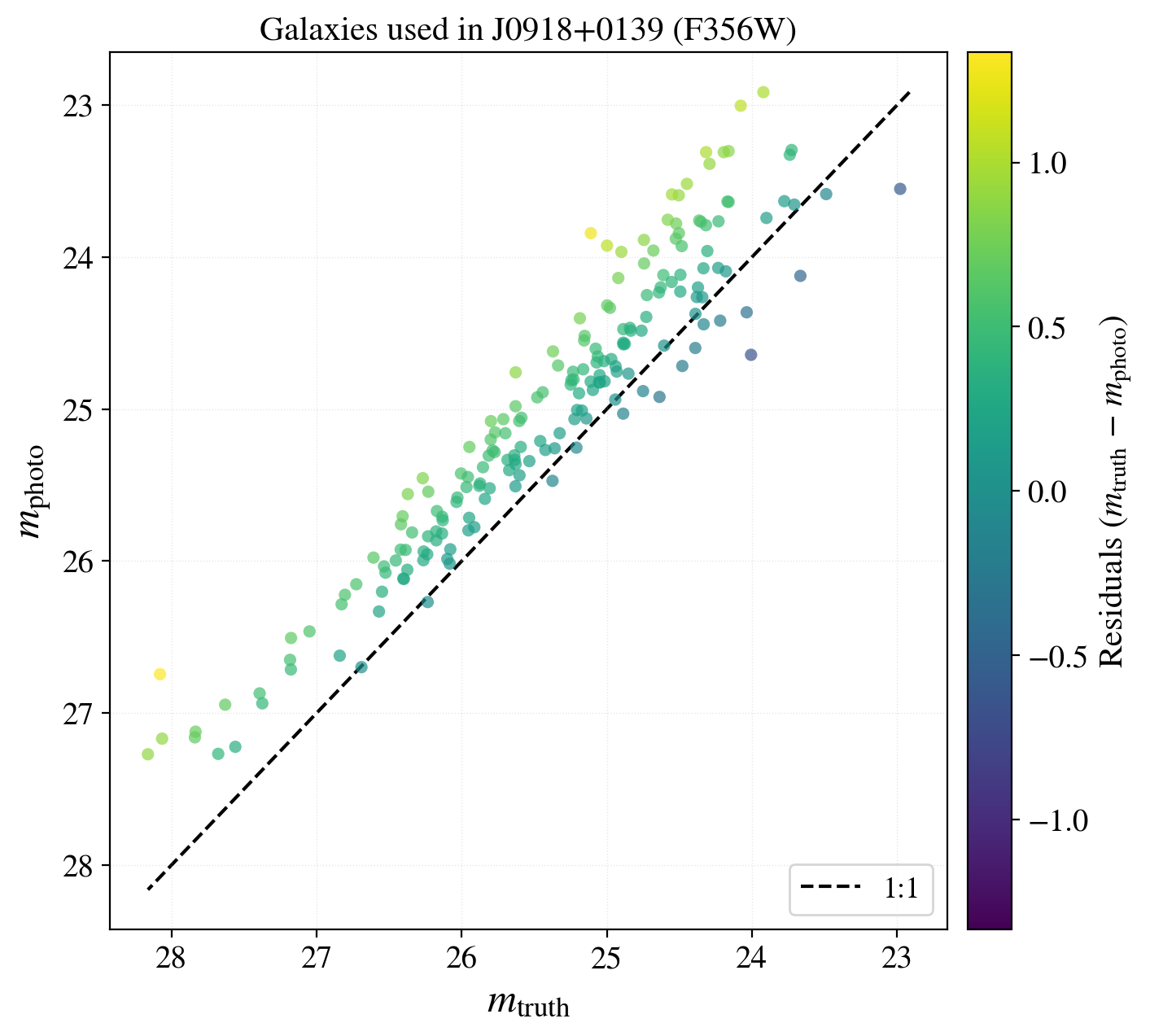}
    \includegraphics[width=0.28\textwidth]{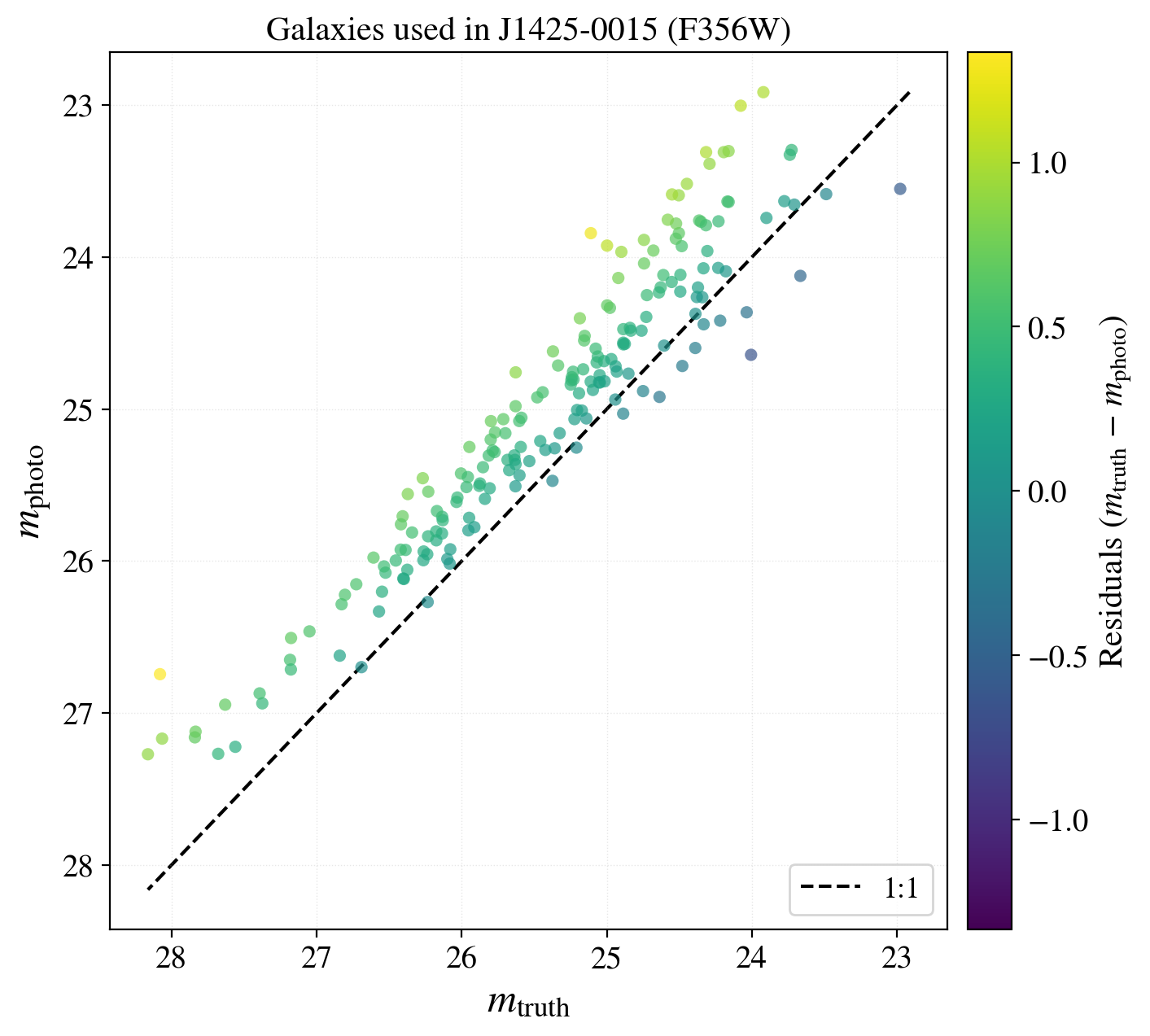}

    \caption{Photometric residuals for all quasar and filter combinations except J0148+0600, which is not possible to consider in this work. Each panel shows the difference between the BlueTides Sérsic galaxy magnitude and the photometric magnitude in the colorbar. We do not add any quasars to these images when calculating these magnitudes.}
    \label{fig:photometric_residuals_panel_page1}
\end{figure*}

\begin{figure*}
    \centering
    \includegraphics[width=0.28\textwidth]{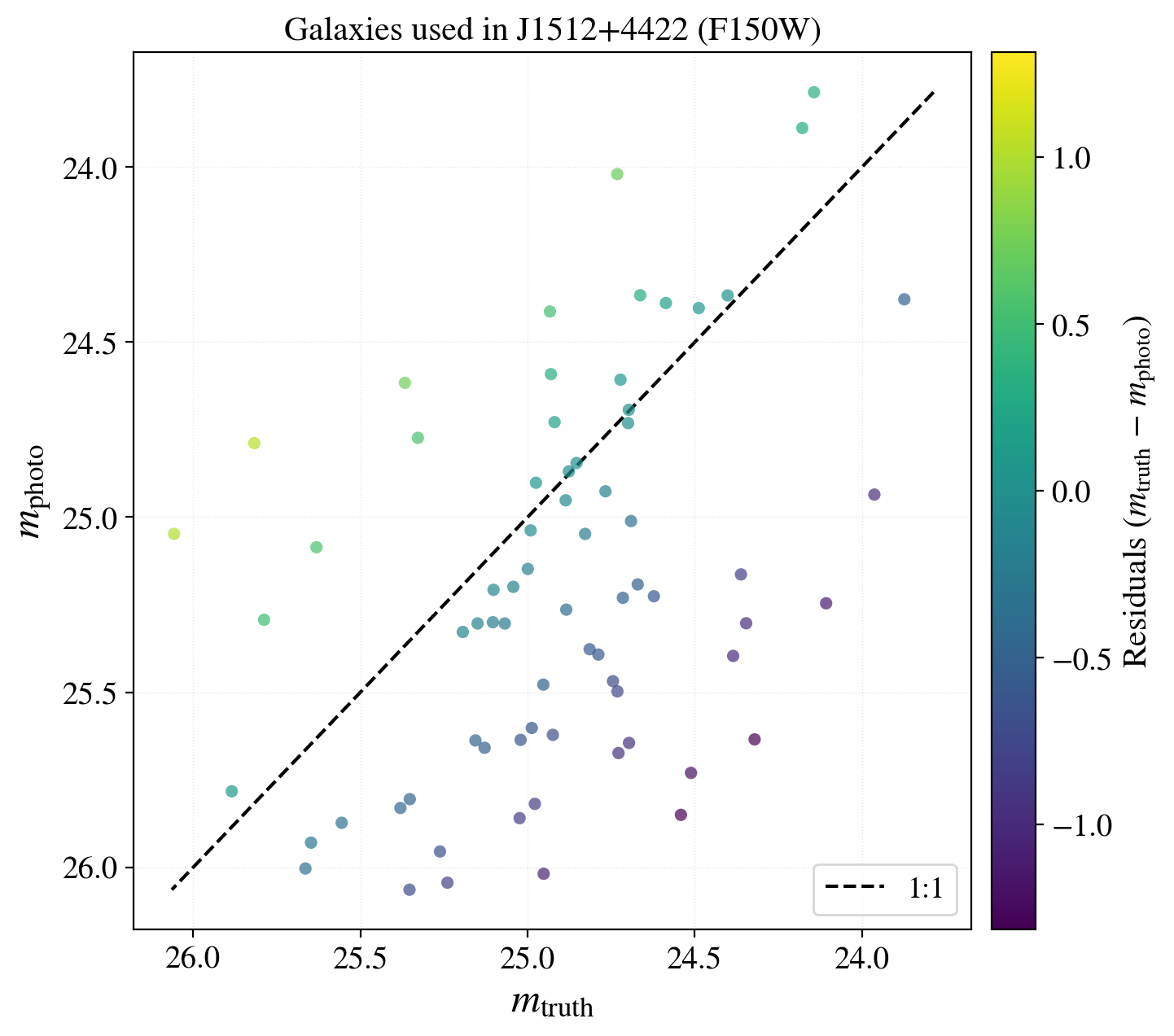}
    \includegraphics[width=0.28\textwidth]{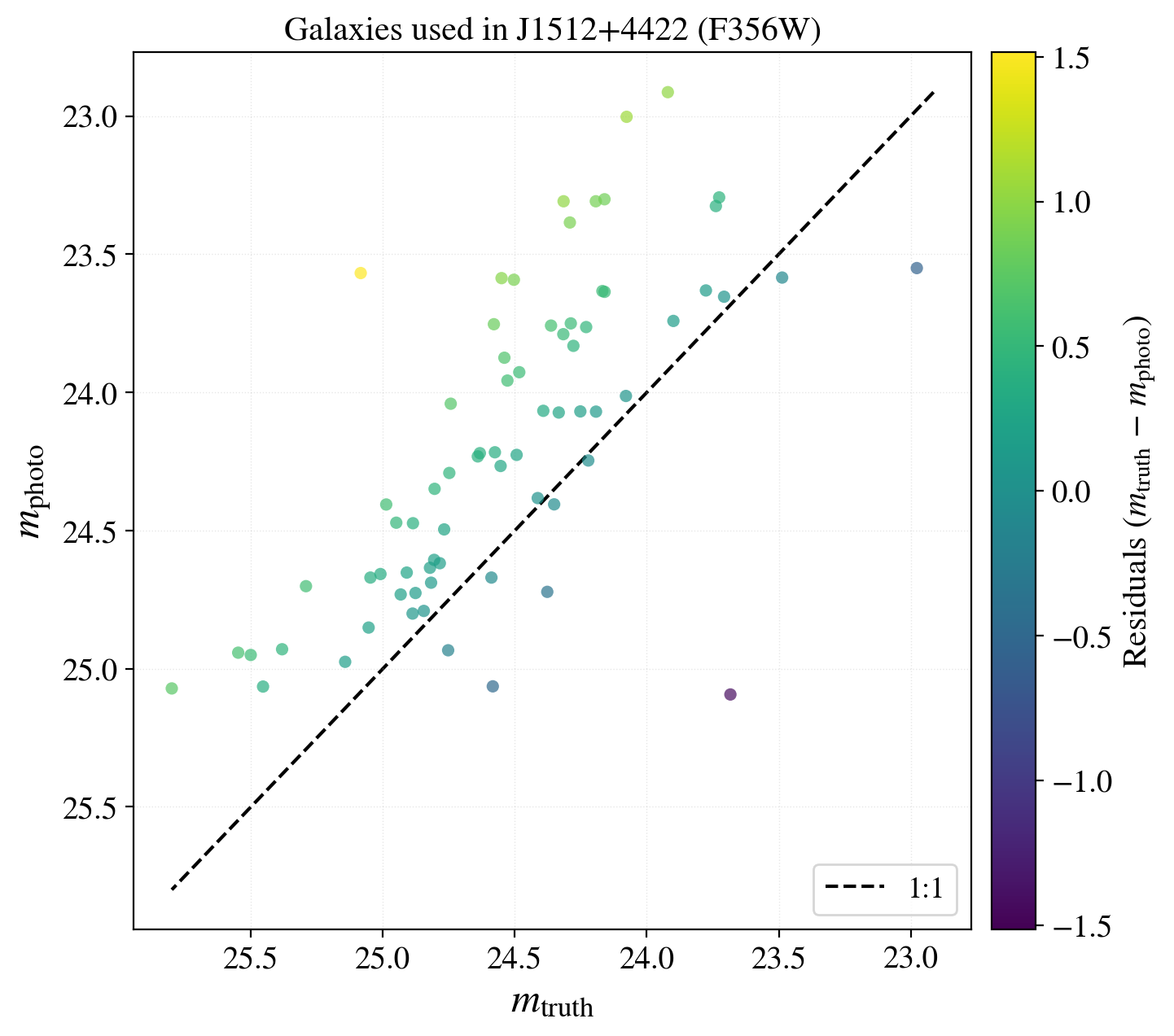}
    \includegraphics[width=0.28\textwidth]{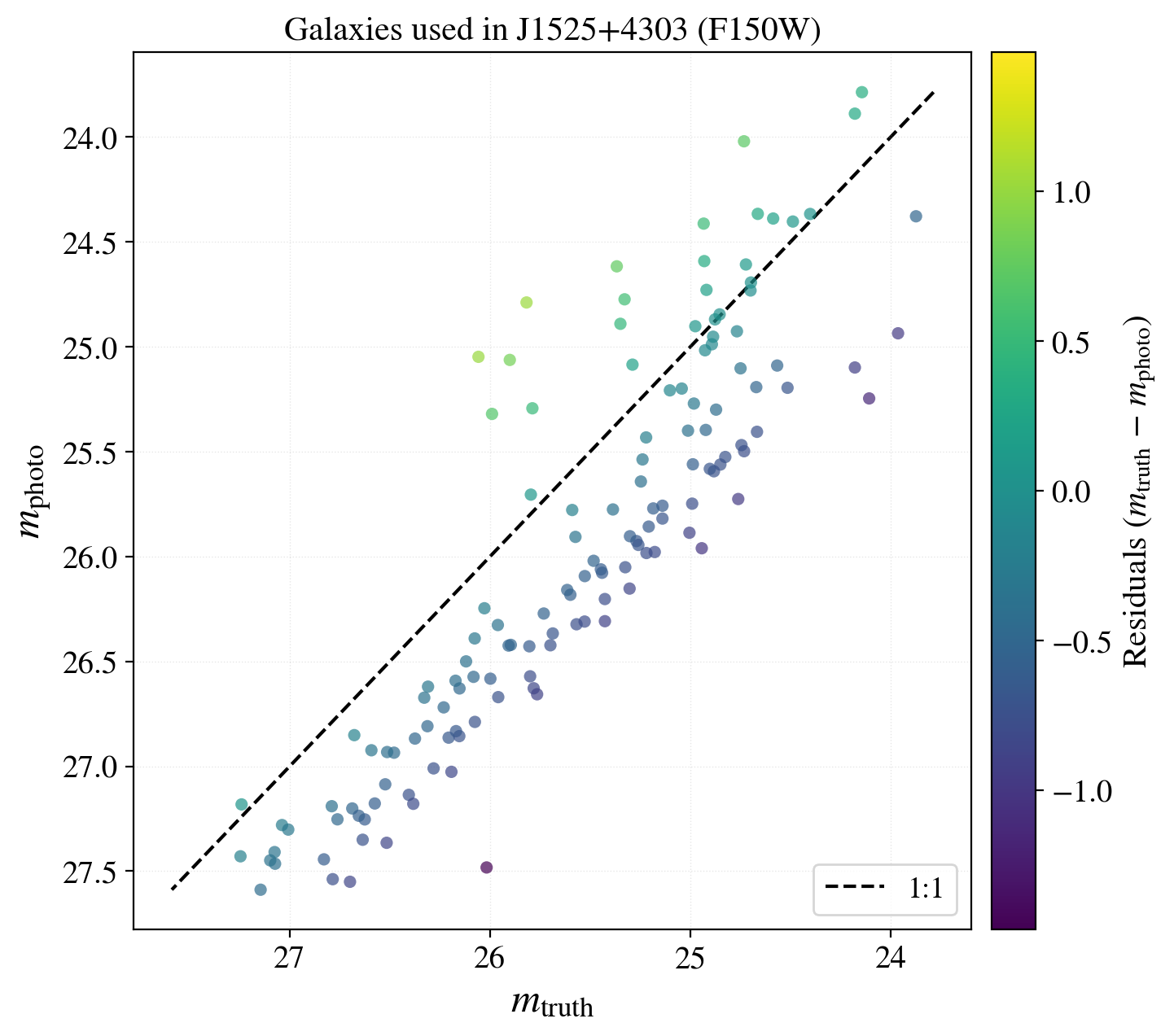}
    \\[0.2cm]
    \includegraphics[width=0.28\textwidth]{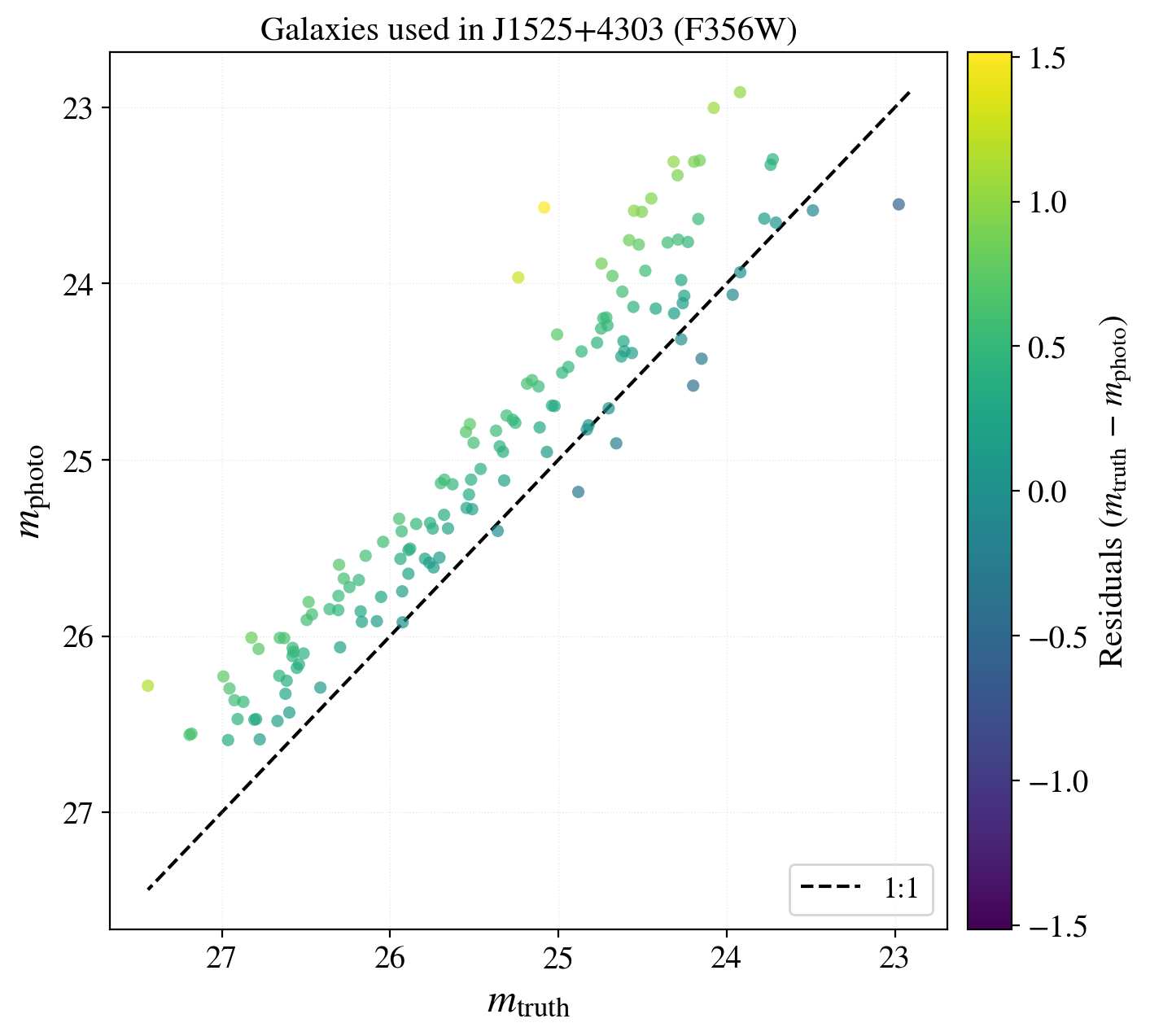}
    \includegraphics[width=0.28\textwidth]{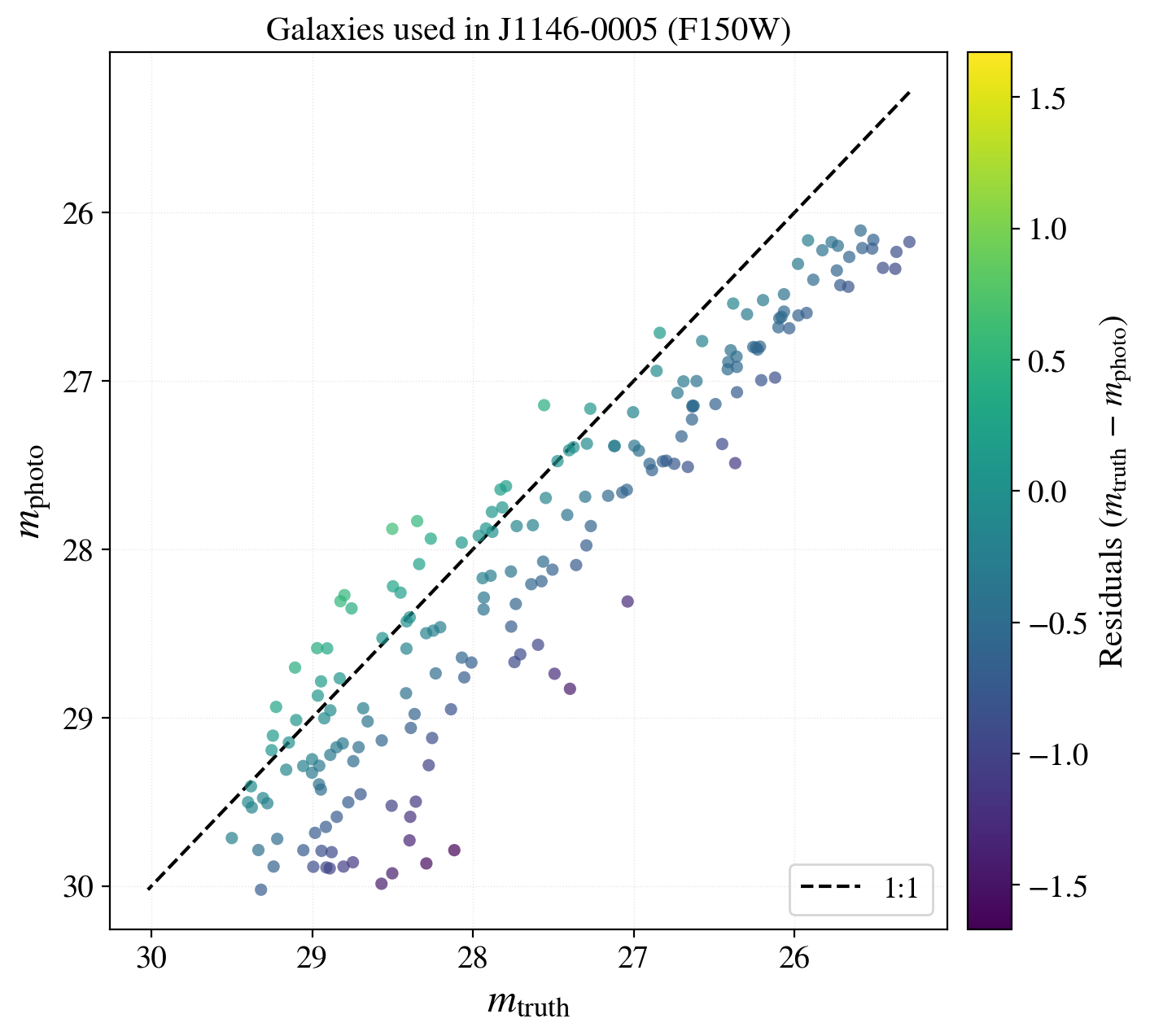}
    \includegraphics[width=0.28\textwidth]{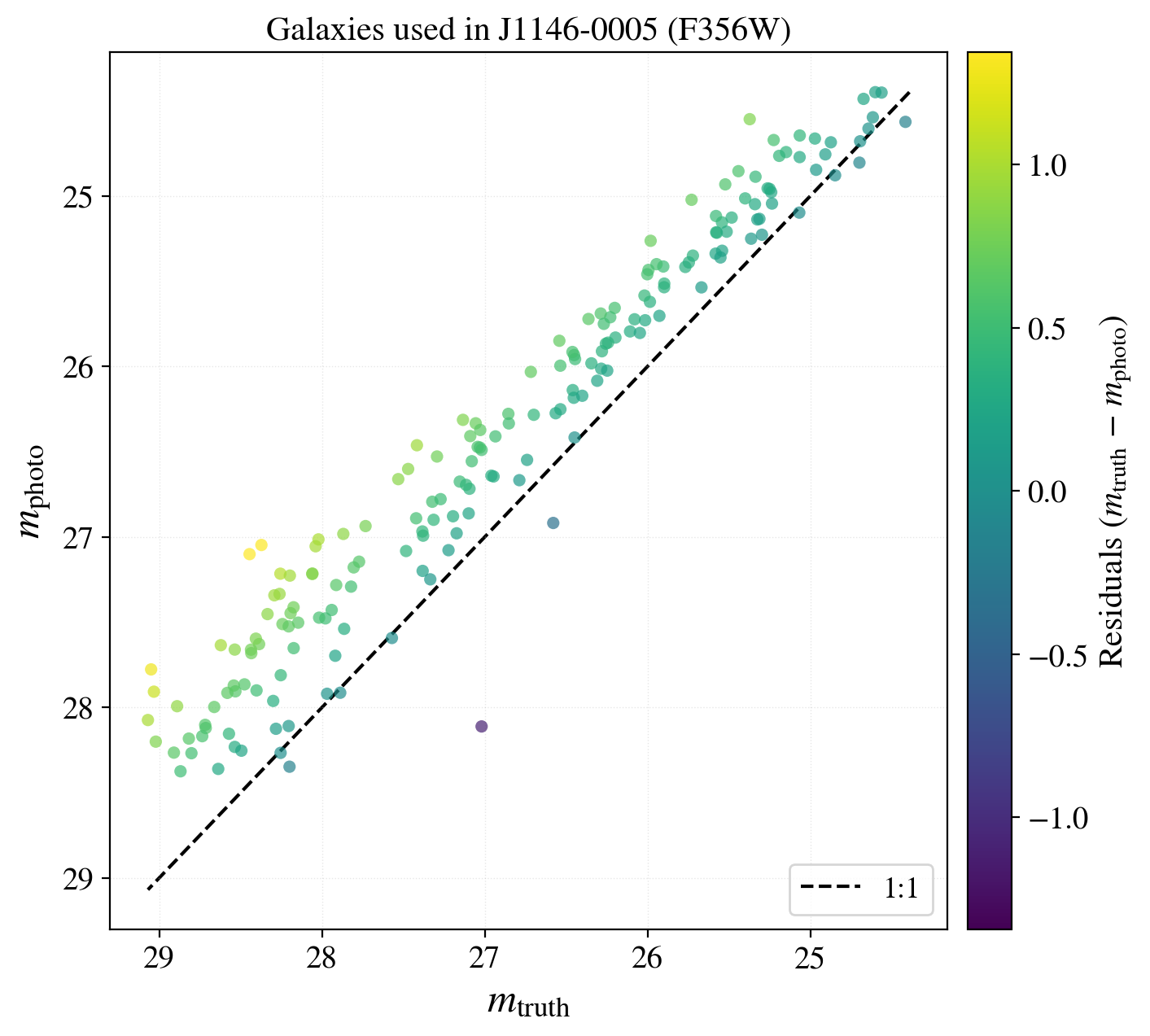}
    \\[0.2cm]
    \includegraphics[width=0.28\textwidth]{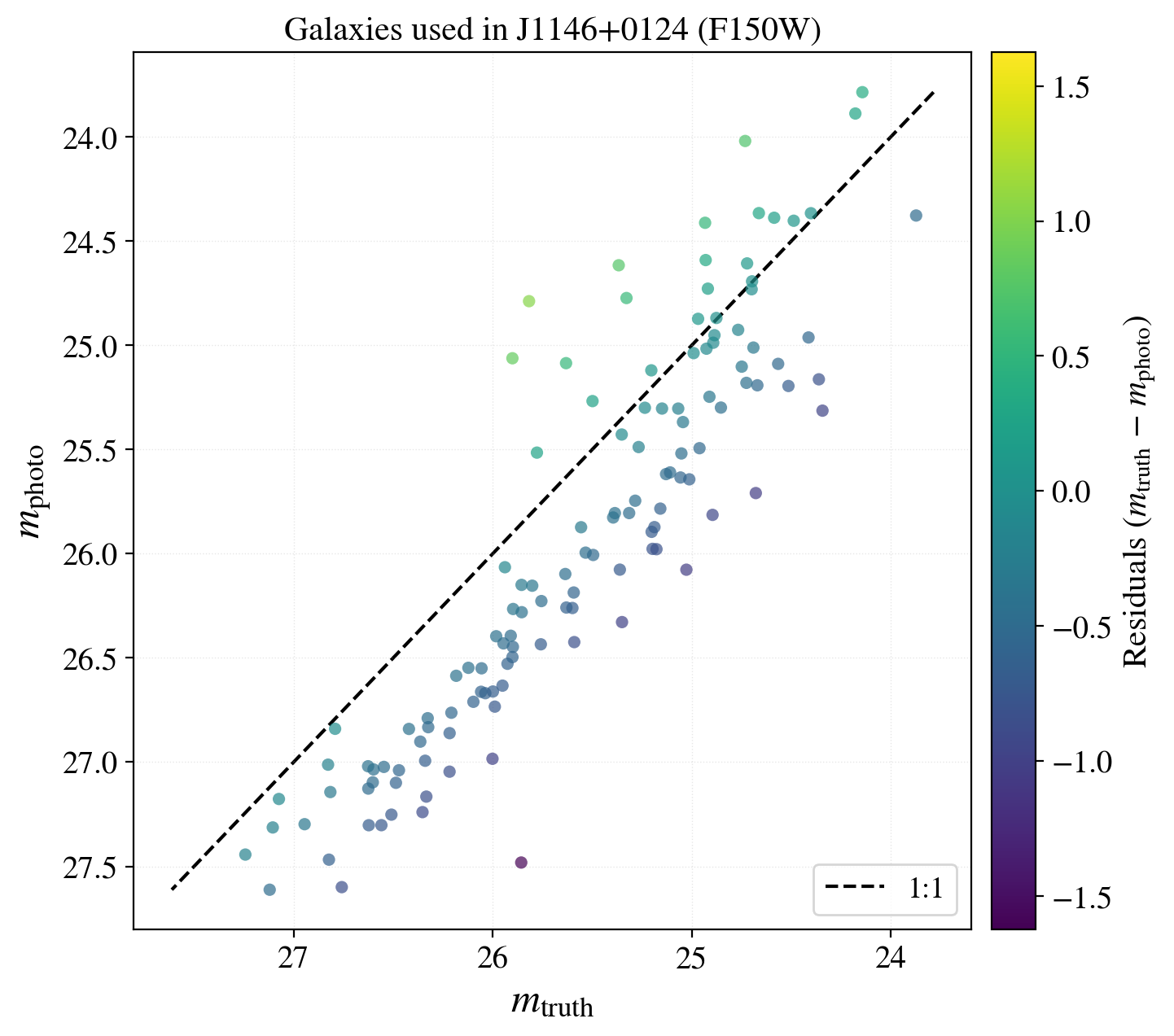}
    \includegraphics[width=0.28\textwidth]{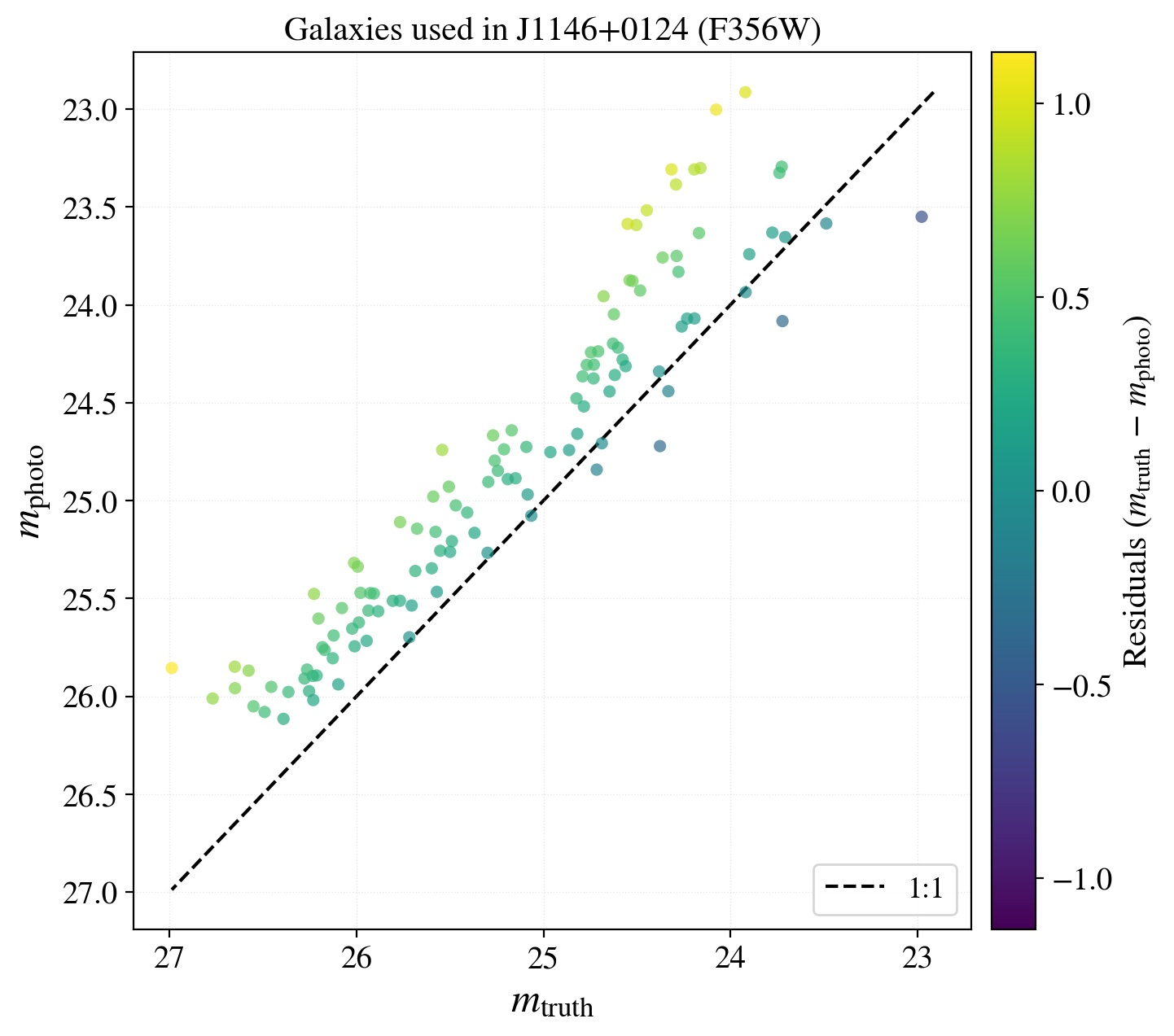}
    \includegraphics[width=0.28\textwidth]{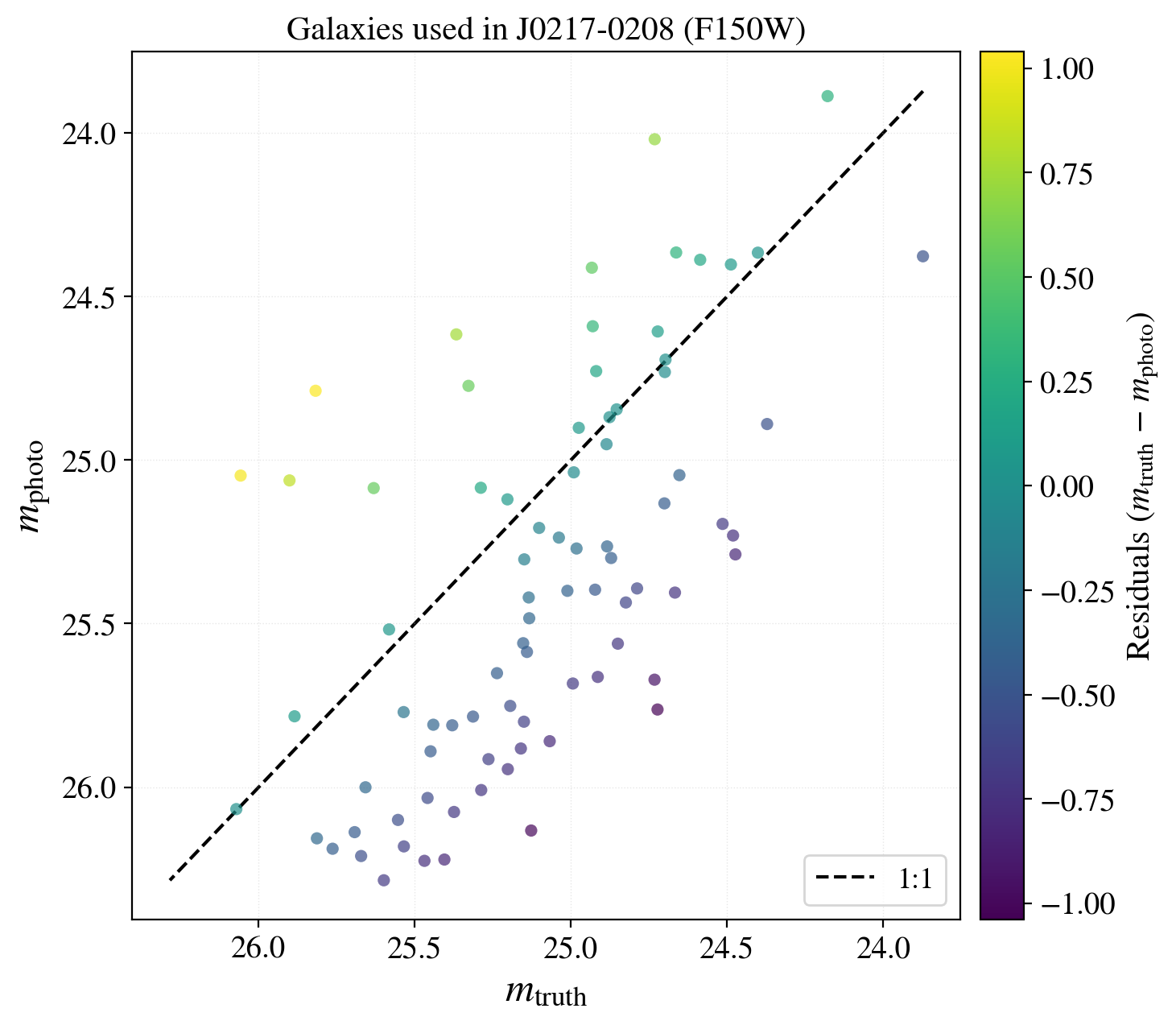}
    \\[0.2cm]
    \includegraphics[width=0.28\textwidth]{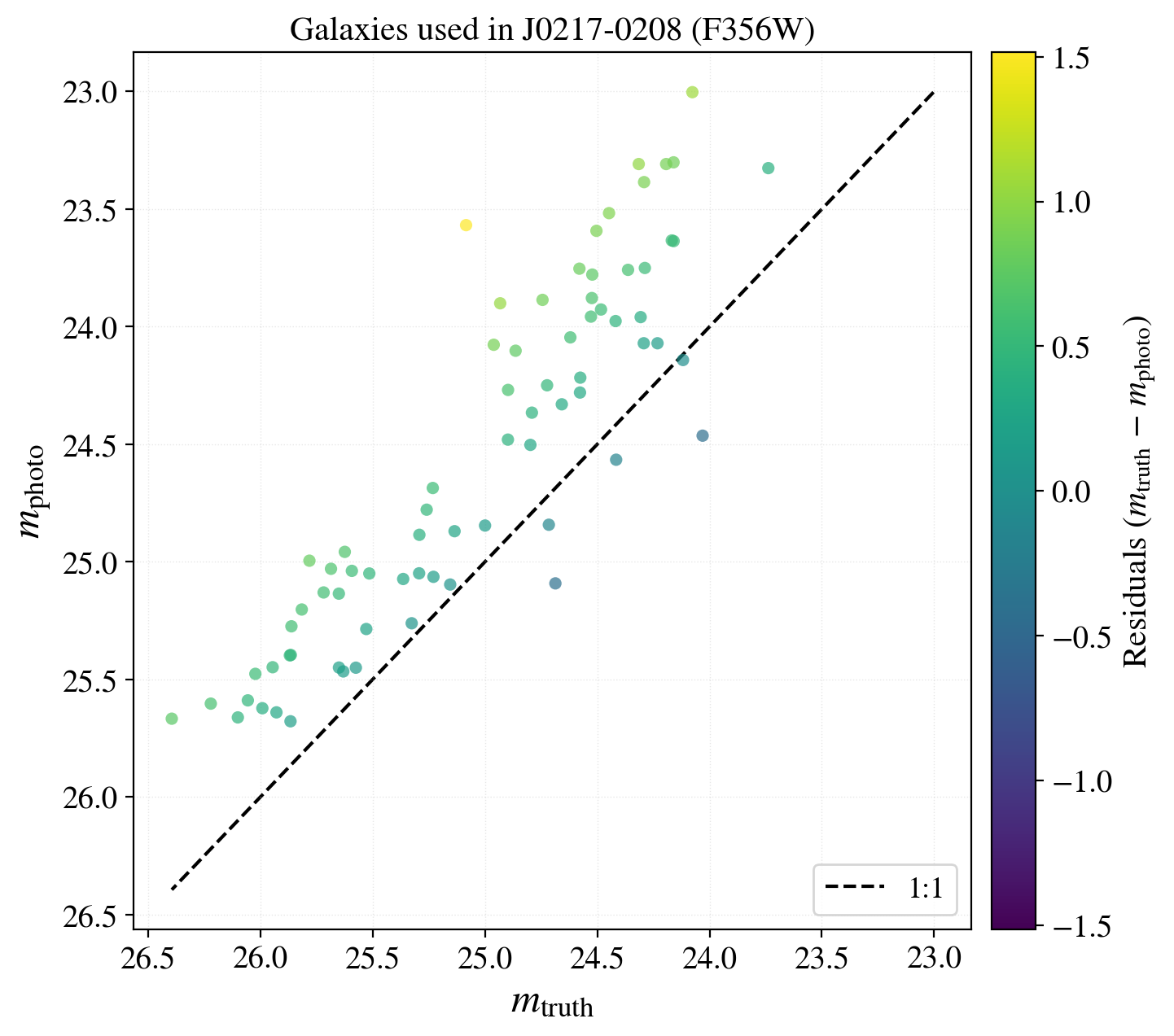}
    \includegraphics[width=0.28\textwidth]{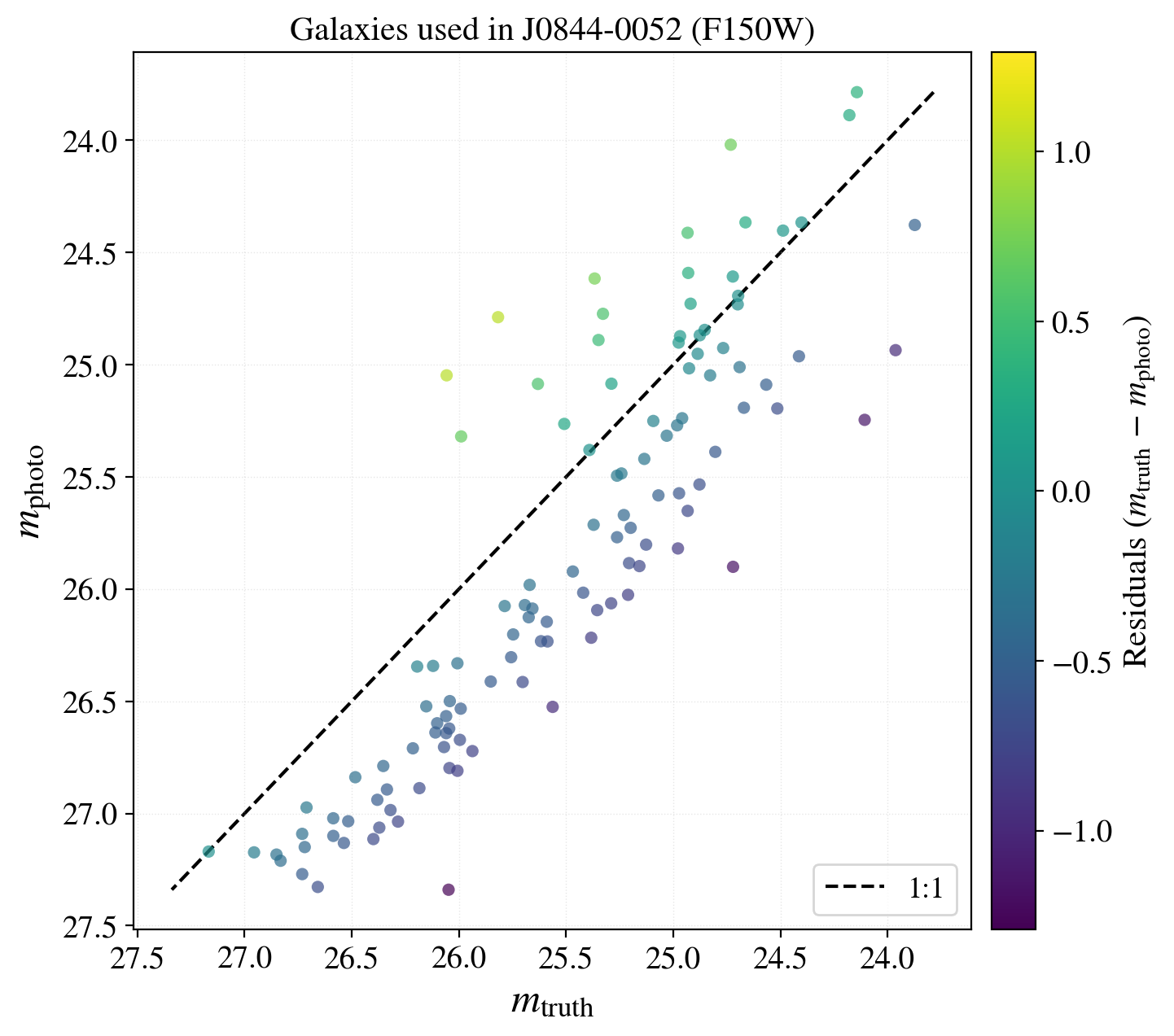}
    \includegraphics[width=0.28\textwidth]{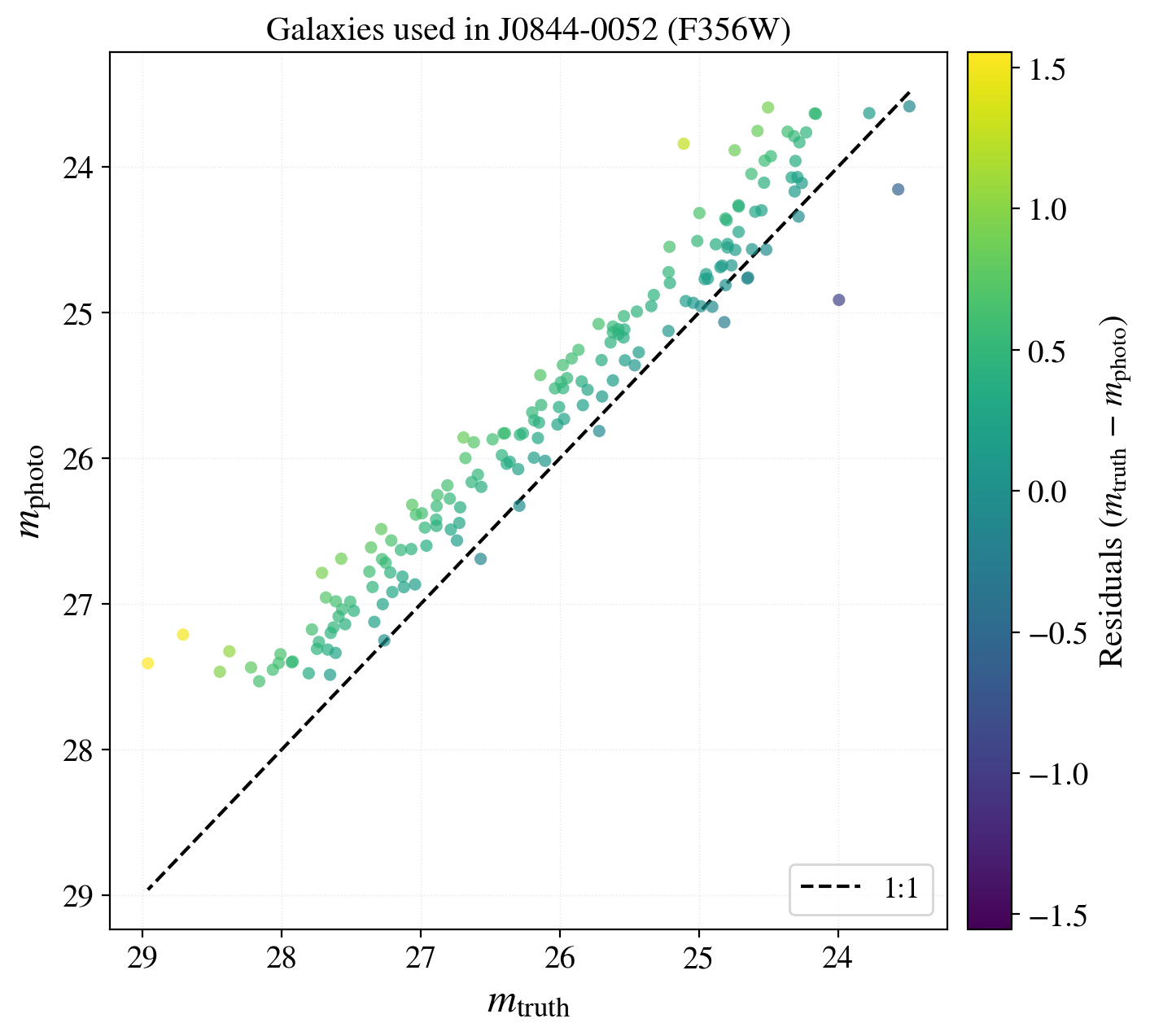}
    \\[0.2cm]
    \includegraphics[width=0.28\textwidth]{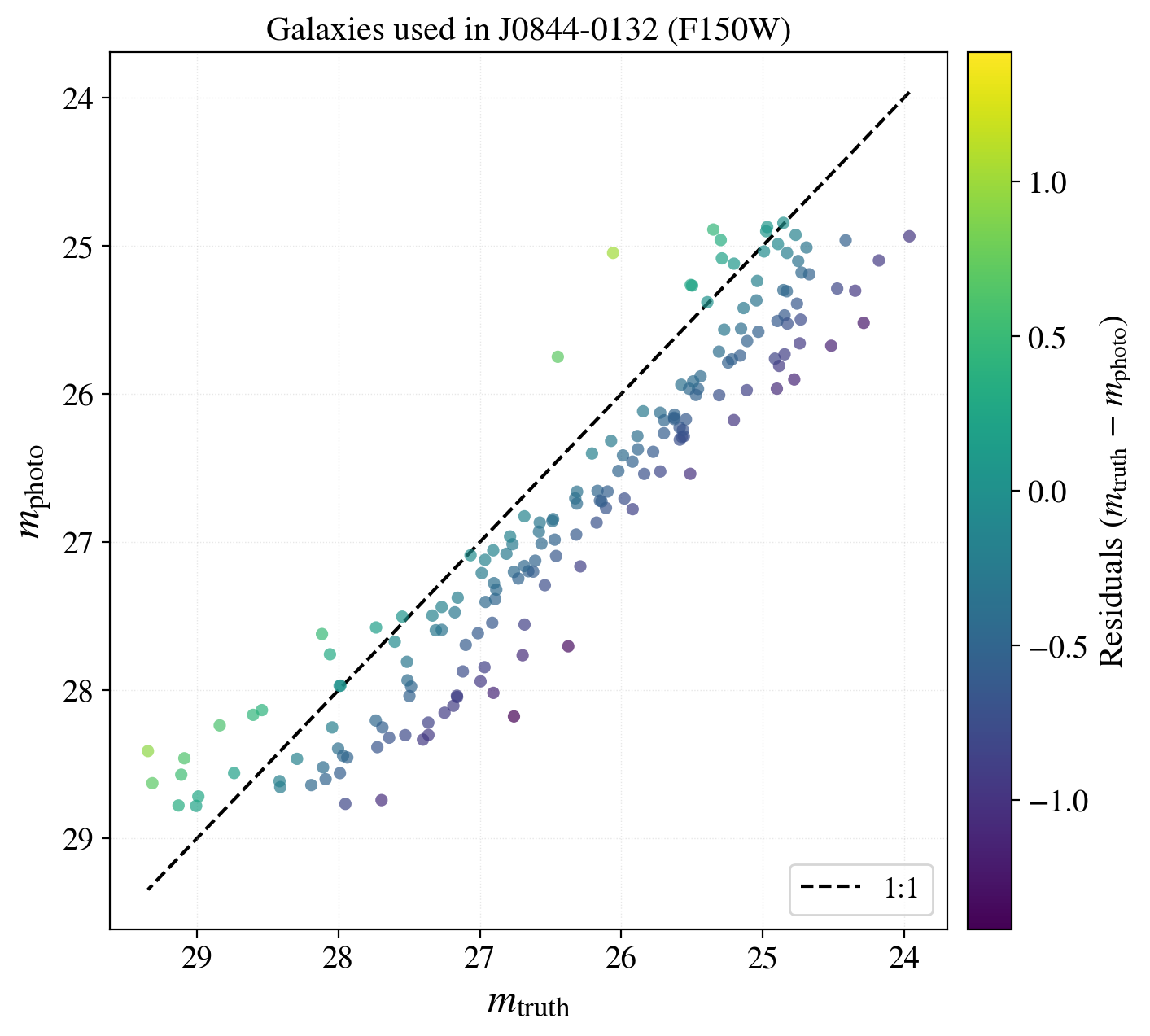}
    \includegraphics[width=0.28\textwidth]{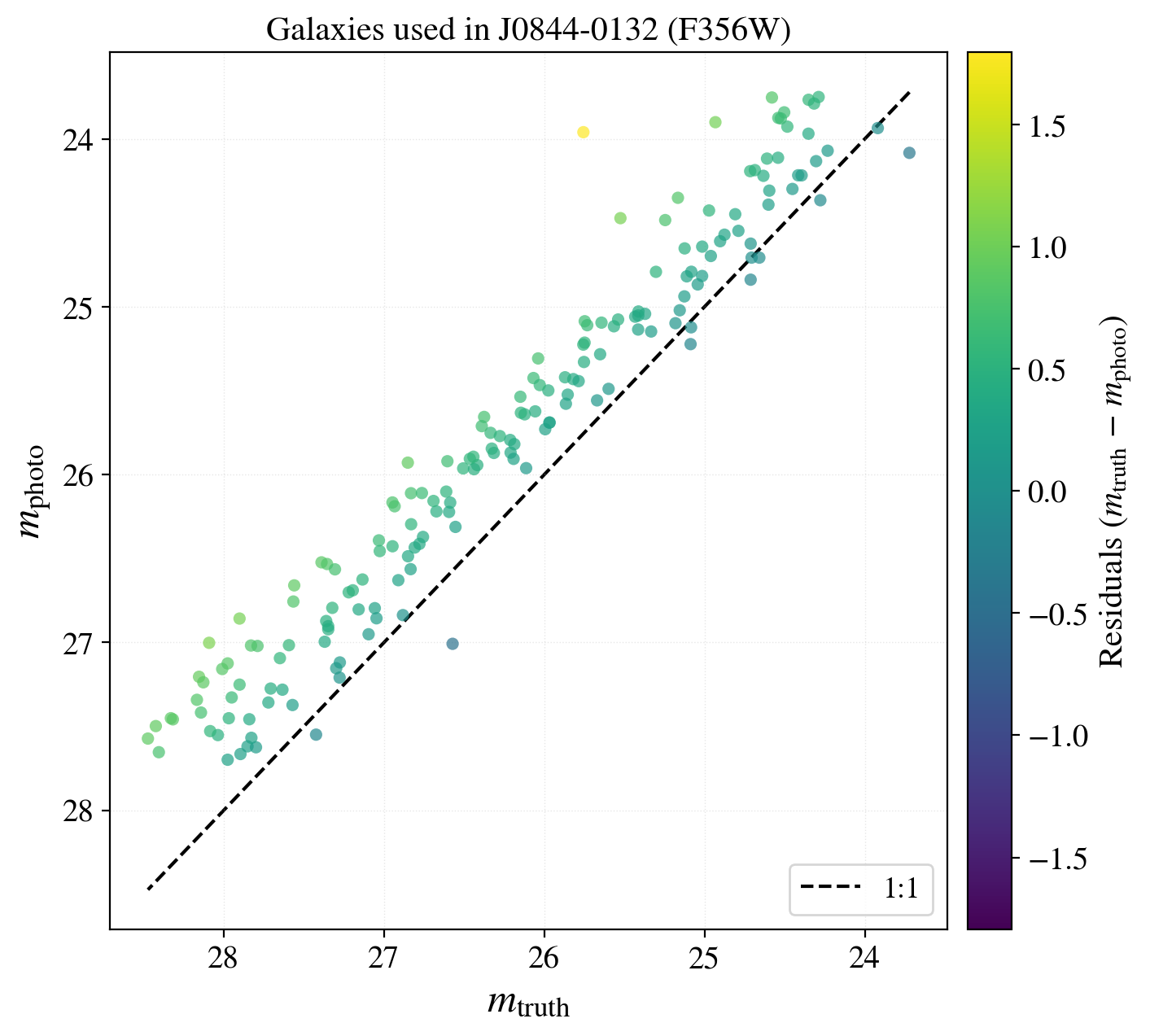}
    \caption{Continuation of photometric residuals. See caption of Figure~\ref{fig:photometric_residuals_panel_page1} for details.}
    \label{fig:photometric_residuals_panel_page2}
\end{figure*}
\label{lastpage}
\end{document}